\documentclass[preprint, aos]{imsart}

\RequirePackage{amsthm,amsmath,amsfonts,amssymb}
\RequirePackage[authoryear]{natbib}
\RequirePackage[colorlinks,citecolor=blue,urlcolor=blue]{hyperref}
\RequirePackage{graphicx}

\RequirePackage{bbm}
\RequirePackage{float}
\RequirePackage {enumitem}
\RequirePackage{subcaption}
\usepackage{todonotes}
\newcommand\independent{\protect\mathpalette{\protect\independenT}{\perp}}
\def\independenT#1#2{\mathrel{\rlap{$#1#2$}\mkern2mu{#1#2}}}

\usepackage{comment}
\RequirePackage{slashbox}

\startlocaldefs
\definecolor{clr1}{RGB}{27,158,119}
\definecolor{clr2}{RGB}{217,95,2}
\definecolor{clr3}{RGB}{117,112,179}
\definecolor{clr4}{RGB}{231,41,138}
\definecolor{clr5}{RGB}{102,166,30}

\theoremstyle{plain}

\newtheorem{theorem}{Theorem}[section]
\newtheorem{lemma}[theorem]{Lemma}
\newtheorem{corollary}[theorem]{Corollary}

\newtheorem{assumption}{Assumption}

\theoremstyle{remark}
\newtheorem{definition}[theorem]{Definition}

\newcommand{\beginsupplement}{%
        \setcounter{table}{0}
        \renewcommand{\thetable}{S\arabic{table}}%
        \setcounter{figure}{0}
        \renewcommand{\thefigure}{S\arabic{figure}}%
        \setcounter{section}{0}
        \renewcommand{\thesection}{S\arabic{section}}
        \setcounter{equation}{0}
        \renewcommand{\theequation}{S\arabic{equation}}
     }

\endlocaldefs

\begin{document}

\begin{frontmatter}
\title{Individual causal effects from observational longitudinal studies with time-varying exposures}
\runtitle{ICE from observational studies with time-varying exposures}

\begin{aug}
\author[A]{\fnms{Richard}~\snm{Post}\ead[label=e1]{r.a.j.post@tue.nl}\orcid{0000-0001-6110-7467}},
\author[A]{\fnms{Zhuozhao}~\snm{Zhan}\ead[label=e2]{z.zhan@tue.nl}\orcid{0000-0001-8364-0983}}
\and
\author[A]{\fnms{Edwin}~\snm{van den Heuvel}\ead[label=e3]{e.r.v.d.heuvel@tue.nl}\orcid{0000-0001-9157-7224}}
\address[A]{Mathematics \& Computer Science,
Eindhoven University of Technology \printead[presep={,\ }]{e1,e2,e3}}

\end{aug}

\begin{abstract}
Causal effects may vary among individuals and can even be of opposite signs. When significant effect heterogeneity exists, the population average causal effect might be uninformative for an individual. Due to the fundamental problem of causality, individual causal effects (ICEs) cannot be retrieved from cross-sectional data. However, in crossover studies, it is accepted that ICEs can be estimated under the assumptions of no carryover effects and time invariance of potential outcomes. A generic potential-outcome formulation with appropriate statistical assumptions to identify ICEs is lacking for other longitudinal data with time-varying exposures. We present a general framework for causal effect heterogeneity in which individual-specific effect modification is parameterized with a latent variable, the receptiveness factor. If the exposure varies over time, then the repeated measurements contain information on an individual's level of this receptiveness factor. Therefore, we study the conditional distribution of the ICE given all an individual's factual information. This novel conditional random variable is called the cross-world causal effect (CWCE). For known causal structures and time-varying exposures, the variability of the CWCE reduces with an increasing number of repeated measurements. The CWCE becomes identifiable from observational data under the causal assumption of cross-world similarity of individual-effect modification (i.e.~there exists an exposure strategy whose effect is affected by all latent causes). We illustrate the theory with examples in which the cause-effect relations can be parameterized as generalized linear mixed assignments.
\end{abstract}


\end{frontmatter}

\section{Introduction}
Estimating the average causal effect (ACE)	from observational data is complicated by all kinds of (unmeasured) confounders. Inference on causal effect heterogeneity is even more complex as it involves the joint distribution of factual and counterfactual outcomes.  In the absence of effect heterogeneity, the ACE equals the individual causal effect (ICE) and is an appropriate measure to evaluate an individual's exposure effect. However, when effect heterogeneity exists, the ACE can deviate significantly from the ICE for some individuals \citep{Kravitz2004, Greenland2019}. For example, when the ICEs have opposite signs, known as `Hand's paradox', the exposure can harm some individuals despite being beneficial on average \citep{Hand1992}. 

The fundamental problem of causal inference is that an individual can only be observed for one level of exposure. In contrast, for other levels, the (counterfactual) outcome of the individual is unknown \citep{Rubin1974, Holland1986}. As a result, 
the ICE is not identifiable from cross-sectional data, where individuals are only exposed to one level \citep{Hernan2004b, Hernan2019}. 
However, when we would collect longitudinal data on individuals under time-varying exposures, we may be able to collect information on different potential outcomes. For instance, in crossover studies, an individual is assigned to a sequence of two or more exposures \citep[Chapter 32]{Rothman2008} and in the absence of both a carryover and period effect, 
the ICE becomes identifiable  
\citep[Fine Point 2.1]{Hernan2019}.
Then, the difference between an individual's repeated measurements equals the ICE. In precision medicine research, there is a growing interest in repeated crossover studies called $n$-of-1 trials \citep{Lillie2011, Duan2013, Kane2021} that may be used to estimate individual effects \citep{Senarathne2020} (although this is not yet common practice \citep{Raman2018}). In the presence of time-varying exposures, we may thus be able to identify ICEs from the repeated measurements of an individual  
under certain assumptions.

The setting of longitudinal data with time-varying exposures or treatments has been studied intensively by James Robins and collaborators \citep[Part~III]{Hernan2019}. They invented the so-called g-methods to make valid marginal causal inferences under the assumption of no direct  unmeasured confounding. These methods have also been used to estimate conditional average treatment effects (CATE) in subpopulations to address potential effect heterogeneity \citep{Murphy2003, Robins2008}. 
However, the estimation of CATEs does not result in the estimation of ICEs when each individual in the subpopulation has its unique causal effect that can still significantly deviate from the CATE. Thus before we can study ICEs in general, it is necessary to quantify the individual-specific effect modification. 

In the field of psychology, it has been suggested to model this individual-specific effect modification using latent variables for designs in which individuals are measured before and after exposure \citep{Steyer2005}. As a result, the authors claim that ICEs can be estimated in the same way as factor scores can be estimated in factor analysis models. However, precise mathematical proofs were lacking, and they did not consider more general designs with repeated measurements, so their idea of modelling ICEs has not received much attention in other fields of causal research. Therefore, we will introduce a general probabilistic framework in which the ICE is defined in terms of latent random individual-specific effect modifiers, which we will refer to as receptiveness factors. In this framework, the longitudinal cause-effect relations for the potential outcomes are first parameterized with a high-level structural causal model (SCM). SCMs are more commonly used the causality literature, but to model the cause-effect relations resulting in the observations \citep{Peters2017}. Subsequently, these relations are reparameterized by including details on measured and unmeasured effect modification at an individual level. Causal effect measures, including the ICE, are defined in terms of the SCM. This new framework is illustrated  with an example in which a mixed (i.e.~multilevel) model describes the data-generating process.

From a population point of view, an arbitrary individual's ICE is considered an unknown random variable, and the conditional distribution of the ICE given the individual's information can be studied. This conditional random variable, which we refer to as the cross-world causal effect (CWCE), is at the core of this chapter. The distribution of the CWCE describes the range of possible or likely values for the ICE, given all an individual's factual observations. The lower the variability of the CWCE, the better we can predict an individual's ICE. We will demonstrate that the CWCE is degenerate in an `ideal' crossover study, where the potential outcome under no exposure and causal effects are time-invariant, and there is no carryover effect. Still, in most examples, the CWCE is not degenerate. The number of repeats and the levels of the time-varying exposure experienced by an individual affect the variability of the CWCE.

In practice, the functional forms and parameters in the SCM are unknown. Thus, due to the fundamental problem of causal inference, the CWCE is not identifiable from observational data without making an additional assumption. We present a cross-world similarity of individual-effect modification assumption, which implies that the number of receptiveness factors is limited such that there exists an exposure strategy for which the potential outcomes are affected by all receptiveness factors. Then, the joint distribution of unmeasured individual features can be derived from observations exposed to that strategy. We will discuss why the validity of the cross-world assumption can be ruled out or be partly supported using data. In the latter case, the assumption can be simplified and discussed with experts in de field. The individual-specific effect modification should be modelled to estimate the CWCE distribution. Individual-specific models, such as mixed models, are commonly used to model individual-specific associations. Empirical Best Linear Unbiased Predictors (EBLUPs) are used to predict individual random effects \citep{Verbeke2000}. Our framework and theorems help to understand under which assumptions the mixed models and EBLUPs could also be used to make inferences on the ICEs. This knowledge is relevant, as recently, Gaussian linear mixed models (LMMs) have been proposed for causal prediction of the person-specific effects in micro-randomized trials \citep{Qian2020} and as a solution to unmeasured direct confounding \citep{Shardell2018}.

The framework for individual causality in longitudinal processes with time-varying exposures is presented in Section \ref{CH4S2}. In Section \ref{CH4CWIT}, the distribution of the CWCE is studied for cause-effect relations for which the SCM is known. In Section \ref{CH4TVC}, we explain how individual-specific effect modification results in time-varying confounding to show that our general framework aligns with the elaborate literature on marginal causal inference from longitudinal data with time-varying exposures. Finally, the identifiability of the CWCE from observational data and individualized inference, when the parameters in the SCM are unknown, are discussed in Section \ref{CH4II}.

\section{Notation and setting of interest}\label{CH4S2}
Before we can study the ICE, we will formalize the longitudinal cause-effect relations under time-varying exposures and the individual-specific effect modification. We consider the cause-effect relations between the exposure (or treatment) strategy, $A_{ji}$, and the outcome process, $Y_{ji}$, given the (possibly time-varying) confounders, $\boldsymbol{L}_{ji}$, at time $j$ ($j~{=}~1,2, \hdots, h$) for individual $i$ ($i~{=}~1, 2, \hdots, n)$. Probability distributions of factual and counterfactual outcomes can be defined in terms of the potential-outcome framework \citep{Neyman1990, Rubin1974} and the do-calculus \citep{Pearl1995}, respectively. In this work, the potential-outcome framework will be used as the focus is on the joint distribution of potential outcomes. The potential outcome of an individual $i$ at time $j$ in the universe where everyone is assigned to exposure process $\overline{a}_{j-1}$, equal to $(a_{1}, a_{2}, \hdots, a_{j-1})$, is referred to as $Y_{ji}^{\overline{a}}$. We focus on binary exposures, i.e.~$a_{j} \in \{0, 1\}$ where zero indicates the absence of exposure. The observed exposure assignment at time $j$, $A_{ji}$, can be caused by the observed outcomes up and until the time of the assignment, $\overline{Y}_{ji}$, equal to $(Y_{1i}, Y_{2i}, \hdots, Y_{ji})$, the previous exposure assignments $\overline{A}_{j-1,i}$ and features that do also cause the next $Y_{j+1,i}$, i.e.~confounders, $\overline{\boldsymbol{L}}_{ji}$ equal to $(\boldsymbol{L}_{1i}, \boldsymbol{L}_{2i}, \hdots, \boldsymbol{L}_{ji})$. Individuals with similar levels of $\overline{Y}_{j}$, $\overline{A}_{j-1}$ and $\overline{\boldsymbol{L}}_{j}$ might still be assigned to different levels of exposure at time $j$, the random variable $N_{Aj}$ will be used to represent these differences. The potential outcome under no exposure, $Y^{\overline{0}}_{ji}$, is affected by $\overline{\boldsymbol{L}}_{j-1,i}$ but can again be different for individuals with similar levels of these confounders as is represented with the random variables $\overline{N}_{Yj}$, equal to $(N_{Y1i}, N_{Y2i}, \hdots, N_{Yji})$, possibly mediated via $\overline{Y}^{\overline{0}}_{j-1,i}$. The potential outcome after the exposure strategy $\overline{a}_{j-1}$, $Y^{\overline{a}}_{ji}$, equals the sum of $Y^{\overline{0}}_{ji}$ and an ICE of the strategy $\overline{a}_{j-1}$ represented by $N_{\overline{a}_{j-1}i}$. Finally, the potential outcome of the exposure assignment after exposure strategy $\overline{a}_{j-1}$, $A^{\overline{a}}_{ji}$, is caused by the exposure, $Y^{\overline{a}}_{ji}$ and $\overline{\boldsymbol{L}}_{ji}$. 

\newpage The causal relations between $\overline{a}_{j-1}$,  $Y^{\overline{a}}_{j}$ and $A^{\overline{a}}_{j}$, including the relation between the observed $\overline{A}_{j-1}$ and $\overline{Y}_{j}$, can be parameterized with a collection of structural assignments $\boldsymbol{f}$ and the probability distribution of all $\boldsymbol{N}$, i.e.~the structural causal model (SCM) $\mathfrak{C}(\{\overline{A}, \overline{\boldsymbol{L}}, \overline{Y}\},\boldsymbol{f}^{*},P_{\boldsymbol{N}^{*}})$\footnote{An SCM as presented in this work is a union of the traditional SCM, for $\overline{a}~{=}~\overline{A}$, and the intervened SCMs for all possible $do(\overline{A}~{=}~\overline{a})$ \citep{Peters2017}. This union forms an example of an intervened multi SCM underlying a multi-network \citep{Shpitser2007} that generalizes the twin-network as proposed in \citet{Balke1994} and is formally defined in \citet{Bongers2021}.}:
\begin{center}
	\fbox{%
		\parbox{0.9\linewidth}{%
			\begin{align}\label{CH4margSCM0}
			Y_{1i}&~{:}{=}~f^{*}_{Y_{1}}(N_{Y1i}^{*}) \\\nonumber
			\boldsymbol{L}_{1i}&~{:}{=}~f^{*}_{L_{1}}(\boldsymbol{N}_{L1i}^{*}) \\ \nonumber
			A_{1i}&~{:}{=}~f^{*}_{A_{1}}(Y_{1i},\boldsymbol{L}_{1i},N_{A1i}^{*})   \\\nonumber
			Y_{ji}^{\overline{0}}&~{:}{=}~f^{*}_{Y_{j}}(\overline{\boldsymbol{L}}_{j-1,i},\overline{N}_{Yji}^{*}) \\\nonumber
			Y_{ji}^{\overline{a}}&~{:}{=}~Y_{hi}^{\overline{0}} + N_{\overline{a}_{j-1}i}^{*} \\\nonumber
			\boldsymbol{L}_{ji}&~{:}{=}~f^{*}_{L_{j}}(\overline{\boldsymbol{L}}_{j-1,i},N_{Lji}^{*})  \\ 	\nonumber		A_{ji}^{\overline{a}}&~{:}{=}~f^{*}_{A_{j}}(\overline{\boldsymbol{L}}_{ji},\overline{Y}_{ji}^{\overline{a}},\overline{a}_{j-1},N_{Aji}^{*}).
			\end{align} }} \end{center} Notice that the random variables $\boldsymbol{N}^{*}$ represent all stochasticity in the system, for example $N^{*}_{\overline{a}_{j-1},i}$ is the ICE of the exposure strategy $\overline{a}_{j-1}$ for individual $i$, and is not just noise as it often represents in traditional SCMs \citep{Pearl2009book, Peters2017}. We have used $\boldsymbol{N}^{*}$ since we will use the notation $\boldsymbol{N}$ in our final parameterization. Note that we omitted $Y^{\overline{0}}_{j-1,i}$ as a cause of $Y^{\overline{0}}_{ji}$, since we don't focus on mediators in this work, but instead allow  $N_{Y_{k}i}^{*}$ and $\boldsymbol{L}_{k}$ to be a cause of $Y^{\overline{0}}_{ji}$ for all $k{<}j$. Throughout this work, we assume that the observed outcome of an individual equals the potential outcome for that exposure assignment, referred to as causal consistency \citep{Hernan2004b}.  
\begin{assumption}{\textbf{Causal consistency}}\label{CH4A2}
	$$Y_{ji}^{\overline{a}}~~{=}~~Y_{ji}{\mid}\overline{A}_{j-1,i}{=}\overline{a}_{j-1}$$
\end{assumption} \noindent By causal consistency, the data-generating mechanism is described with the SCM since $Y_{ji}~{=}~Y_{ji}^{\overline{A}_{ji}}$ and $A_{ji}~{=}~A_{ji}^{\overline{A}_{j-1,i}}$. 

SCM \eqref{CH4margSCM0} is a saturated model as the structural assignments $\boldsymbol{f}^{*}$ are not restricted, and the random variables $\boldsymbol{N}^{*}$ can be dependent. However, in conventional models, the $\boldsymbol{N}^{*}$ are often limited to be independent. Otherwise, direct unmeasured confounding might be present as defined next \citep{ Pearl2009book, Peters2017}.
\begin{definition}{\textbf{Unmeasured confounding}}\label{CH4directconf}
	There is unmeasured confounding of the effect of $\overline{A}_{j-1}$ on the outcome $\overline{Y}_{j}$ when $\exists \overline{a}{:}~\overline{A}_{j-1} \not\independent \overline{Y}_{j}^{\overline{a}} \mid \overline{\boldsymbol{L}}_{j-1}$. In SCM \eqref{CH4margSCM0} unmeasured confounding is called indirect when there is dependency among $\left(\overline{N}^{*}_{Y}, \boldsymbol{N}^{*}_{\overline{a}}\right)$, and direct when $\overline{N}^{*}_{A} \not\independent \left(\overline{N}^{*}_{Y}, \boldsymbol{N}^{*}_{\overline{a}}\right)$. 
\end{definition} \noindent In this work, we assume that the confounding process $\overline{\boldsymbol{L}}_{j}$ is fully observed so that direct unmeasured confounding is absent. This assumption has also been referred to as sequential conditional exchangeability \citep{Hernan2019}. 
\begin{assumption}{\textbf{Sequential conditional exchangeability}}\label{CH4A1}
	$$\forall \overline{a}{:}~A_{j-1} \independent Y_{j}^{\overline{a}} \mid A_{j-2}{=}\overline{a}_{j-2}, \overline{\boldsymbol{L}}_{j-1}, \overline{Y}_{j-1}$$
\end{assumption} \noindent In fact, sequential conditional exchangeability extends absence of direct unmeasured confounding as it also requires that $Y_{1}$ is observed before any exposure assignment or that $Y_{1}$ is caused by an exposure $A_{0}$ that is randomly assigned (otherwise $Y_{1}^{a_{0}} \not \independent A_{0}$). 

In practice, it is unreasonable to assume that the elements of $\left(\overline{N}^{*}_{Y}, \boldsymbol{N}^{*}_{\overline{a}}\right)$ are independent as there will be features of an individual that are (unmeasured) common causes of the outcome process. In the literature, this dependence is often represented by latent factors $\boldsymbol{U}$ that represent the individual's level of baseline features that influence the outcome process without directly affecting the exposure assignment \citep[Chapter 19]{Hernan2019}. In a population-averaged analysis, these latent factors complicate the analysis (as will be the topic of Section \ref{CH4TVC}) since they are also indirect unmeasured confounders \citep{Hernan2019}. Part of the $\boldsymbol{U}$, say $\boldsymbol{U}_{0}$ can represent
features that cause $\overline{Y}^{\overline{0}}$ of an individual to deviate in a particular way from the population mean, $\mathbb{E}[$\scalebox{0.8}{$\overline{Y}^{\overline{0}}$}$]$ and result in dependence among $\overline{N}_{Y}^{*}$. Also, the individual causal effects $(\boldsymbol{N}_{\overline{a}}^{*})$ of different exposure strategies will be dependent as a result of individual features. Such features that affect the exposure effects are more commonly known as modifiers. 
\begin{definition}{\textbf{Effect modification}}\label{CH4mod}
	A single factor $M$ modifies the effect of exposure $\overline{A}$ on outcome $\overline{Y}$~when there exists a level of the exposure $\overline{a}$ and a time point $j$ such that 
	\begin{equation*}
	Y_{j}^{\overline{a}}-Y_{j}^{\overline{0}} \not \independent M.
	\end{equation*}
\end{definition} \noindent Modification in the sense of Defintion \ref{CH4mod} can be induced by a continuous (e.g.~age) or categorical (e.g.~gender) factor that may or may not be observed.\footnote{Note that a variable that affects the causal effect distribution but does not affect the expected causal effect is a modifier in the sense of Defintion \ref{CH4mod}. Therefore, this definition extends the commonly used definition of a modifier where a variable $M$ is a modifier when there exists a level of the exposure $\overline{a}$ and different levels $m_{1}, m_{2}$ of $M$ such that \begin{equation*}
\exists j:\mathbb{E}\left[Y_{j}^{\overline{a}}-Y_{j}^{\overline{0}}\mid M{=}m_{1} \right] \neq \mathbb{E}\left[Y_{j}^{\overline{a}}-Y_{j}^{\overline{0}}\mid M{=}m_{2}\right], 
\end{equation*} see, e.g.~\citet[Appendix 9.2]{VanderWeele2016}.}

While some modifiers are measured, there will also be individual features that affect the causal effect but are not observed. To emphasize the difference between measured and unmeasured modifiers, we will use the word modifier for observed modifiers, $\boldsymbol{M}$, and refer to latent modifiers as receptiveness factors, $\boldsymbol{U}_{\text{AY}}$. Both result in individual-specific effect modification since causal effects will differ for individuals with different levels of $\boldsymbol{M}$, and the effects of individuals with similar levels of $\boldsymbol{M}$ can still differ as a result of $\boldsymbol{U}_{\text{AY}}$. The latter remaining variability can, in turn, also depend on $\boldsymbol{M}$ and is captured by the distribution of $\boldsymbol{U}_{\text{AY}}~{\mid}~\boldsymbol{M}{=}\boldsymbol{m}$. Moreover, as confounders, modifiers may be time-varying, and $\overline{\boldsymbol{M}}_{j}$ refers to the levels of the modifiers until time $j$. To accommodate these individual features, the cause-effect relations in the longitudinal process can be reparameterized with the SCM $\mathfrak{C}(\{\overline{A}, \overline{\boldsymbol{L}}, \overline{Y}, \overline{\boldsymbol{M}}, \boldsymbol{U}_{0}, \boldsymbol{U}_\text{AY}\},\boldsymbol{f}, P_{\boldsymbol{N}})$ that equals 
\begin{center}
	\fbox{%
		\parbox{0.9\linewidth}{%
			\begin{align}\label{CH4indSCM}
			\boldsymbol{U}_{0i}&~{:}{=}~f_{U}(\boldsymbol{N}_{0i}) \\ \nonumber
			\boldsymbol{M}_{1i}&~{:}{=}~{f}_{{M}}(\boldsymbol{N}_{M1i})\\ \nonumber
			Y_{1i}&~{:}{=}~f_{Y_{1}}(\boldsymbol{U}_{0i},N_{Y1i}) \\\nonumber
			\boldsymbol{L}_{1i}&~{:}{=}~f_{L_{1}}(\boldsymbol{N}_{L1i})  \\ \nonumber
			A_{1i}&~{:}{=}~f_{A_{1}}(Y_{1i},\boldsymbol{L}_{1i},N_{A1i})   \\\nonumber
			\boldsymbol{U}_{\text{AY}i}&~{:}{=}~f_{{U}_{\text{AY}}}(\boldsymbol{N}_{{{U}_{\text{AY}}}i})\\ \nonumber
			Y_{ji}^{\overline{a}}&~{:}{=}~f_{Y_{j}}(\boldsymbol{U}_{0i},\boldsymbol{U}_{\text{AY}i},\overline{\boldsymbol{M}}_{j-1,i},   \overline{a}_{j-1},\overline{\boldsymbol{L}}_{j-1,i},\overline{N}_{Yji}) \\\nonumber
             \boldsymbol{M}_{ji}&~{:}{=}~f_{M_{j}}(\overline{\boldsymbol{M}}_{j-1,i},\boldsymbol{N}_{Mji})  \\ 	\nonumber
			\boldsymbol{L}_{ji}&~{:}{=}~f_{L_{j}}(\overline{\boldsymbol{L}}_{j-1,i},\boldsymbol{N}_{Lji})  \\ 	\nonumber	
			A_{ji}^{\overline{a}} &~{:}{=}~f_{A_{j}}(\overline{\boldsymbol{L}}_{ji},\overline{Y}_{ji}^{\overline{a}},\overline{a}_{j-1},N_{Aji}),
			\end{align} 
		}%
} \end{center} 

such that the elements of $\boldsymbol{N}_{Y}$ are mutually independent. 
SCM \eqref{CH4margSCM0} can always be parameterized as SCM \eqref{CH4indSCM}, for $\boldsymbol{U}_{\text{AY}} ~{=}~ \boldsymbol{N}^{*}_{\overline{a}}$ and $\boldsymbol{U}_{0} ~{=}~ \boldsymbol{N}^{*}_{Y}$. However, in the parameterization as SCM \eqref{CH4indSCM}, $\boldsymbol{U}_{0}$ and $\boldsymbol{U}_{\text{AY}}$ will be typically lower dimensional. In the remainder of this chapter we will refer to $(\boldsymbol{U}_{0}, \boldsymbol{U}_{\text{AY}})$ as $\boldsymbol{U}$. 

To study the ICE, we will parameterize cause-effect relations as SCM \eqref{CH4indSCM} so that the ICE is presented in Definition \ref{CH4def:ICE}. \begin{definition}{\textbf{Individual causal effect}}\label{CH4def:ICE}
	In a parameterization of the cause-effect relations as SCM \eqref{CH4indSCM}, the ICE of exposure strategy $\overline{a}$ at time point $j$, for individual $i$, $Y_{ji}^{\overline{a}}-Y_{ji}^{\overline{0}}$, equals
	\begin{equation*}
	f_{Y_{j}}(\boldsymbol{U}_{0i},\boldsymbol{U}_{\mathrm{AY}i},\overline{\boldsymbol{M}}_{j-1,i}, \overline{a},\overline{\boldsymbol{L}}_{j-1,i},\overline{N}_{Y_{ji}}) -f_{Y_{j}}(\boldsymbol{U}_{0i},\boldsymbol{U}_{\mathrm{AY}i},\overline{\boldsymbol{M}}_{j-1,i},  \overline{0},\overline{\boldsymbol{L}}_{j-1,i},\overline{N}_{Y_{ji}}).
	\end{equation*}
\end{definition} \noindent This ICE is the total effect of exposure strategy $\overline{a}$. By incorporating mediators of interest in the SCM, direct and indirect individual effects can be defined similarly,  but this is beyond the scope of this work.

Marginalized effect measures can also be expressed in terms of the SCM \eqref{CH4indSCM}. The conditional average causal effect (CACE), more commonly referred to as the CATE, $\mathbb{E}[Y^{\overline{a}}_{j}-Y^{\overline{0}}_{j}\mid \overline{\boldsymbol{M}}_{j-1}{=}\boldsymbol{m}]$, follows by the marginalization as described in Definition  \ref{CH4def:CACE}.  \begin{definition}{\textbf{Conditional average causal effect}}\label{CH4def:CACE}
	In a parameterization of the cause-effect relations as SCM \eqref{CH4indSCM}, the $\overline{\boldsymbol{M}}_{j-1}{=}\boldsymbol{m}$-CACE of exposure strategy $\overline{a}$ equals
	\begin{align*}
	\int \left(Y_{j}^{\overline{a}}-Y_{j}^{\overline{0}} \mid \overline{\boldsymbol{M}}_{j-1}{=}\boldsymbol{m}\right)~dF_{\left(\boldsymbol{U}_{0},\boldsymbol{U}_{\text{AY}},\boldsymbol{N}_{Y}\right)\mid \overline{\boldsymbol{M}}_{j-1}{=}\boldsymbol{m}},
	\end{align*}
	where $Y_{j}^{\overline{a}}-Y_{j}^{\overline{0}}$ as defined in \ref{CH4def:ICE} for individual $i$.
\end{definition} \noindent Throughout this work, we abbreviate the Lebesque-Stieltjes integral of a function $g$ with respect to probability law $F_{X}$, $\int g(x) dF_{X}(x)$, as $\int g(X) dF_{X}$.  Finally, the ACE, $\mathbb{E}[Y^{\overline{a}}_{j}-Y^{\overline{0}}_{j}]$ can be obtained by further marginalization over the modifiers as presented in Definition \ref{CH4def:ACE}. \newpage \begin{definition}{\textbf{Average causal effect}}\label{CH4def:ACE}
	In a parameterization of the cause-effect relations as SCM \eqref{CH4indSCM},
	\begin{align*}
	\int \left(Y_{j}^{\overline{a}}-Y_{j}^{\overline{0}}\right) dF_{\left(\boldsymbol{U}_{0},\boldsymbol{U}_{\text{AY}},\boldsymbol{N}_{Y},\overline{\boldsymbol{M}}_{j-1}\right)}, 
	\end{align*}
	where $Y_{j}^{\overline{a}}-Y_{j}^{\overline{0}}$ as defined in \ref{CH4def:ICE} for individual $i$.
\end{definition} \noindent Note that in definitions \ref{CH4def:CACE} and \ref{CH4def:ACE} integration with respect to the law of $\overline{
	\boldsymbol{L}}_{j-1}$ is not necessary as those confounders that also modify the causal effect will intersect with $\overline{\boldsymbol{M}}_{j-1}$ and are thus already considered.  

\begin{figure}[h]
	\centering
	\captionsetup{width=\textwidth}
	\begin{subfigure}{.45\textwidth}
		\centering
		\resizebox{1\textwidth}{!}{\includegraphics{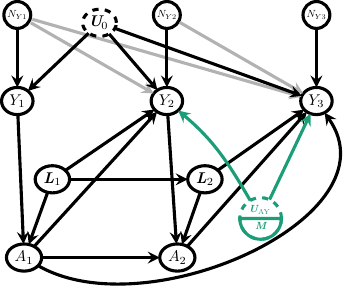}}
		\caption{}\label{CH4modvdw}	
	\end{subfigure} \hfill
	\begin{subfigure}{.45\textwidth}
		\centering
		\resizebox{1\textwidth}{!}{\includegraphics{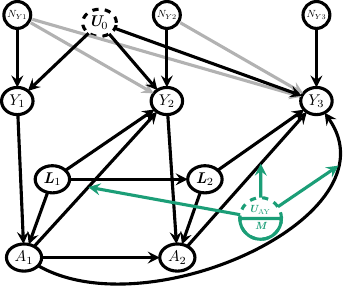}}
		\caption{}\label{CH4modalt}
	\end{subfigure}
	\caption{\small{The traditional representation of effect modification in a DAG as described by \citet{VanderWeele2007} (a) and the alternative arrow-on-arrow representation (b). The grey arrows from $N_{Yk}$ to $Y_{j}$ are only present when the history of the outcome causes $Y_{j}$ since common features are already covered by $\boldsymbol{U}_{0}$. }}
\end{figure}

A causal graph is induced by an SCM and visualizes the relations between exposure assignments, confounders and observed outcomes at a population 
level and can help during discussions on causal assumptions with experts \citep{Pearl2009book}. 
In a traditional visualization of the SCM \eqref{CH4indSCM}, the nodes of $\overline{\boldsymbol{M}}_{j-1}$ and $\boldsymbol{U}_{\text{AY}}$ would point to the outcome as has been proposed by \citet{VanderWeele2007} for cross-sectional settings. However, from the directed acyclic graph  based on this convention, as presented in Figure \ref{CH4modvdw} for time-invariant modifiers $\boldsymbol{M}$, it is impossible to disentangle the role of $\boldsymbol{M}$ and $\boldsymbol{U}_{\text{AY}}$ from the role of $\boldsymbol{U}_{0}$. Instead, one could directly point the nodes to the arrow representing the relation that is modified as demonstrated in Figure \ref{CH4modalt} and was proposed before by \citet{Weinberg2007}. This representation allows us to illustrate that the effect size of causal relations in the SCM can change individually. 

   In the remainder of this section, we will present the SCM underlying an `ideal' crossover trial and a Gaussian linear mixed assignment example. Two other related examples can be found in Section \ref{CH4ex2} and Section \ref{CH4ex3} of the Supplementary Material. 

\newpage
\subsection{The `ideal' crossover trial}\label{CH4ex0}
To illustrate the framework, we start by parameterizing the cause-effect relations behind a randomized crossover study with binary exposure, where the potential outcome under no exposure as well as the causal effects are time-invariant, and there is no carryover effect, 
with the following SCM:  
\begin{align}\label{CH4SCM:crossover}
U_{0i}&~{:}{=}~f_{U}(N_{0i}) \\ \nonumber
Y_{1i} &~{:}{=}~ f_{Y_{1}}(U_{0i})\\ \nonumber
A_{1i} &~{:}{=}~ f_{A}(N_{A1i})\\ \nonumber
A_{2i} &~{:}{=}~ 1-A_{1i}\\ \nonumber
U_{\text{AY}i} &~{:}{=}~ f_{U_{\text{AY}}}(N_{\text{AY}i})\\ \nonumber
Y_{ji}^{\overline{a}}&~{:}{=}~ Y_{1i} + U_{\text{AY}i}\cdot a_{j-1}, \text{ for } j~{=}~2 \text{ and } j~{=}~3.
\end{align} The randomization results in the absence of confounders so that $U_{0}, N_{\text{AY}} \independent A_{1}$. The distribution of $A_{1}$ is Bernoulli with probability $p_{A}$. The outcome of a participant before the start of the study, in the absence of the exposure, equals $Y_{1}$ and depends on some latent features of the individual represented by $U_{0}$. The study starts after assigning the exposure at the first time point $A_{1}$ (often a treatment) randomly. The exposure level for the second time point $A_{2}$ is then fixed. The observed outcomes $Y_{2}$ equals the potential outcome $Y_{2}^{A_{1}}$, similarly $Y_{3}$ equals $Y_{3}^{(A_{1}, 1-A_{1})}$. Due to the individual level of the receptiveness factor $U_{\text{AY}}$, the exposure effect is different for each participant but constant over time. The ICE can still depend on $Y^{\overline{0}}_{j}$ as $U_{0}$ and $U_{\text{AY}}$ can be dependent. Follow-up in the randomized trial starts after the first exposure assignment. Thus $Y_{1}$ is not observed, while $Y_{2}$ and $Y_{3}$ are. The ICE equals $U_{{\text{AY}}}$, while the ACE equals $\mathbb{E}\left[U_{{\text{AY}}}\right]$.  Measured modifiers $\overline{\boldsymbol{M}}$ (e.g.~gender or age), when any exist, are not included in this SCM but would affect the distribution of $U_{{\text{AY}}}$ so that the $\overline{\boldsymbol{M}}{=}\boldsymbol{m}$-CACE equals $\mathbb{E}\left[U_{{\text{AY}}} \mid \overline{\boldsymbol{M}}{=}\boldsymbol{m} \right]$. Note that for this system $Y_{3i}^{(0,1)}{-}Y_{3i}^{(0,0)} ~{=}~ Y_{3i}^{(1,1)}{-}Y_{3i}^{(0,0)} ~{=}~ Y_{2i}^{(1)}{-}Y_{2i}^{(0)}$, while $Y_{3i}^{(1,0)}{-}Y_{3i}^{(0,0)}~{=}~0$, i.e.~in a parameterization in the form of SCM \eqref{CH4margSCM0}, $N^{*}_{(1,1)i}~{=}~N^{*}_{(0,1)i}~{=}~N^{*}_{(1)i}~{=}~U_{\text{AY}i}$ and $N^{*}_{(1,0)i}~{=}~0$.

\subsection{Gaussian linear mixed assignment}\label{CH4ex1}

\begin{figure}[H]
	\centering
	\begin{minipage}[b]{.45\linewidth}
			\centering
			\resizebox{1\textwidth}{!}{\includegraphics{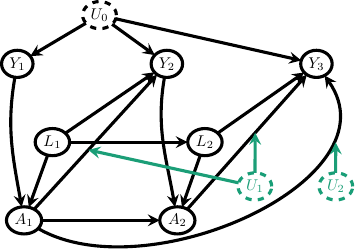}}
			
			\captionof{figure}{\small{Causal graph induced by the SCM \eqref{CH4SCM:ex1}. }}\label{CH4fig:GDAG}
	\end{minipage}
	\hspace{0.5cm}
	\begin{minipage}[b]{.45\linewidth}
			\centering
			\resizebox{0.9\textwidth}{!}{\includegraphics{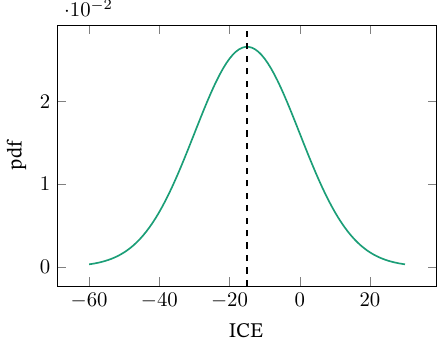}}
			\captionsetup{width=\linewidth}
			
			\captionof{figure}{\small{ICE distribution for the causal Gaussian linear mixed assignment. }}\label{CH4fig:ICE1}
	\end{minipage}
\end{figure}

As the running example in this chapter we consider a linear mixed-effects assignment with Gaussian-distributed random effects and residuals. The $\overline{N}_{Y}$ in SCM \eqref{CH4indSCM} equal the residuals, and the latent variables $\boldsymbol{U}$ equal the random effects. In this example, we assume that we start with an unexposed cohort, implying that the exposure cannot affect $Y_{1}^{\overline{a}}$. For simplicity, we omitted any measured modifiers $\overline{\boldsymbol{M}}$ and used a confounder that was not also a modifier. Only the last two levels of exposure affect the potential outcomes. Moreover, previous outcomes do not mediate the effect of the exposures, so subsequent outcomes are independent conditional on the random effects $(\boldsymbol{U})$. To be precise, we study cause-effect relations that can be parameterized as
\begin{align}\label{CH4SCM:ex1}
Y_{1i}&~{:}{=}~\theta_{0}+U_{0i}+N_{Y1i} \\ \nonumber
A_{1i}&~{:}{=}~\mathbbm{1}_{\{\text{logit}^{-1}\left(\alpha_{0}+\alpha_{1}Y_{1i}+\alpha_{3}L_{1i}\right)>N_{A1i}\}} \\ \nonumber
Y_{2i}^{\overline{a}}&~{:}{=}~\theta_{0}+U_{0i}+L_{1i}\theta_{L}+(\theta_{1}+U_{1i})a_{1}+N_{Y2i}\\	\nonumber
\forall j{>}1{:}~A_{ji}^{\overline{a}}&~{:}{=}~\mathbbm{1}_{\{\text{logit}^{-1}\left(\alpha_{0}+\alpha_{1}Y_{ji}^{\overline{a}}+\alpha_{2}a_{j-1}+\alpha_{3}L_{ji}\right)>N_{Aji}\}} \\	\nonumber
\forall j{>}2{:}~Y_{ji}^{\overline{a}}&~{:}{=}~\theta_{0}+U_{0i}+L_{j-1,i}\theta_{L}+(\theta_{1}+U_{1i})a_{j-1}+(\theta_{2}+U_{2i})a_{j-2}+N_{Yji}.
\end{align}	Here $U_{0}$ and the receptiveness factors ($U_{1}$ and $U_{2}$) are multivariate Gaussian distributed, $\boldsymbol{U}~{\sim}~\mathcal{N}(\boldsymbol{0}, \Sigma)$ with covariance matrix $\Sigma$. The time-varying confounder is Bernoulli distributed, $\forall j{\geq} 0{:}~L_{ji} \sim \text{Ber}(p_{Lji})$, where the probability $p_{Lji}$ can depend on the history of confounders $\overline{L}_{j-1,i}$. The noise variables are Gaussian or uniformly distributed, $N_{Yj} \sim \mathcal{N}(0,\sigma^{2})$ and $N_{Aj} \sim \text{Uni}[0,1]$. For this example, assumptions \ref{CH4A1} applies when $N_{Yj}\independent N_{Ak}$, and $N_{Yj}\independent N_{Yk}$ for all $j$ and $k$. Furthermore, assuming independence among the latent factors, the SCM \eqref{CH4SCM:ex1} induces the causal graph presented as Figure \ref{CH4fig:GDAG} for the first three measurements. Note that the independence of $\boldsymbol{U}$ is just for convenience. 

By Definition \ref{CH4def:ICE}, the ICE of $\overline{a}$ at time point $j$ for individual $i$ equals \\ $\left( \theta_{1} + U_{1i} \right) a_{j-1} + \left( \theta_{2} + U_{2i} \right) a_{j-2}$. For this example, the CACE equals the ACE as we did assume the absence of modifiers and equals $a_{j-1}\theta_{1} + a_{j-2}\theta_{2}$. The random intercept $U_{0}$ and the noise variables $\overline{N}_{Y} $ do not affect the causal-effect measures. The same applies for $\overline{L}$ as this does not include any modifiers. In Section \ref{CH4ex2} of the Supplementary Material, we present the example where the outcome of interest is $\exp\left(Y^{\overline{a}}_{j}\right)$, in which case $U_{0}$ and  $\overline{N}_{Y}$ do affect the causal effect.  For the current example, although $U_{0}$ is not a direct cause of the ICE, the ICE can depend on $U_{0}$ (if $U_{0}$ is correlated with $U_{1}$ or $U_{2}$). 

This example will be hypothetical to illustrate the introduced framework. However, for the remainder of the chapter, we have chosen numerical values for the parameters in SCM \eqref{CH4SCM:ex1} so that the cause-effect system could represent the effect of a treatment on systolic blood-pressure (see e.g.~\citet{SPRINT2015}). The mean outcome of $Y^{\overline{0}}$ equals $\theta_{0}{=}120$, the effect of the treatment on the potential outcome at time point $j$ equals $\theta_{1}{=}-10$, and $\theta_{2}{=}-5$ for the treatment at time point $j-1$ and $j-2$ respectively. The causal effect of the confounder $L_{j-1}$ equals $\theta_{L}{=}5$. The effect of the potential outcome $Y_{j}^{\overline{a}}$ on the log odds of receiving treatment at time $j$ equals $\alpha_{1}{=}0.05$, the effect of the previous treatment (at time point $j-1$) equals $\alpha_{2}{=}1$, the effect of the confounder $L_{j}$ equals $\alpha_{3}{=}0.7$ and the intercept of the log odds equals $\alpha_{0}{=}-3$. Furthermore, the covariance matrix of $U_{0}$ and the receptiveness factors $U_{1}$ and $U_{2}$, $\Sigma$, is equal to a diagonal matrix with diagonal $(\sigma_{0}^{2}, \sigma_{1}^{2}, \sigma_{2}^{2})$, where $\sigma_{0}{=}5$, $\sigma_{1}{=}10$ and $\sigma_{2}{=}5$. The standard deviation of the $N_{Yj}$ equals $\sigma{=}1$. Furthermore, the confounder at time $j$, $L_{j}$, is independent of $\overline{L}_{j-1}$ and equals $0.7$ with probability $0.3$ or $-0.3$ otherwise (zero mean).

The ICE of $\overline{a}{=}\overline{1}$, at an arbitrary point in time, is distributed as $\mathcal{N}(-15,15^2)$ and is presented together with the ACE (equal to $-15$) in Figure \ref{CH4fig:ICE1}. We can observe the heterogeneity in causal effect, and with a probability of approximately $0.15$, the effect is opposite to the lowering ACE. We will return to this example later in this chapter, where we will consider simulated data. The simulation code is written in \texttt{SAS 9.4}, and all programming codes used in this chapter can be found online at \url{https://github.com/RAJP93/ICE}. 

\newpage
\section{The cross-world causal effect}\label{CH4CWIT}
The ICE of an arbitrary individual from a population is considered a random variable. In this section, we introduce the cross-world causal effect (CWCE) that equals the random ICE conditioned on all observed information of an individual. The distribution of the CWCE indicates the probable values for the ICE given an individual's factual-world observations. A precise definition of the CWCE is presented as Definition \ref{CH4def:CWCE}. 
\begin{definition}{\textbf{Cross-world causal effect}}\label{CH4def:CWCE} 	In a parameterization of the cause-effect relations as SCM \eqref{CH4indSCM}, the cross-world causal effect at time point $j$ given factual-world data up until time $h$ equals
	\begin{equation*}
	Y_{j}^{\overline{a}}-Y_{j}^{\overline{0}} \mid \overline{Y}_{h}, \overline{\boldsymbol{M}}_{h},\overline{\boldsymbol{L}}_{h}, \overline{A}_{h}.
	\end{equation*} 
\end{definition} \noindent The CWCE describes the random variable obtained as a hypothetical draw from the ICE of individuals sharing $\overline{Y}_{h}, \overline{\boldsymbol{M}}_{h},\overline{\boldsymbol{L}}_{h}$, and $\overline{A}_{h}$. Knowing the distribution of the CWCE or specific properties of this distribution will be the ultimate goal when interested in individual causal effects, as we will never have access to more information on the individual than measured in the factual world. 

\noindent In the `ideal' crossover study, as introduced in Section \ref{CH4ex0}, the outcomes are measured for $j~{=}~2$ and $j~{=}~3$, such that \begin{equation*}
Y_{j}^{\overline{1}}-Y_{j}^{\overline{0}}~{=}~ Y_{j}^{a_{j-1}{=}1}-Y_{j}^{a_{j-1}{=}0}~{=}~ U_{\text{AY}} ~{=}~ \begin{cases}
Y_{2}-Y_{3}, \text{ if $A_{1}~{=}~1$}\\
Y_{3}-Y_{2}, \text{ if $A_{1}~{=}~0$}
\end{cases},
\end{equation*} and the CWCE equals
\begin{equation*}
U_{\text{AY}} \mid Y_{2}, Y_{3}, A_{1}, A_{2}  ~{=}~ \begin{cases}
Y_{2}-Y_{3}, \text{ if $A_{1}~{=}~1$}\\
Y_{3}-Y_{2}, \text{ if $A_{1}~{=}~0$}
\end{cases}. 
\end{equation*} Now, when we observe  $Y_{2i}~{=}~y_{2}$, $Y_{3i}~{=}~y_{3}$ and $\overline{A}_{2i}~{=}~(1,0)$ for individual $i$, the ICE is known and equals $y_{2}-y_{3}$. The degenerate CWCE equals the ICE, which is thus known after two measurements (as $\overline{N}_{Y}$ is absent in SCM \eqref{CH4SCM:crossover}). In the case of an additive effect of $N_{Yj}$ on $Y_{j}^{\overline{0}}$ (time-varying residuals), the CWCE is non-degenerate but would inform on the likely values of the ICE given the individual's observations. 

The conditional distribution of $Y_{j}^{\overline{a}}-Y_{j}^{\overline{0}} \mid \overline{Y}_{h},\overline{\boldsymbol{M}}_{h},\overline{\boldsymbol{L}}_{h},\overline{A}_{h}$ is a cross-world distribution as both the factual and counterfactual worlds are involved.  When $j{>}h$, the distribution of the CWCE can be used for the prediction of future causal effects for an individual. Otherwise, when we are retrospectively interested in the causal effect at time point $j{\leq}h$, observations on future values in the factual world can still provide information on the levels of the individual's $\boldsymbol{U}$, as was the case for the ideal crossover study evaluating the causal effect $Y_{2}^{\overline{a}}-Y_{2}^{\overline{0}}$ conditioned on the observed outcome $Y_{3}$. 

In this section, we assume that the distributions of $(\boldsymbol{U},\overline{N}_{Y})$ and the structural assignments in SCM \eqref{CH4indSCM} are known (in particular, the values of the parameters involved). We will show how 
the distribution of the CWCE can be expressed in terms of the data-generating distribution. In Section \ref{CH4II}, we will discuss under what assumption the CWCE distribution becomes identifiable from observational data.

\newpage
\subsection{The induced dependency problems}\label{CH4sec:inddep}
We want to express the distribution of the CWCE in terms of the observed outcome distribution. As with standard g-computation, it is necessary to find a set of variables $\boldsymbol{X}$ so that the potential outcome is independent of the exposure assignment given $\boldsymbol{X}$, i.e.~$Y_{j}^{\overline{a}} \independent \overline{A}_{j-1}~{\mid}~\boldsymbol{X}$ \citep[Chapter 13]{Hernan2019}. Then, the potential outcome would be equal in distribution to a conditional random variable that can be observed in the factual world, i.e.~
\begin{equation*}
Y_{j}^{\overline{a}} {\mid} \boldsymbol{X}~\overset{d}=~Y_{j}^{\overline{a}} {\mid} \boldsymbol{X}, \overline{A}_{j-1}{=}\overline{a}_{j-1}~\overset{d}=~Y_{j}{\mid} \boldsymbol{X}, \overline{A}_{j-1}{=} \overline{a}_{j-1},
\end{equation*} where the last equality follows from Assumption \ref{CH4A2}. 

Expressing the distribution of the CWCE in terms of the distributions that can be observed in the factual world is complex. First of all, by definition of SCM \eqref{CH4indSCM}, \begin{equation*}\resizebox{\linewidth}{!}{$Y_{j}^{\overline{a}}{\mid} \overline{\boldsymbol{M}}_{j-2},\overline{\boldsymbol{L}}_{j-2}, \overline{Y}_{j-1}~\overset{d}=~Y_{j}^{\overline{a}}{\mid} \overline{\boldsymbol{M}}_{j-2},\overline{\boldsymbol{L}}_{j-2}, 
(\boldsymbol{U}, \overline{N}_{Yj},  \overline{A}_{j-2}){\in}\mathcal{S}\left(\overline{\boldsymbol{M}}_{j-2},\overline{\boldsymbol{L}}_{j-2}, \overline{Y}_{j-1}\right)
,$}\end{equation*} where~\scalebox{0.93}{$\mathcal{S}\left(\boldsymbol{m},\boldsymbol{l}, \boldsymbol{y}\right) ~{=}~ \left \{ \boldsymbol{U}, \overline{N}_{Yj},  \overline{A}_{j-2} \mid  \forall k{<}j{:}~y_{k}~{=}~f_{Yk}\left(\boldsymbol{U},\boldsymbol{m}_{1{:}k{-}1},   \overline{A}_{k{-}1},\boldsymbol{l}_{1{:}k{-}1},\overline{N}_{Yk} \right) \right \}$,} i.e.~the set of $\boldsymbol{U}, \overline{N}_{Yj},  \overline{A}_{j-2}$ that given $\overline{\boldsymbol{M}}_{j-2},\overline{\boldsymbol{L}}_{j-2}$ result in the observed outcomes. Additionally conditioning on $\overline{A}_{j-2}$  would thus inform on the levels of $(\boldsymbol{U},\overline{N}_{Y,j-1})$. So, $$Y_{j}^{\overline{a}} \not \independent \overline{A}_{j-2} \mid \overline{\boldsymbol{M}}_{j-2},\overline{\boldsymbol{L}}_{j-2}, \overline{Y}_{j-1},$$
while $$Y_{j}^{\overline{a}} \independent A_{j-1} \mid \overline{\boldsymbol{M}}_{j-2},\overline{\boldsymbol{L}}_{j-2}, \overline{Y}_{j-1}.$$ This is also the key issue in systems with treatment-confounder feedback that leads to time-varying confounding, as will be discussed in Section \ref{CH4TVC}. Similarly, 
$$Y_{j}^{\overline{a}} \not \independent A_{j-1} \mid \overline{\boldsymbol{M}}_{j-1},\overline{\boldsymbol{L}}_{j-1}, \overline{Y}_{j},$$
as $A_{j-1}$ then informs on $(\boldsymbol{U},\overline{N}_{Yj})$ via the realization of $Y_{j}$. Secondly, since the CWCE is a function of two potential outcomes, the distribution will depend on the joint distribution of the potential outcomes. To express the joint distribution in terms of the distribution of the observations, we need to factorize the joint distribution in parts, each involving only one potential outcome as we cannot condition on the mutually exclusive events $\{\overline{A}~{=}~\overline{a}\}$ and $\{\overline{A}~{=}~\overline{0}\}$ at once. For this we need conditional independence between $Y_{j}^{\overline{a}}$ and $Y_{j}^{\overline{0}}$, while 
\begin{equation*}
Y_{j}^{\overline{a}}\not \independent Y_{j}^{\overline{b}} \mid \overline{\boldsymbol{M}}_{h-1},\overline{\boldsymbol{L}}_{h-1}, \overline{Y}_{h},
\end{equation*} as $Y_{j}^{\overline{b}}$ will additionally inform on $(\boldsymbol{U},\overline{N}_{Yj})$ when we do not condition on $\overline{A}_{h-1}$.

Conditioning on all latent variables $(\boldsymbol{U}, \overline{N}_{Y})$ will trivially result in independence between the potential outcome and the exposure assignment and other potential outcomes, respectively, as stated in Lemma \ref{CH4l1}. 
\begin{lemma}\label{CH4l1}
	Using a valid parameterization of the cause-effect relations of interest as SCM \eqref{CH4indSCM}, \begin{equation*}
	Y_{j}^{\overline{a}}\independent Y_{j}^{\overline{b}}, \overline{A}_{j-1} \mid \boldsymbol{U},\overline{\boldsymbol{M}}_{h},\overline{\boldsymbol{L}}_{h},\overline{Y}_{h}, \overline{N}_{Yj}.\end{equation*}
\end{lemma} \noindent These independence relations will be critical in the proof of Theorem \ref{CH4th4.2}.

\subsection{The CWCE distribution for a known SCM}
For the sake of notation, we will refer to the history of measured variables 
at time point $j$, $\left(\overline{\boldsymbol{M}}_{j},\overline{\boldsymbol{L}}_{j}, \overline{Y}_{j}, \overline{A}_{j}\right)$ as $\mathcal{H}_{j}$. 
By Lemma \ref{CH4l1}, $Y_{j}^{\overline{a}} \independent Y_{j}^{\overline{0}} \mid \mathcal{H}_{h} \backslash \overline{A}_{h} ,\boldsymbol{U}, \overline{N}_{Yj}$. The cross-world joint distribution of two potential outcomes can be expressed in terms of the parameters of the distribution of the observed data and the conditional distributions of the latent variables as shown in Theorem \ref{CH4th4.2}. 

\begin{theorem}\label{CH4th4.2}
	Using a valid parameterization SCM \eqref{CH4indSCM} of the cause-effect relations of interest, the joint probability density function (pdf) of the cross-world potential outcomes under exposures $\overline{a}$ and $\overline{0}$ at time point $j$ given history up until time $h$, $f\left((Y_{j}^{\overline{a}}, Y_{j}^{\overline{0}})\mid \mathcal{H}_{h} \right)$, equals
	\begin{equation*}	
	\resizebox{\textwidth}{!}{$\int f\left(Y_{j}\mid \boldsymbol{U}, \overline{\boldsymbol{M}}_{j-1},\overline{\boldsymbol{L}}_{j-1},  \overline{N}_{Yj}, \overline{A}_{j-1}{=}\overline{a}_{j-1} \right)f\left(Y_{j}\mid \boldsymbol{U}, \overline{\boldsymbol{M}}_{j-1},\overline{\boldsymbol{L}}_{j-1},  \overline{N}_{Yj}, \overline{A}_{j-1}{=}\overline{0} \right)
		dF_{(\boldsymbol{U}, \overline{N}_{Yj}) \mid \mathcal{H}_{h}}$,}
	\end{equation*}
	where $Y_{j}\mid \boldsymbol{U}, \overline{\boldsymbol{M}}_{j-1},\overline{\boldsymbol{L}}_{j-1},  \overline{N}_{Yj}, \overline{A}_{j-1}{=}\overline{a}_{j-1}$ is degenerate and its value depends on the parameters of the distribution of the observed data, so $f(Y_{j}\mid \boldsymbol{U}, \overline{\boldsymbol{M}}_{j-1},\overline{\boldsymbol{L}}_{j-1},  \overline{N}_{Yj}, \overline{A}_{j-1}{=}\overline{a}_{j-1})$ is an indicator function.
\end{theorem} \noindent Examples of the expression in Theorem \ref{CH4th4.2} are presented in Section \ref{CH4sec:31}, Section \ref{CH4ex2} and in Section \ref{CH4ex3} of the Supplementary Material. 
If each $f_{Y_{j}}$ in SCM \eqref{CH4indSCM} is an injective function of $N_{Yj}$, then for $j{\leq}h$, $N_{Yj}{\mid}\mathcal{H}_{h}, \boldsymbol{U}{=}\boldsymbol{u}$ is degenerate, and the joint pdf of the cross-world potential outcomes can be simplified as shown in Corollary \ref{CH4cor:331}.
\begin{corollary}\label{CH4cor:331}
	Using a valid parameterization SCM \eqref{CH4indSCM} of the cause-effect relations of interest, 
	in which $\forall j{\leq}h$, $f_{Y_{j}}$ is an injective function of $N_{Yj}$, the joint pdf of the cross-world potential outcomes under exposures $\overline{a}$ and $\overline{0}$ for $j{\leq}h$, $f\left((Y_{j}^{\overline{a}},Y_{j}^{\overline{0}})\mid \mathcal{H}_{h}\right )$, equals
	\begin{align*}	
	\int & \scalebox{0.9}{$f\left(Y_{j}\mid \boldsymbol{U}, \overline{\boldsymbol{M}}_{j-1},\overline{\boldsymbol{L}}_{j-1}, 
  \forall k{\leq}j{:}~~ N_{Yk}~{=}~f_{Y_{k}}^{-1}(\boldsymbol{U}, \overline{\boldsymbol{M}}_{k-1},\overline{\boldsymbol{L}}_{k-1},\overline{A}_{k-1}, \overline{N}_{Y,k-1}, \circ)(Y_{k}), \overline{A}_{j-1}{=}\overline{a}_{j-1} \right)$}\\
	&\scalebox{0.9}{$f\left(Y_{j}\mid \boldsymbol{U},\overline{\boldsymbol{M}}_{j-1},\overline{\boldsymbol{L}}_{j-1}, 
  \forall k{\leq}j{:}~~ N_{Yk}~{=}~f_{Y_{k}}^{-1}(\boldsymbol{U}, \overline{\boldsymbol{M}}_{k-1},\overline{\boldsymbol{L}}_{k-1},\overline{A}_{k-1}, \overline{N}_{Y,k-1}, \circ)(Y_{k}), \overline{A}_{j-1}{=}\overline{0} \right)dF_{\boldsymbol{U}\mid \mathcal{H}_{h}},$}
	\end{align*} where $f_{Y_{k}}^{-1}(\boldsymbol{U}, \overline{\boldsymbol{M}}_{k-1},\overline{\boldsymbol{L}}_{k-1},\overline{A}_{k-1}, \overline{N}_{Y,k-1}, \circ)(Y_{k})$ is the inverse function of $f_{Y_{k}}$ w.r.t. to $N_{Yk}$. 
\end{corollary} \noindent The pdf of the CWCE $f\left(Y_{j}^{\overline{a}}-Y_{j}^{\overline{0}}~{=}~d\mid \mathcal{H}_{h}\right)
$ can be trivially derived from the cross-world joint distribution as
\begin{equation}
\int_{y\in\mathcal{Y}} f\left((Y_{j}^{\overline{a}}~{=}~y,Y_{j}^{\overline{0}}~{=}~y-d)\mid \mathcal{H}_{h}\right) dy,
\end{equation}
where $\mathcal{Y}$ is the support of $Y_{j}^{\overline{a}}$.

In summary, given a parameterization of the cause-effect relations as SCM \eqref{CH4indSCM} with known parameters and distributions, the joint distribution of the cross-world potential outcomes and the distribution of the CWCE can be expressed in terms of the associations in the observed data and the conditional distribution of $(\boldsymbol{U},\overline{N}_{Y})$ given the individual's history. The expressions of these distributions simplify when all $f_{Yj}$ are injective functions of $N_{Yj}$, and the CWCE distribution is fully captured by the conditional distribution of the $\boldsymbol{U}$. We will continue by presenting an example of the latter case.

\subsection{Gaussian linear mixed assignment continued}\label{CH4sec:31}
Let us consider the causal Gaussian linear mixed assignment introduced in Section \ref{CH4ex1}. For this example,
\begin{equation*}
\begin{pmatrix} \boldsymbol{U} \\ \overline{Y}_{h} \end{pmatrix} \mid \overline{L}_{h}, \overline{A}_{h} \sim
\mathcal{N}\left(\begin{pmatrix}
\mathbf{0} \\ \mu_{Yh}
\end{pmatrix}, \begin{pmatrix}
\Sigma_{11} & \Sigma_{12} \\
\Sigma_{21} & \Sigma_{22}
\end{pmatrix} \right),
\end{equation*} 
where $\mu_{Yj} ~{=}~ \theta_{0} + \theta_{L}L_{j-1} + \theta_{1}A_{j-1} + \theta_{2}A_{j-2}$ (and $A_{0}~{=}~A_{-1}~{=}~0$). 
Therefore, 
\begin{equation*}
\boldsymbol{U} \mid \mathcal{H}_{h}  \sim
\mathcal{N}\left(
\Sigma_{12}\Sigma_{22}^{-1}(\overline{Y}_{h}-\mu_{Yh}), \Sigma_{11}-\Sigma_{12}\Sigma_{22}^{-1}\Sigma_{21}
\right).
\end{equation*} The $f_{Yj}$ are injective functions of $N_{Yj}$, so that, by Corollary \ref{CH4cor:331}, the joint pdf of the cross-world potential outcomes, $f\left((Y_{j}^{\overline{a}}~{=}~y_{1},Y_{j}^{\overline{0}}~{=}~y_{2})\mid \mathcal{H}_{h} \right)$ for $j{\leq}h$,  equals
$$\resizebox{\textwidth}{!}{$\int \mathbbm{1}_{\left\{\left(Y_{j}+(\theta_{1}-U_{1})(a_{j-1}-A_{j-1})+(\theta_{2}-U_{2})(a_{j-2}-A_{j-2})\right)~{=}~y_{1}\right\}}\mathbbm{1}_{\left \{\left(Y_{j}-(\theta_{1}+U_{1})A_{j-1}-(\theta_{2}+U_{2})A_{j-2}\right)~{=}~y_{2}\right \}}dF_{\boldsymbol{U} \mid \mathcal{H}_{h}}$.}$$
As a result, the pdf of the CWCE, $f\left((Y_{j}^{\overline{a}}-Y_{j}^{\overline{0}}~{=}~d)\mid \mathcal{H}_{h} \right)$,  equals
$$ \int \mathbbm{1}_{\left\{(\theta_{1}+U_{1})a_{j-1}+(\theta_{2}+U_{2})a_{j-2}~{=}~d\right\}}dF_{\boldsymbol{U} \mid \mathcal{H}_{h}},$$ so that 
$$
\resizebox{\linewidth}{!}{$Y_{j}^{\overline{a}}-Y_{j}^{\overline{0}}\mid \mathcal{H}_{h} \sim \mathcal{N}\left(a_{j-1}\theta_{1} + a_{j-2}\theta_{2} + \boldsymbol{x} \left(\Sigma_{12}\Sigma_{22}^{-1}(\overline{Y}_{h}-\mu_{Yh})\right), \boldsymbol{x} \left(\Sigma_{11}-\Sigma_{12}\Sigma_{22}^{-1}\Sigma_{21} \right) \boldsymbol{x}^{T}\right)$,} 
$$ where $\boldsymbol{x}~{=}~(0, a_{j-1}, a_{j-2})$. The variance of the CWCE thus depends on the exposure history of an individual.

\begin{figure}[H]
	\centering
	\centering
	\resizebox{0.8\textwidth}{!}{\includegraphics{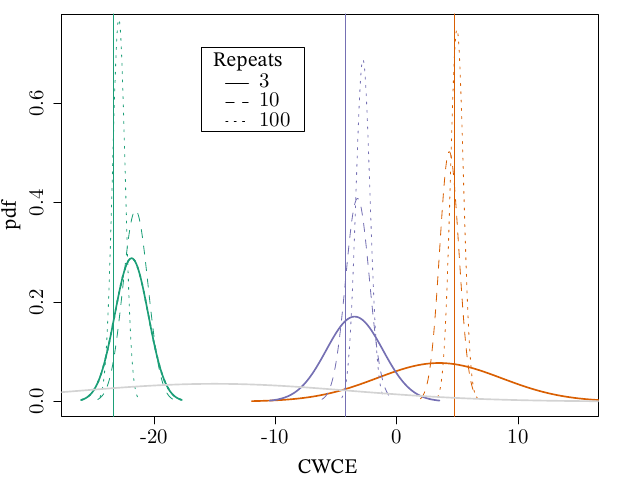}}
	\captionsetup{width=\textwidth}
	\caption{\small{Distribution of the CWCE of $\overline{a}{=}\overline{1}$, at the third repeat, for three individuals, based on information from $3$ (solid), $10$ (dashed) and $100$ (dotted) repeats respectively for the Gaussian example. The exposure assignment at the first two time points equal $(1,1)$, $(1,0)$ and $(0,1)$ for the green, blue and orange curves, respectively. Furthermore, the actual ICE for each individual (vertical lines) and the population ICE distribution (grey) are presented. }}		\label{CH4fig8a}
\end{figure}

For the parameter values introduced in Section \ref{CH4ex1}, where $U_{0}\independent U_{1}, U_{2}$, the CWCE for $j~{=}~3$ of unexposed (during the first two time points) individuals equals the population distribution of the ICE (grey line, Figure \ref{CH4fig8a}) as nothing can be learned on their $U_{1}$ and $U_{2}$ values.  For the simulated data, the distributions of the CWCE, $Y_{3}^{\overline{a}}-Y_{3}^{\overline{0}}\mid \overline{L}_{h}{=}\overline{\ell}_{h}, \overline{Y}_{h}{=}\overline{y}_{h}, \overline{A}_{h}{=}\overline{a}_{h}$, are presented for three individuals with different exposure strategies in Figure \ref{CH4fig8a} together with their actual ICE of $\overline{a}{=}\overline{1}$. The individual with an ICE of $-23.3$ (green line, Figure \ref{CH4fig8a}) has been exposed at both time points. Therefore the CWCE (using three repeats) is less variable than the other two individuals. The individual with an ICE of $-4.2$ (blue line, Figure \ref{CH4fig8a}) has only been exposed once, at the first time point, so that $Y_{2}$ informs on $U_{1}$ and only $Y_{3}$ informs on $U_{2}$. On the other hand, the individual with a positive ICE of $4.8$ (orange line, Figure \ref{CH4fig8a}) has only been exposed at the second time point and is thus the only one for which nothing can be learned on its $U_{2}$ value, and only $Y_{3}$ informs on the level of $U_{1}$. The latter results in the CWCE distribution that has the largest variability. The CWCE of these individuals using data from three measurements thus differ in variance due to the assigned exposure. 

When more repeats are used, the shapes of the CWCE distributions becomes more alike for these three individuals and lower in variability (see the dotted and dashed lines in Figure \ref{CH4fig8a}). The software code to derive the CWCE was written in \texttt{R}. 

\section{Heterogeneity as the source of time-varying confounding}\label{CH4TVC}

Causal inference for longitudinal processes is complicated due to the presence of time-varying confounding (also known as time-depending confounding), such that 
$\mathbb{E}\left[Y_{j}^{\overline{a}}\right]\neq \mathbb{E}\left[Y_{j}\mid \overline{A}_{j-1}{=}\overline{a}_{j-1}\right]$. If the time-varying confounding is only the result of measured confounders of the exposure-outcome relation that may vary over time, then valid (marginal) causal inference can be performed by adjusting for the history of the measured confounders $\overline{\boldsymbol{L}}$, and \begin{equation*}\forall j{:}~\mathbb{E}\left[Y_{j}^{\overline{a}}\mid \overline{\boldsymbol{L}}_{j-1}\right]~{=}~ \mathbb{E}\left[Y_{j}\mid \overline{A}_{j-1}{=}\overline{a}_{j-1},\overline{\boldsymbol{L}}_{j-1}\right]. \end{equation*} So far, we have not discussed making causal inferences from data. This requires that the probability of exposure at any point in time is bounded away from 0 and 1 irrespective of the history (levels of past outcomes, exposure,  modifiers and confounders), referred to as positivity \citep[Chapter 3]{Hernan2019}.
\begin{assumption}{\textbf{Positivity}}\label{CH4A3}
	$$\exists \eta>0 \text{ such that } \forall j, \boldsymbol{m}, \boldsymbol{\ell}, \boldsymbol{y}{:}~\eta \leq \mathbb{P}(A_{j}~{=}~1 \mid \overline{\boldsymbol{M}}_{j-1}{=}\boldsymbol{m}, \overline{\boldsymbol{L}}_{j-1}{=}\boldsymbol{\ell}, \overline{Y}_{j-1}{=}\boldsymbol{y}) \leq 1-\eta $$\end{assumption} 
 
Presence of receptiveness factors $\boldsymbol{U}_{\text{AY}}$ and common latent causes $\boldsymbol{U}_{0}$ result in time-varying confounding of the exposure effect on the outcome. There exist measured variables (the past of the outcome process) that are on the causal path from $\boldsymbol{U}_{\text{AY}}$ and $\boldsymbol{U}_{0}$ to $A_{j}$ (and could thus be adjusted for to account for the indirect unmeasured confounding) but at the same time are caused by $A_{k}$ for $k{<}j$ (so that adjustment would result in collider bias of the relation between $A_{k}$ and the outcome of interest, as shown in Section \ref{CH4sec:inddep}). This phenomenon is referred to as treatment-confounder feedback and leads to a challenging form of time-varying confounding that cannot be accounted for with standard methods \citep[Chapter 20]{Hernan2019}. 

\begin{figure}[H]
	\centering
	\begin{subfigure}{.325\textwidth}
		\centering
		\resizebox{1\textwidth}{!}{\includegraphics[page=1]{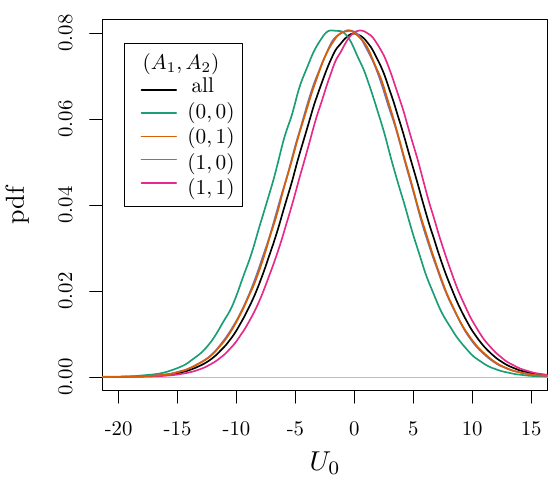}}
		\caption{}\label{CH4TVCUa}	
	\end{subfigure} \hfill
	\begin{subfigure}{.325\textwidth}
			\centering
		\resizebox{1\textwidth}{!}{\includegraphics[page=1]{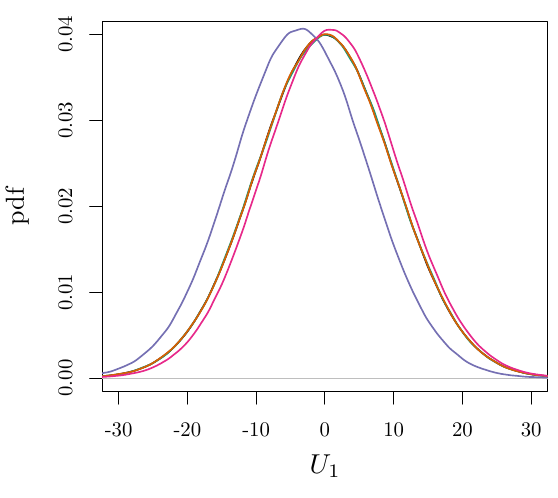}}
		\caption{}\label{CH4TVCUb}	
	\end{subfigure} \hfill
	\begin{subfigure}{.325\textwidth}
			\centering
		\resizebox{1\textwidth}{!}{\includegraphics[page=1]{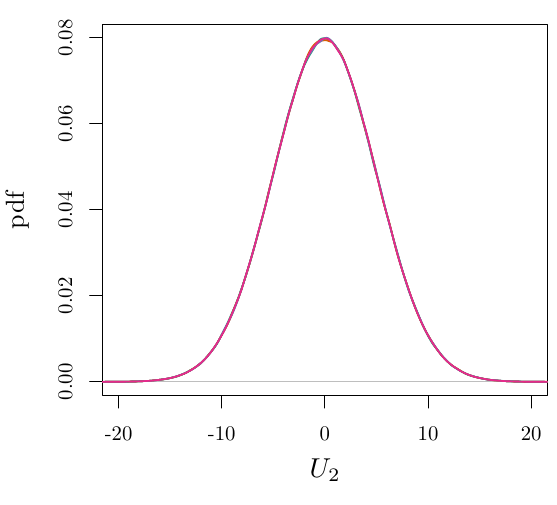}}
		\caption{}\label{CH4TVCUc}	
	\end{subfigure}
	\captionsetup{width=\textwidth}
	\caption{\small{Conditional distributions of $U_{0} {\mid} A_{1}, A_{2}$ (a), $U_{1} {\mid} A_{1}, A_{0}$ (b) and $U_{2} {\mid} A_{1}, A_{0}$ (c) and the population distribution of these latent variables (black lines) for the Gaussian linear mixed assignment example. The exposure levels can be found in the legend. The distributions are derived empirically from a simulation with $10{,}000{,}000$ individuals. The lines for $(0,1)$ and $(1,0)$ in (a), for $(0,0)$, $(0,1)$ and the population distribution in (b) and all lines in (c) do overlap.}}\label{CH4TVCU}
\end{figure}

By Lemma \ref{CH4l1}, $Y_{j}^{\overline{a}} {\independent} \overline{A}_{j-1} {\mid} \boldsymbol{U}, \boldsymbol{M}, \overline{\boldsymbol{L}}_{j-1}, \overline{N}_{Yj}$, so that the distribution of $Y_{j}^{\overline{a}}{\mid}  \overline{\boldsymbol{M}}_{j-1},$ $\overline{\boldsymbol{L}}_{j-1}$ can be expressed in terms of the data generating distribution by marginalization over $\boldsymbol{U}$ and $\overline{N}_{Yj}$. The resulting expression is presented in Theorem \ref{CH4th:latent}. 
\begin{theorem}\label{CH4th:latent}
	Using a valid parameterization of the cause-effect relations as SCM \eqref{CH4indSCM}, the joint pdf of the potential outcomes until time point $j$ for exposure strategy $\overline{a}$, $f\left(\overline{Y}_{j}^{\overline{a}}\mid \overline{\boldsymbol{M}}_{j-1}, \overline{\boldsymbol{L}}_{j-1} \right)$ equals
	\begin{equation*}	
 \int f\left(\overline{Y}_{j}\mid \boldsymbol{U}, \overline{\boldsymbol{M}}_{j-1},\overline{\boldsymbol{L}}_{j-1},  \overline{N}_{Yj}, \overline{A}_{j-1}{=}\overline{a}_{j-1} \right)
		dF_{(\boldsymbol{U},\overline{N}_{Yj}) \mid \overline{\boldsymbol{M}}_{j-1},\overline{\boldsymbol{L}}_{j-1}}
	\end{equation*} 
\end{theorem} \noindent There is time-dependent confounding because this distribution does deviate from $\overline{Y}_{j} \mid \overline{\boldsymbol{M}}_{j-1}, \overline{\boldsymbol{L}}_{j-1}, \overline{A}_{j-1}{=}\overline{a}_{j-1}$, which equals
\begin{equation*}	
\int f\left(\overline{Y}_{j}\mid \boldsymbol{U}, \overline{\boldsymbol{M}}_{j-1},\overline{\boldsymbol{L}}_{j-1},  \overline{N}_{Yj}, \overline{A}_{h-1}{=}\overline{a}_{j-1} \right)
dF_{(\boldsymbol{U},\overline{N}_{Y})\mid \overline{\boldsymbol{M}}_{j-1},\overline{\boldsymbol{L}}_{j-1},\overline{A}_{j-1}{=}\overline{a}_{j-1}}.
\end{equation*}Since $\boldsymbol{U}$ and $\overline{N}_{Y}$ cause outcomes that affect the probability of receiving exposure at that time point, elements of $(\boldsymbol{U},\overline{N}_{Yj})$ depend on $\overline{A}_{j}$, so that $$F_{\left(\boldsymbol{U}, \overline{N}_{Yj}\right)\mid \overline{\boldsymbol{M}}_{j-1},\overline{\boldsymbol{L}}_{j-1},}\neq F_{\left(\boldsymbol{U}, \overline{N}_{Yj}\right) \mid \overline{\boldsymbol{M}}_{j-1},\overline{\boldsymbol{L}}_{j-1},\overline{A}_{j}{=}\overline{a}_{j}}.$$ For the Gaussian linear mixed assignment example (see Section \ref{CH4ex1} and Section \ref{CH4sec:31}), the distributions of $U_{0}, U_{1}$ and $U_{2}$ given the first two exposures are presented  in Figure \ref{CH4TVCU}. In expectation, lower outcomes result in a lower probability of receiving exposure the next time. Therefore, individuals not exposed at the first two time points are expected to have lower $U_{0}$ values, while those exposed twice are expected to have higher $U_{0}$.  The distribution of $U_{1}$ will differ for those individuals that have $A_{1}~{=}~1$ as then $U_{1}$ is a cause of $Y_{2}$ and, in turn, affects $A_{2}$. On the other hand, $U_{2}$ is independent of $A_{1}$ and $A_{2}$ as $U_{2}$ is not a cause of $Y_{1}$ and $Y_{2}$. However, if $A_{1}~{=}~1$, then $U_{2}$ will depend on $A_{3}$. 
  \noindent Also $N_{Y1}$ and $N_{Y2}$ depend on the value of $A_{1}$ and $A_{2}$ respectively as is shown in the Supplementary Figure \ref{CH4TVCN}. However, for the Gaussian linear mixed assignment example, this does not result in time-varying confounding since $N_{Yj}$ only causes the outcome at time $j$ ($Y^{\overline{a}}_{j}$) and $N_{Yj} \independent \overline{A}_{j-1}$.
 
 Under Assumption \ref{CH4A1} (and Assumption \ref{CH4A2}), the so-called g-methods have been proposed to account for the time-varying confounding and thus to make valid marginal causal inference in the presence of treatment-confounder feedback \citep[Chapter 21]{Hernan2019}. These three methods, the g-computation formula, the inverse probability weighting (IPW) of marginal structural models (MSM, see, e.g.~\citet{Robins2000}) and the g-estimation of structural nested models (SNMMs, see, e.g.~\citet{Lok2004, Vansteelandt2015}) are summarized by \citet{Daniel2013} and by \citet{Naimi2017}. IPW estimation of MSM and g-estimation of SNMMs  account for the time-varying confounding by modelling the exposure assignment adjusted for covariate and exposure histories \citep{Daniel2013}. On the contrary, the g-formula directly expresses the expectations of potential outcomes in terms of the conditional expectations of the observed data \citep[Section 21.1]{Hernan2019}. Furthermore, besides the expectations, the g-formula can be used to express the distribution of potential outcomes in terms of the conditional distributions of observations, see \citet[Technical Point 21.1]{Hernan2019}, and thus implies that the expression in Theorem \ref{CH4th:latent} is identifiable from observational data (under assumptions \ref{CH4A2}, \ref{CH4A1} and \ref{CH4A3}).

Let $(\boldsymbol{U})_{\overline{a}_{j-1}}$ equal those elements of $\boldsymbol{U}$ that affect $\overline{Y}_{j}^{\overline{a}}$. So, $(\boldsymbol{U})_{\overline{a}_{j-2}} {\subseteq}~(\boldsymbol{U})_{\overline{a}_{j-1}}$ and we let  $(\boldsymbol{U})_{a_{j-1}}{=}~ (\boldsymbol{U})_{\overline{a}_{j-1}}{\backslash}(\boldsymbol{U})_{\overline{a}_{j-2}}$. Under Assumption \ref{CH4A1}, and if  $(\boldsymbol{U})_{\overline{a}_{j-1}} \independent \overline{\boldsymbol{M}}_{j-1},\overline{\boldsymbol{L}}_{j-1}$ (for simplicity), then $(\boldsymbol{U})_{\overline{a}_{j-1}} \independent \overline{A}_{j-1} \mid (\boldsymbol{U})_{\overline{a}_{j-2}}$, as then the elements of $\boldsymbol{U}$ are independent of the exposure assignments that were caused by the outcomes unaffected by those elements. For example, for the Gaussian linear mixed example we just illustrated that $U_{2} \independent A_{1}, A_{2}$.  Therefore, the distribution of $(\boldsymbol{U})_{\overline{a}_{j-1}}$ can be expressed as a factorization over the conditional distributions of the receptiveness factors in the observed data as shown in Lemma \ref{CH4lemma:fUN}. 
\begin{lemma}\label{CH4lemma:fUN} 	Using a valid parameterization SCM \eqref{CH4indSCM} of the cause-effect relations of interest, under Assumption \ref{CH4A1}, the distribution of $\left((\boldsymbol{U})_{\overline{a}_{j-1}}, \overline{N}_{Yj}\right)$ given $\overline{\boldsymbol{M}}_{j-1}$ and $\overline{\boldsymbol{L}}_{j-1}$ equals
  $$\prod_{k~{=}~1}^{j} f\left((\boldsymbol{U})_{a_{k-1}},N_{Yk} \mid (\boldsymbol{U})_{\overline{a}_{k-2}},\overline{N}_{Y,k-1}, \overline{A}_{k-1}{=}\overline{a}_{k-1},\overline{\boldsymbol{M}}_{j-1},\overline{\boldsymbol{L}}_{j-1}\right). 
  $$
\end{lemma} \noindent 
Following the same reasoning, under Assumption \ref{CH4A1}, the expression in Theorem \ref{CH4th:latent} is identifiable from observational data and equals the g-formula as shown in Corollary \ref{CH4th:latent2}. 
\begin{corollary}\label{CH4th:latent2}
	Using a valid parameterization SCM \eqref{CH4indSCM} of the cause-effect relations of interest, under Assumption \ref{CH4A1}, the joint pdf of the potential outcomes until time point $j$ for exposure strategy $\overline{a}$, $f\left(\overline{Y}_{j}^{\overline{a}}\mid \overline{\boldsymbol{M}}_{j-1}, \overline{\boldsymbol{L}}_{j-1} \right)$ equals
	$$\prod_{k~{=}~1}^{j} f(Y_{k} \mid \overline{Y}_{k-1}, \overline{A}_{k-1}{=}\overline{a}_{k-1}, \overline{\boldsymbol{M}}_{j-1}, \overline{\boldsymbol{L}}_{j-1}).$$
\end{corollary} \noindent For each $\overline{a}_{j-1}$, under Assumption \ref{CH4A1}, the joint distribution of $\overline{Y}^{\overline{a}}_{j}$ is thus identifiable from the observed data. Therefore, $(\boldsymbol{U})_{\overline{a}_{j-1}}$ or a transformation of $(\boldsymbol{U})_{\overline{a}_{j-1}}$ is identifiable, e.g.~for the Gaussian linear mixed assignment example the joint distribution of $U_{0} + a_{k-1} U_{1} + a_{k-2} U_{2}+N_{Yk}$ for $1{\leq} k {\leq} j$ is identifiable for each $\overline{a}$. There exist systems of cause-effect relations where the heterogeneity is relatively simple such that the cardinality of $\boldsymbol{U}$ is low $(|\boldsymbol{U}| < h)$. Particularly when many repeats are available, there might exist an exposure strategy $\overline{a}_{j}$ such that $(\boldsymbol{U})_{\overline{a}_{j-1}}$ is identifiable and equal to $\boldsymbol{U}$. Then, also the CWCE becomes identifiable and could be estimated from observational data, as we will elaborate on in the next section. 

\section{Individual inference}\label{CH4II}
By Theorem \ref{CH4th4.2}, the CWCE distribution 
is equal to the marginalized joint posterior distribution of $Y_{j}{\mid}\boldsymbol{U},\overline{\boldsymbol{M}}_{h},\overline{\boldsymbol{L}}_{h},  \overline{N}_{Yj}, \overline{A}_{h}{=}\overline{a}$ and $Y_{j}{\mid}\boldsymbol{U}, \overline{\boldsymbol{M}}_{h}, \overline{\boldsymbol{L}}_{h},$  
 $\overline{N}_{Yj}, \overline{A}_{h}{=}\overline{0}$  using $F_{(\boldsymbol{U},\overline{N}_{Y} \mid \mathcal{H}_{h})}$ as a prior. In practice,  $f\left(Y_{j}{\mid}\boldsymbol{U}, \overline{\boldsymbol{M}}_{h},\overline{\boldsymbol{L}}_{h}, \overline{N}_{Y}, \overline{A}_{h}{=}\overline{a} \right)$ as well as $F_{(\boldsymbol{U},\overline{N}_{Y})}$, and thus the prior, depend on unknown (hyper-) parameters and should be estimated from the observed data using Empirical Bayes methods to derive the distribution of the CWCE. As mentioned in the previous section, we cannot observe an individual's outcomes under different exposure strategies. Therefore, the CWCE can only be estimated for systems where the entire joint distribution of latent variables $(\boldsymbol{U}, \overline{N}_{Y})$ can be learned from individuals exposed to a particular strategy, i.e.~$\exists \overline{a}{:}~(\boldsymbol{U},\overline{N}_{Y})_{\overline{a}_{j-1}}~{=}~(\boldsymbol{U}, \overline{N}_{Y})$ where  $(\boldsymbol{U},\overline{N}_{Y})_{\overline{a}_{j-1}}~{=}~\left((\boldsymbol{U})_{\overline{a}_{j-1}},\overline{N}_{Yj})\right)$. 
 We can never verify such similarity between worlds with the observed data, and thus need to make a cross-world assumption on the joint distribution of the $(\boldsymbol{U})_{\overline{a}_{j-1}}$ for different exposure strategies: \begin{assumption}{\textbf{Cross-world similarity of individual-effect modification}}\label{CH4A4} 	In a parameterization of the cause-effect relations as SCM \eqref{CH4indSCM},
	$$\exists \overline{a}{:}~ (\boldsymbol{U},\overline{N}_{Y})_{\overline{a}_{j-1}} ~{=}~ (\boldsymbol{U}, \overline{N}_{Y}).$$ 
\end{assumption} \noindent Assumption \ref{CH4A4} can be translated into case-specific assumptions that experts should review. One can always think of theoretical cases that lead to the observed data but won't meet the assumption. However, for systems with a limited amount of complexity, i.e.~when the cardinality $|\boldsymbol{U}|$ is small, while relatively many repeats per individual are obtained, these theoretical cases might be very unreasonable. At the end of this section, we will elaborate on how one could deal with Assumption \ref{CH4A4} in practice. In Theorem \ref{CH4th41}, we show that under Assumption \ref{CH4A4}, the CWCE distribution can be accurately estimated from the observational data when consistent estimators for the (hyper-) parameters exist. The convergence of the estimator of the CWCE distribution function thus depends on the convergence of the estimators of the structural assignments $f_{Y_{j}}$ and of the distribution function of $(\boldsymbol{U}, \overline{N}_{Y})_{\overline{a}_{h-1}}$, $F_{(\boldsymbol{U}, \overline{N}_{Y})_{\overline{a}_{h-1}}}$. This thus also implies that $F_{(\boldsymbol{U}, \overline{N}_{Y})_{\overline{a}_{h-1}}}$ needs to be identifiable from the observed data.

The maximum a posteriori probability (MAP) of the estimated CWCE distribution could be used to estimate the ICE. If Corollary \ref{CH4cor:331} applies, the actual CWCE distribution converges to a degenerate random variable for an increasing number of repeats $h$ (as $(\boldsymbol{U}, \overline{N}_{Y})_{\overline{a}_{h-1}}{\mid}\mathcal{H}_{h}$ converges to a degenerate random vector), so that the MAP estimate will converge to the ICE. However, for cause-effect relations where the $f_{Y_{j}}$ are not injective functions of $N_{Yj}$, it will be impossible to consistently estimate all receptiveness factors so that the ICE cannot be retrieved for all individuals as demonstrated with the logistic linear mixed assignment example presented in Section \ref{CH4ex3}. 

\newpage
\begin{theorem}\label{CH4th41}
	Consider cause-effect relations parameterized as SCM \eqref{CH4indSCM}, under assumptions \ref{CH4A2}, \ref{CH4A1}, \ref{CH4A3} and Assumption \ref{CH4A4} (for some $j\leq h$). Let the number of repeats per individual, $h$, be large enough so that for all $\overline{a}$ 
	the estimated distribution function of $(\boldsymbol{U}, \overline{N}_{Y})_{\overline{a}_{h-1}}$, $\hat{F}_{(\boldsymbol{U},\overline{N}_{Y} )_{\overline{a}_{h-1}}} (\boldsymbol{x})~{=}~\hat{\mathbb{P}}_{n}\left((\boldsymbol{U},\overline{N}_{Y} )_{\overline{a}_{h-1}}\leq \boldsymbol{x}\right)$, 
	satisfies
	\begin{enumerate}[label=\roman*]
		\item $\hat{F}_{(\boldsymbol{U},\overline{N}_{Y} )_{\overline{a}_{h-1}}} (\boldsymbol{x}) \xrightarrow[n \to \infty]{} F_{(\boldsymbol{U},\overline{N}_{Y} )_{\overline{a}_{h-1}}} (\boldsymbol{x})$, pointwise for all points $\boldsymbol{x}$ where $F_{(\boldsymbol{U},\overline{N}_{Y} )_{\overline{a}_{h-1}}}$ is continuous, 
	\end{enumerate} 
	and the estimates of the structural assignments $f_{Y_{j}}$, $\hat{f}_{Y_{j}}(\boldsymbol{u}_{0},  \boldsymbol{u}_{\text{AY}}, \boldsymbol{m}, \boldsymbol{a}, \boldsymbol{\ell}, \boldsymbol{n}_{Y})$, satisfy
	\begin{enumerate}[resume, label=\roman*]
		\item 
		$\hat{f}_{Y_{j}}(\boldsymbol{u}_{0}, \boldsymbol{u}_{\text{AY}}, \boldsymbol{m}, \boldsymbol{a}, \boldsymbol{\ell}, \boldsymbol{n}_{Y}) \xrightarrow[n \to \infty]{} f_{Y_{j}}(\boldsymbol{u}_{0},  \boldsymbol{u}_{\text{AY}}, \boldsymbol{m}, \boldsymbol{a}, \boldsymbol{\ell}, \boldsymbol{n}_{Y})$ pointwise.
	\end{enumerate} 
	Then, 
	$$ \int_{\mathcal{D}_{n}(d,j,\overline{a},\mathcal{H}_{h})} 1 d\hat{F}_{(\boldsymbol{U},\overline{N}_{Y} )_{\overline{a}_{h-1}}\mid \mathcal{H}_{h}} 
 \xrightarrow[n \to \infty]{} 
 F_{Y_{j}^{\overline{a}}-Y_{j}^{\overline{0}} \mid \mathcal{H}_{h}}(d),$$ pointwise for all $d$ that are continuity points of $F_{Y_{j}^{\overline{a}}-Y_{j}^{\overline{0}} \mid \mathcal{H}_{h}}$, 
	where $$ \mathcal{D}_{n}(d,j,\overline{a},\mathcal{H}_{h}) ~{=}~ \left\{ (\boldsymbol{u},\boldsymbol{n}_{Y}){:}~\hat{f}_{Y_{j}}(\boldsymbol{u}, \overline{\boldsymbol{M}}_{j-1}, \overline{a}, \overline{\boldsymbol{L}}_{j-1}, \boldsymbol{n}_{Y}) - \hat{f}_{Y_{j}}(\boldsymbol{u}, \overline{\boldsymbol{M}}_{j-1}, \overline{0}, \overline{\boldsymbol{L}}_{j-1}, \boldsymbol{n}_{Y}) \leq d \right\},$$ and
	$$\hat{F}_{(\boldsymbol{U},\overline{N}_{Y} )_{\overline{a}_{j-1}}\mid \mathcal{H}_{h}} (\tilde{{}\boldsymbol{x}}) ~{=}~ \frac{\int_{\mathcal{X}(\tilde{{}\boldsymbol{x}})\cap \mathcal{A}_{n}(\mathcal{H}_{h})} 1 d\hat{F}_{(\boldsymbol{U},\overline{N}_{Y})_{\overline{a}_{h-1}}} }{\int_{\mathcal{X}(\boldsymbol{\infty})\cap \mathcal{A}_{n}(\mathcal{H}_{h})} 1 d\hat{F}_{(\boldsymbol{U},\overline{N}_{Y})_{\overline{a}_{h-1}}}},
	$$ for $\mathcal{X}(\tilde{{}\boldsymbol{x}})~{=}~\{(\boldsymbol{u},\boldsymbol{n}_{Y}){:}~(\boldsymbol{u},\boldsymbol{n}_{Y})\leq \tilde{{}\boldsymbol{x}} \}$ and 
	$$		\mathcal{A}_{n}(\mathcal{H}_{h})~{=}~\left \{(\boldsymbol{u},\boldsymbol{n}_{Y}){:}~\forall 1{\leq} j {\leq} h{:}~\hat{f}_{Y_{j}}(\boldsymbol{u}, \overline{\boldsymbol{M}}_{j-1}, \overline{A}_{j-1}, \overline{\boldsymbol{L}}_{j-1}, \boldsymbol{n}_{Y}) ~{=}~ Y_{j} \right \}.
	$$
\end{theorem}  
On a side note, Assumption \ref{CH4A4} is sufficient but not necessary. It would suffice when there exists an exposure strategy $\overline{a}$ and a time point $j$ such that $$\forall \overline{b}, \forall \overline{c}{\neq}\overline{b}, \forall k{\geq} j{:}~(\boldsymbol{U})_{\overline{b}_{k-1}}{\independent}(\boldsymbol{U})_{\overline{c}_{k-1}}{\mid}(\boldsymbol{U})_{\overline{a}_{k-1}}{\cap}~ (\boldsymbol{U})_{\overline{b}_{k-1}}.$$ So, there exists a strategy which effect is modified by all the elements of $\boldsymbol{U}$ that affect the outcome in multiple worlds, while all other elements only affect the outcome in a single world. Then, the joint distribution of $(\boldsymbol{U})_{\overline{a}_{k-1}}$ can be learned from the observations with $\overline{A}_{k-1}~{=}~\overline{a}_{k-1}$, and the distribution of  the other elements can be learned from the observations with $\overline{A}_{k-1}~{=}~\overline{b}_{k-1}$.  This sufficient cross-world independence is less intuitive and will be harder to discuss with experts than Assumption \ref{CH4A4}.

\subsection{Gaussian linear mixed assignment closing}\label{CH4Ex1I}
Let us return to SCM \eqref{CH4SCM:ex1} introduced in Section \ref{CH4ex1}, where assumptions \ref{CH4A2}, \ref{CH4A1} and \ref{CH4A3} apply. Furthermore, $\forall \overline{a}$ such that $\sum_{k=1}^{j-1} a_{k} \geq 2$, $ (\boldsymbol{U}, \overline{N}_{Yj})_{\overline{a}_{j-1}}~{=}~ (U_{0}, U_{1}, U_{2}, \overline{N}_{Yj})$ $~{=}~ (\boldsymbol{U}, \overline{N}_{Yj})$, so that Assumption \ref{CH4A4} applies.  $F_{\boldsymbol{U}}$ is identifiable for $h{\geq} 3$ (when the functional form of the SCM is known) since $\overline{N}_{Y} \independent U_{0}, U_{1}, U_{2}$. 

\begin{figure}[H]
	\centering
	\begin{subfigure}{.325\textwidth}
		\resizebox{1\textwidth}{!}{\includegraphics[page=1]{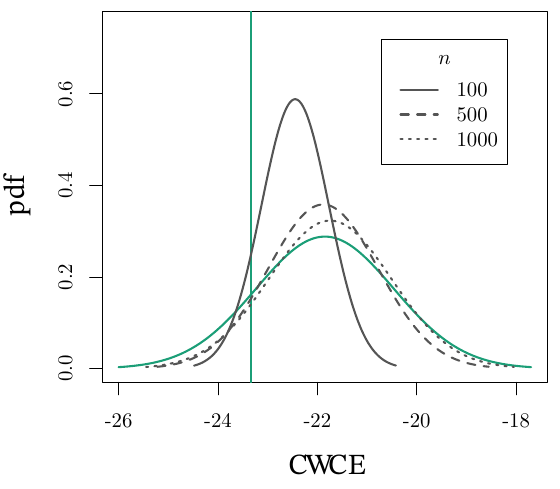}}
		\caption{}\label{CH4CWCEa}	
	\end{subfigure}
	\begin{subfigure}{.325\textwidth}
		\resizebox{1\textwidth}{!}{\includegraphics[page=1]{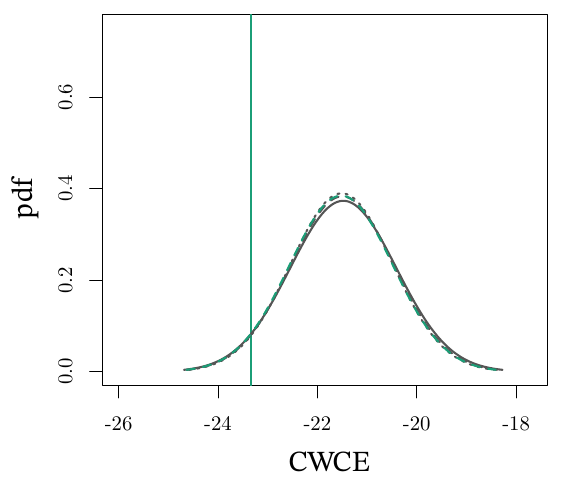}}
		\caption{}\label{CH4CWCEb}	
	\end{subfigure}
	\begin{subfigure}{.325\textwidth}
		\resizebox{1\textwidth}{!}{\includegraphics[page=1]{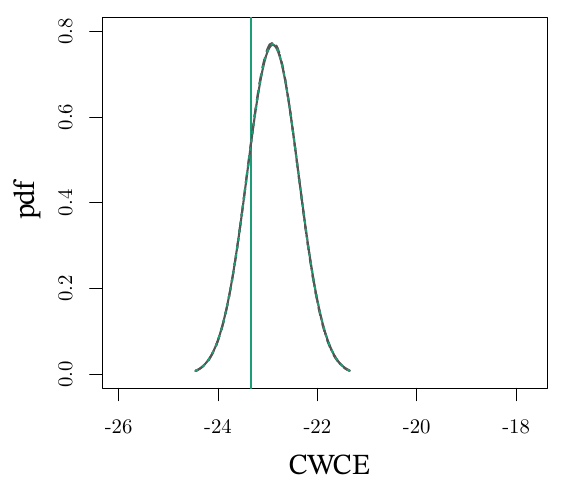}}
		\caption{}\label{CH4CWCEc}	
	\end{subfigure}
	\captionsetup{width=\textwidth}
	\caption{\small{Estimated CWCE distribution, at the third repeat, for a specific individual, based on a sample containing $100$ (solid grey), $500$ (dashed grey) and $1000$ (dotted grey) individuals	for $3$ (a), $10$ (b) and $100$ (c) repeats. The actual CWCE distribution (green lines) and ICE (vertical lines) were already presented in Figure \ref{CH4fig8a}. In Figure \ref{CH4CWCEb} and Figure \ref{CH4CWCEc}, the lines do (practically) overlap.}}\label{CH4fig8est}
\end{figure}

 Estimators of different parameters in latent variable models can converge at different rates depending on the number of repeats and individuals \citep{Miller1977}. Consistent estimators for mixed models are generally available but are not guaranteed in all cases \citep{Jiang2017}. Consistency of restricted maximum likelihood (REML) estimators under LMM has been shown for Gaussian distributed random effects \citep{Das1979}. 
 Consistency of the REML estimators 
 remains when the number of fixed effects increases with the sample size. On the contrary, the maximum likelihood estimator is then not consistent, as shown by the Neyman-Scott problem \citep{Jiang2017}. To estimate the parameters $\theta_{0}, \theta_{1}, \theta_{2}$, $\sigma_{0}, \sigma_{1}$ and $ \sigma_{2}$, 
 we fit a linear mixed model using REML estimation for the outcomes $Y_{1}, Y_{2}$ up to $Y_{h}$,
\begin{equation}\label{CH4model:latent}
Y_{ji}~{=}~ (Z_{0i}+\beta_{0}) + (Z_{1i}+\beta_{1})A_{j-1,i} + (Z_{2i}+\beta_{2})A_{j-2,i} + \beta_{L}L_{j-1,i} + \epsilon_{ji},
\end{equation} where $A_{ki}~{=}~0$ for $k~{<}~1$, $Z_{1i}~{\sim}~\mathcal{N}(0,\tau_{1}^2)$, $Z_{2i}~{\sim}~\mathcal{N}(0,\tau_{2}^2)$, $Z_{0i}~{\sim}~\mathcal{N}(0,\tau_{0}^2)$ and $\epsilon_{hi}~{\sim}~\mathcal{N}(0,\tau^2)$. For this LMM, REML estimation thus gives rise to consistent estimates of $(\theta_{0},\theta_{1},\theta_{2})$ and $(\sigma_{0},\sigma_{1},\sigma_{2})$ \citep{Verbeke2000, Jiang2017} so that Theorem \ref{CH4th41} applies. REML fitting of the mixed model was performed with \texttt{SAS~9.4}. 
We have estimated the distribution of the CWCE at the time of the third repeat $(j=3)$. For a specific individual (with $A_{1}=1$ and $A_{2}=1$) that was already highlighted in Figure \ref{CH4fig8a}, the estimated pdf of the CWCE is compared to the actual pdf for a varying number of individuals and number of repeats respectively in Figure \ref{CH4fig8est}. For $j=3$, the convergence presented in Theorem \ref{CH4th41} is already applicable for $h=3$ and $n=1000$ and $h=10$ and $n=100$. 

\begin{figure}[H]
	\centering
		\resizebox{\textwidth}{!}{\includegraphics{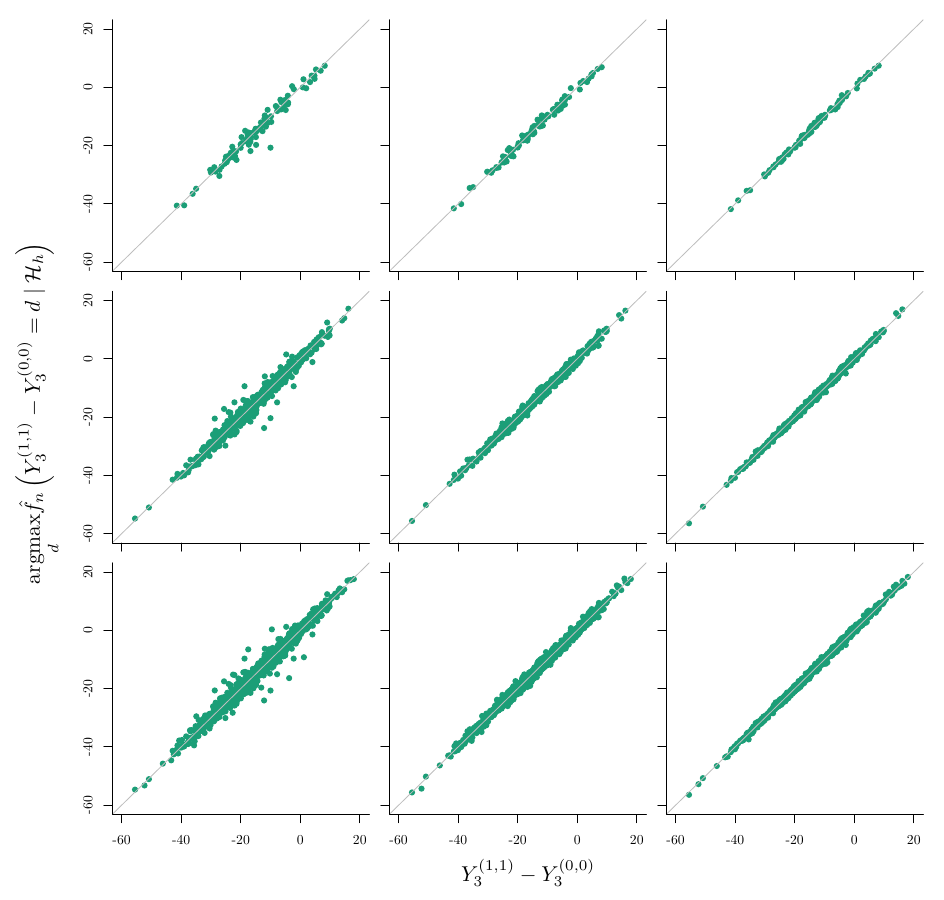}}
	\captionsetup{width=\linewidth}
	\caption{\small{The ICE of $\overline{a}{=}\overline{1}$ at the time of the third repeat versus the estimated ICE  based on different subsets of the data. The 
 rows correspond to the sample sizes $(100, 500, 1000)$ and the columns to the number of repeated measurements $(3,10,100)$. 
	}}\label{CH4GICE}
\end{figure}

\noindent Furthermore, we have estimated the ICE of $\overline{a}{=}\overline{1}$ for all individuals in the sample using the mode of the CWCE distribution (now equal to the expectation as the CWCE distribution was shown to be Gaussian in Section \ref{CH4sec:31}). In Figure \ref{CH4GICE}, we present the actual ICE versus the estimated ICE for different subsets of the simulated data. By fitting the random effects model to the observed data, we can accurately estimate the ICE for this example. For the parameter values introduced in Section \ref{CH4ex1}, an individual's ICE can already be accurately estimated given one hundred individuals with three repeated measurements each (Figure \ref{CH4GICE}). To accurately estimate $\exp\left(Y^{\overline{a}}_{j}\right)-\exp\left(Y^{\overline{0}}_{j}\right)$ given $\mathcal{H}_{h}$ one needs more repeats as shown in Figure \ref{CH4LNICE} in Section \ref{CH4ex2} of the Supplementary Material.

\noindent Notice that when $\beta_{1}$ and $\beta_{2}$ are estimated by fitting a model without random effects (using the full dataset), even the ACE estimate is biased as a consequence of the time-varying confounding; for this example, $\beta_{1}+\beta_{2}$ then equals $-14.6$ ($h~{=}~3$), $-13.6$ ($h~{=}~10$) and $-9.8$ ($h~{=}~100$) which deviate from the ACE, $\theta_{1}+\theta_{2}$, equal to $-15$. 
Moreover, when individuals with a particular exposure strategy are missing, the latent factors might be unidentifiable by Lemma \ref{CH4lemma:fUN}. For example, for $h~{=}~3$ when fitting the model with random effects on the sample without those individuals that were not exposed at the first two time points ($A_{1}~{=}~0$ or $A_{2}~{=}~0$), $\beta_{1}+\beta_{2}$ equals $-15.6$ $(h~{=}~3)$. 

The ICE distribution can be obtained by marginalizing the CWCE distribution over all individuals. For one hundred individuals, the resulting density was rather bumpy without additional smoothing, as shown in Figure \ref{CH4fig:ex1marginalicedistribution} in the Supplementary Material. Instead, one could apply a kernel density estimator to the estimated expected CWCE, which estimator is already quite accurate for data on one hundred individuals with each three repeated measurements, as illustrated in the figure. 
	
\subsection{Validity of the cross-world similarity assumption}\label{A4val}
Assumption \ref{CH4A4} that is sufficient for the identifiability of the CWCE is, like Assumption \ref{CH4A1}, a causal assumption that cannot be verified with data. Experts in the field of interest should review the validity of such assumptions. 
The repeated measurements under time-varying exposures can still be helpful when assessing the validity of Assumption \ref{CH4A4}. Under Assumption \ref{CH4A1}, one should be able to find a latent variable model that uniquely fits the repeated measurements for all exposure strategies so that there is a strategy whose potential outcome is affected by all $\boldsymbol{U}$. 
If we consider the Gaussian linear assignment example, under Assumption \ref{CH4A2}, then when the number of repeated measurements $h > 4$, one can verify that \begin{equation}\label{exampleA4} Y_{j}~\overset{d}=~ \beta_{0}+Z_{0}+\epsilon_{j} +\beta_{L}L_{j-1} +(\beta_{1}+Z_{1}) A_{j-1} + (\beta_{2}+Z_{2}) A_{j-2}, \end{equation} where all random effects are Gaussian distributed and $\overline{\epsilon} \independent Z_{0}, Z_{1}, Z_{2}$. The distributions of these random effects can be appropriately derived from the observed data assuming Assumption \ref{CH4A3} and following the reasoning presented in Section \ref{CH4TVC}. Note that for $h~{=}~3$, the independence between $Z_{1}$ and $\epsilon_{1}$ is not yet identifiable. This is identifiable for $h~{=}~4$, but the problem remains for $Z_{2}$ and $\epsilon_{2}$. For $h \leq 4$, the joint distribution of $(Z_{0}, Z_{1}, Z_{2}, \overline{\epsilon})$ is thus not yet identifiable,  but this is not the `causal' issue as it could be solved by increasing the number of repeats. The causal issue is that knowing the distribution of the observations is not enough to verify Assumption \ref{CH4A4}. Nevertheless, when the number of receptiveness factors is small compared to the number of repeated measurements, part of Assumption \ref{CH4A4} can be verified as demonstrated in Theorem \ref{CH4prop1} when the equality in distribution in Equation \eqref{exampleA4} and assumptions \ref{CH4A2} and \ref{CH4A1} hold. 

\begin{theorem}\label{CH4prop1}
	Under assumptions \ref{CH4A2} and \ref{CH4A1}, 
 $$Y_{j}~\overset{d}=~ \beta_{0}+Z_{0}+\epsilon_{j} +\beta_{L}L_{j-1} +(\beta_{1}+Z_{1}) A_{j-1} + (\beta_{2}+Z_{2}) A_{j-2},$$ implies  
		that the joint distribution of the ICEs of $\overline{a}$ at $q$ time points satisfies
  $$	
\resizebox{\linewidth}{!}{$(Y_{j_{1}}^{\overline{a}}-Y_{j_{1}}^{\overline{0}}, \hdots, Y_{j_{q}}^{\overline{a}}-Y_{j_{q}}^{\overline{0}})  ~\overset{d}=~ 
((\beta_{1}+Z_{1})a_{j_{1}-1} + 
(\beta_{2}+Z_{2})a_{j_{1}-2}, \hdots, (\beta_{1}+Z_{1})a_{j_{q}-1} + 
(\beta_{2}+Z_{2})a_{j_{q}-2}).$}$$ 
So, if $\forall p \in \{1,2, \hdots, q\}{:}~a_{j_{p}-1}~{=}~b_{k_{p}-1},$ and $a_{j_{p}-2}~{=}~b_{k_{p}-2}$, then 
	\begin{equation}\label{CH4eqd}(Y_{j_{1}}^{\overline{a}}-Y_{j_{1}}^{\overline{0}}, \hdots, Y_{j_{q}}^{\overline{a}}-Y_{j_{q}}^{\overline{0}}) 
	~\overset{d}=~ (Y_{k_{1}}^{\overline{b}}-Y_{k_{1}}^{\overline{0}}, \hdots, Y_{k_{q}}^{\overline{b}}-Y_{k_{q}}^{\overline{0}}).
 \end{equation}
	Moreover, if
	$a_{j-1} ~{=}~ a_{k-1}$ and $a_{j-2} ~{=}~ a_{k-2}$ then \begin{equation}\label{CH4eq}Y_{ji}^{\overline{a}}-Y_{ji}^{\overline{0}}~{=}~Y_{ki}^{\overline{a}}-Y_{ki}^{\overline{0}}.\end{equation}
\end{theorem}\noindent When the equality in distribution in Equation \eqref{exampleA4} applies, we can thus identify that $$( Y_{3}^{(0,1)}-Y_{3}^{(0,0)}, Y_{5}^{(0,1,0,1)}-Y_{5}^{(0,0,0,0)}) ~\overset{d}=~ (Z_{1}, Z_{1}) ~\overset{d}=~ (Y_{2}^{(1)}-Y_{2}^{(0)}, Y_{4}^{(1,0,1)}-Y_{4}^{(0,0,0)}),$$ implying that $Y_{3i}^{(0,1)}-Y_{3i}^{(0,0)} ~{=}~ Y_{5i}^{(0,1,0,1)}-Y_{5i}^{(0,0,0,0)}$. On the contrary, theoretically, $Y_{2i}^{(1)}-Y_{2i}^{(0)}$ can still differ from $Y_{3i}^{(0,1)}-Y_{3i}^{(0,0)}$, i.e.~two different (dependent) realizations from the distribution of $Z_{1}{\mid}Z_{0}$. Assumption \ref{CH4A4} thus simplifies to the assumption that the equality in distribution in Equation \eqref{CH4eqd} in Theorem \ref{CH4prop1} is an actual equality. Again, we can never verify this equality due to the fundamental problem of causal inference. However, the equality in Equation \eqref{CH4eq} stated at the end of Theorem \ref{CH4prop1} will be important for an expert to review the validity of Assumption \ref{CH4A4}. This equality, e.g.~$Y_{2i}^{(1)}-Y_{2i}^{(0)} ~{=}~ Y_{4i}^{(1,0,1)}-Y_{4i}^{(0,0,0)}$, excludes the existence of time-varying causal effects. Time-varying causal effects could, for instance, arise when the effect depends on the value of $Y^{\overline{0}}$ (as is the case for the log-normal linear mixed assignment as presented in Section \ref{CH4ex2}). The expert can most likely rule out the theoretical examples in which $Y_{3i}^{(0,1)}-Y_{3i}^{(0,0)} \neq Y_{2i}^{(1)}-Y_{2i}^{(0)}$, while $Y_{3i}^{(0,1)}-Y_{3i}^{(0,0)} ~{=}~ Y_{5i}^{(0,1,0,1)}-Y_{5i}^{(0,0,0,0)}$. 
 
In summary, with repeated measurements under time-varying exposures, we may learn the joint distributions of $(\boldsymbol{U}, \overline{N}_{Y})_{\overline{a}_{h-1}}$ for all ${\overline{a}}_{h-1}$. When a low dimensional latent variable model can fit the observations, we can identify equality in distribution of the causal effects in different worlds. Finally, to fit the CWCE distribution, one should discuss with an expert whether this cross-world equality in distribution is an actual equality. 

\section{Discussion}
In the era of personalized medicine, there is a need to understand the effect of a treatment on an individual. So far, causal inference is, at best, focused on estimating the conditional average treatment effects (CATEs). 
For cross-sectional data, individual-specific effect modification cannot be disentangled from the variability between individuals. However, from longitudinal data with time-varying exposures,  one can observe how an individual responds to a change in exposure. We proposed a general framework to parameterize the cause-effect relations with an SCM in which receptiveness factors (i.e.~unmeasured modifiers) explain individual differences in causal effects. Subsequently, we have studied the CWCE, which represents the distribution of ICEs of individuals who share the same set of observed values $\overline{Y}_{h}, \overline{\boldsymbol{M}}_{h}, \overline{\boldsymbol{L}}_{h}, \overline{A}_{h}$. Certain levels of the unmeasured receptiveness factors become more likely given the repeated measurements.  In the case of injective outcome-assignment functions in the SCM, the CWCE converges the ICE with an increasing number of repeats. The presented theory extends existing theories by studying the joint distributions of potential outcomes while conditioning (across worlds) on the history of the real world. As a result, the distribution of the CWCE for individuals can be studied. The factual-world outcomes can improve the precision for estimating and predicting the ICE in individuals. For cause-effect relations where the individual-effect modification is relatively simple (e.g.~time-invariant), the resulting CWCE is low in variability and thus strongly informs about the ICE. Furthermore, from the individual CWCE distributions, marginal or stratified causal-effect distributions can be obtained.  For instance, the set of ICEs for hypertension treatment in men with volatile temporal blood-pressure patterns may show large positive and negative effects. Or the distribution of ICEs in women with a specific exposure strategy differs from that of men with the same exposure strategy (on average, in variability, or any other distributional characteristic). In such examples, the expectation of the conditional ICE distribution links our framework to existing causal theory for CATEs. Our framework was illustrated by presenting a linear mixed-effects assignment as an example, but the framework is more generic, and the receptiveness factors' distributions are not restricted to these classes. 

Most of this chapter focused on quantifying heterogeneity in longitudinal cause-effect relations via the distribution of the CWCE when the SCM is known. In practice, the parameters and functional forms in the SCM are unknown and should be estimated from data. As a result of the fundamental problem of causal inference, the joint distribution of the receptiveness factors for different exposure strategies is not identifiable. Therefore, the CWCE is only identifiable from observational data when the joint distribution of the receptiveness factors can be identified from observations with a particular exposure strategy. As the common assumption of absence of direct unmeasured confounding, this causal assumption cannot be tested and should be discussed with experts. Under assumptions \ref{CH4A2}, \ref{CH4A1}, \ref{CH4A3}, and \ref{CH4A4}, by Theorem \ref{CH4th41}, valid inference on the CWCE might be made possible from observational data. Then, the ICE could be estimated with the mode of this estimated CWCE distribution. For the linear mixed-effects assignment, only a small number of repeats was needed to estimate the ICEs accurately, while for non-linear mixed-effects assignment, a larger number of repeats may be necessary to reduce the variability of the distribution of the CWCE significantly (see, e.g.~Figure \ref{CH4LNICE} in the Supplementary Material). Innovative data collection methods \citep{Dias2018} and specific longitudinal data designs, e.g.~micro-randomized trials \citep{Li2020, Qian2020}, may help provide such large numbers of repeats under varying exposures. Using the mode of the CWCE distribution to estimate the ICE may not be consistent for some individuals when there are time-varying receptiveness factors that can never be retrieved from the repeated measurements with full certainty, e.g.~for the logistic linear mixed-effects assignment as presented in Section \ref{CH4ex3} of the Supplementary Material. 

Mixed models are often used to estimate (average) treatment effect parameters from longitudinal data.  If these models are well-specified, the distribution of the CWCEs based on their fits makes mixed models suitable for estimating ICEs. Moving towards such practice requires sufficient data to be able to validate the model for the marginal distribution of 
$\overline{Y}_{j}\mid \overline{\boldsymbol{M}}_{j-1}, \overline{\boldsymbol{L}}_{j-1}, \overline{A}_{j-1}$.
The latter is crucial as different combinations of conditional models for 
$\overline{Y}_{j}\mid \boldsymbol{U}, \overline{\boldsymbol{M}}_{j-1}, \overline{\boldsymbol{L}}_{j-1}, \overline{A}_{j-1}, \overline{N}_{Yj}$
and distributions of $(\boldsymbol{U}, \overline{N}_{Y})$ can result in the same marginal distribution. The repeated measurements with time-varying exposures can be used to make an inference on the distribution of $((\boldsymbol{U})_{\overline{a}_{h-1}}, \overline{N}_{Yh})$, 
which might be identifiable from the observational data. Assumption \ref{CH4A4} cannot be completely verified with data. However, if the observed data is appropriately described with a low-dimensional latent variable model (see Section \ref{A4val}), the assumption becomes more realistic and can be translated into a weaker assumption that can be better discussed with field experts. Formalization of procedures for validation of the latent variable model should thus be the topic of future research to make our framework operational. Furthermore, integration of mixed models with machine learning techniques (see, e.g.~\citep{Hajjem2014, Hajjem2017}) could extend recently developed flexible techniques to estimate CATEs \citep{Wendling2018, Green2012, Hill2011, Wager2018, Athey2019, Lu2018, Bica2021}, and might allow our framework to be used for individual causal inference in a broad range of applications. 

\bibliographystyle{imsart-nameyear} 
\bibliography{ReferencesICE}   

\begin{thebibliography}{52}

\bibitem[\protect\citeauthoryear{Athey, Tibshirani and Wager}{2019}]{Athey2019}
\begin{barticle}[author]
\bauthor{\bsnm{Athey},~\bfnm{Susan}\binits{S.}},
  \bauthor{\bsnm{Tibshirani},~\bfnm{Julie}\binits{J.}} \AND
  \bauthor{\bsnm{Wager},~\bfnm{Stefan}\binits{S.}}
(\byear{2019}).
\btitle{{Generalized random forests}}.
\bjournal{Annals of Statistics}
\bvolume{47}
\bpages{1179--1203}.
\bdoi{10.1214/18-AOS1709}
\end{barticle}
\endbibitem

\bibitem[\protect\citeauthoryear{Balke and Pearl}{1994}]{Balke1994}
\begin{binproceedings}[author]
\bauthor{\bsnm{Balke},~\bfnm{Alexander}\binits{A.}} \AND
  \bauthor{\bsnm{Pearl},~\bfnm{Judea}\binits{J.}}
(\byear{1994}).
\btitle{Probabilistic Evaluation of Counterfactual Queries}.
In \bbooktitle{Proceedings of the Twelfth AAAI National Conference on
  Artificial Intelligence}.
\bseries{AAAI'94}
\bpages{230–237}.
\bpublisher{AAAI Press}.
\bdoi{10.5555/2891730.2891765}
\end{binproceedings}
\endbibitem

\bibitem[\protect\citeauthoryear{Bica et~al.}{2021}]{Bica2021}
\begin{barticle}[author]
\bauthor{\bsnm{Bica},~\bfnm{Ioana}\binits{I.}},
  \bauthor{\bsnm{Alaa},~\bfnm{Ahmed~M.}\binits{A.~M.}},
  \bauthor{\bsnm{Lambert},~\bfnm{Craig}\binits{C.}} \AND \bauthor{\bsnm{{van
  der Schaar}},~\bfnm{Mihaela}\binits{M.}}
(\byear{2021}).
\btitle{From Real-World Patient Data to Individualized Treatment Effects Using
  Machine Learning: Current and Future Methods to Address Underlying
  Challenges}.
\bjournal{Clinical Pharmacology \& Therapeutics}
\bvolume{109}
\bpages{87--100}.
\bdoi{10.1002/cpt.1907}
\end{barticle}
\endbibitem

\bibitem[\protect\citeauthoryear{Bongers et~al.}{2021}]{Bongers2021}
\begin{barticle}[author]
\bauthor{\bsnm{Bongers},~\bfnm{Stephan}\binits{S.}},
  \bauthor{\bsnm{Forr{\'e}},~\bfnm{Patrick}\binits{P.}},
  \bauthor{\bsnm{Peters},~\bfnm{Jonas}\binits{J.}} \AND
  \bauthor{\bsnm{Mooij},~\bfnm{Joris~M.}\binits{J.~M.}}
(\byear{2021}).
\btitle{{Foundations of structural causal models with cycles and latent
  variables}}.
\bjournal{The Annals of Statistics}
\bvolume{49}
\bpages{2885 -- 2915}.
\bdoi{10.1214/21-AOS2064}
\end{barticle}
\endbibitem

\bibitem[\protect\citeauthoryear{Daniel et~al.}{2013}]{Daniel2013}
\begin{barticle}[author]
\bauthor{\bsnm{Daniel},~\bfnm{Rhian~M.}\binits{R.~M.}},
  \bauthor{\bsnm{Cousens},~\bfnm{Simon~N.}\binits{S.~N.}}, \bauthor{\bsnm{{De
  Stavola}},~\bfnm{Bianca~L.}\binits{B.~L.}},
  \bauthor{\bsnm{Kenward},~\bfnm{Michael~G.}\binits{M.~G.}} \AND
  \bauthor{\bsnm{Sterne},~\bfnm{Jonathan A.~C.}\binits{J.~A.~C.}}
(\byear{2013}).
\btitle{{Methods for dealing with time-dependent confounding}}.
\bjournal{Statistics in Medicine}
\bvolume{32}
\bpages{1584--1618}.
\bdoi{10.1002/sim.5686}
\end{barticle}
\endbibitem

\bibitem[\protect\citeauthoryear{Das}{1979}]{Das1979}
\begin{barticle}[author]
\bauthor{\bsnm{Das},~\bfnm{Kalyan}\binits{K.}}
(\byear{1979}).
\btitle{Asymptotic Optimality of Restricted Maximum Likelihood Estimates for
  the Mixed Model}.
\bjournal{Calcutta Statistical Association Bulletin}
\bvolume{28}
\bpages{125--142}.
\bdoi{10.1177/0008068319790108}
\end{barticle}
\endbibitem

\bibitem[\protect\citeauthoryear{Dias and {Silva Cunha}}{2018}]{Dias2018}
\begin{barticle}[author]
\bauthor{\bsnm{Dias},~\bfnm{Duarte}\binits{D.}} \AND \bauthor{\bsnm{{Silva
  Cunha}},~\bfnm{João~Paulo}\binits{J.~P.}}
(\byear{2018}).
\btitle{Wearable Health Devices—Vital Sign Monitoring, Systems and
  Technologies}.
\bjournal{Sensors}
\bvolume{18}.
\bdoi{10.3390/s18082414}
\end{barticle}
\endbibitem

\bibitem[\protect\citeauthoryear{Duan, Kravitz and Schmid}{2013}]{Duan2013}
\begin{barticle}[author]
\bauthor{\bsnm{Duan},~\bfnm{N.}\binits{N.}},
  \bauthor{\bsnm{Kravitz},~\bfnm{R.~L.}\binits{R.~L.}} \AND
  \bauthor{\bsnm{Schmid},~\bfnm{C.~H.}\binits{C.~H.}}
(\byear{2013}).
\btitle{Single-patient (n-of-1) trials: A pragmatic clinical decision
  methodology for patient-centered comparative effectiveness research}.
\bjournal{Journal of Clinical Epidemiology}
\bvolume{66}
\bpages{S21-S28}.
\bdoi{10.1016/j.jclinepi.2013.04.006}
\end{barticle}
\endbibitem

\bibitem[\protect\citeauthoryear{Green and Kern}{2012}]{Green2012}
\begin{barticle}[author]
\bauthor{\bsnm{Green},~\bfnm{Donald~P.}\binits{D.~P.}} \AND
  \bauthor{\bsnm{Kern},~\bfnm{Holger~L.}\binits{H.~L.}}
(\byear{2012}).
\btitle{{Modeling heterogeneous treatment effects in survey experiments with
  bayesian additive regression trees}}.
\bjournal{Public Opinion Quarterly}
\bvolume{76}
\bpages{491--511}.
\bdoi{10.1093/poq/nfs036}
\end{barticle}
\endbibitem

\bibitem[\protect\citeauthoryear{Greenland et~al.}{2019}]{Greenland2019}
\begin{barticle}[author]
\bauthor{\bsnm{Greenland},~\bfnm{Sander}\binits{S.}},
  \bauthor{\bsnm{Fay},~\bfnm{Michael~P.}\binits{M.~P.}},
  \bauthor{\bsnm{Brittain},~\bfnm{Erica~H.}\binits{E.~H.}},
  \bauthor{\bsnm{Shih},~\bfnm{Joanna~H.}\binits{J.~H.}},
  \bauthor{\bsnm{Follmann},~\bfnm{Dean~A.}\binits{D.~A.}},
  \bauthor{\bsnm{Gabriel},~\bfnm{Erin~E.}\binits{E.~E.}} \AND
  \bauthor{\bsnm{Robins},~\bfnm{James~M.}\binits{J.~M.}}
(\byear{2019}).
\btitle{{On Causal Inferences for Personalized Medicine: How Hidden Causal
  Assumptions Led to Erroneous Causal Claims About the D-Value}}.
\bjournal{The American Statistician}
\bvolume{74}
\bpages{243--248}.
\bdoi{10.1080/00031305.2019.1575771}
\end{barticle}
\endbibitem

\bibitem[\protect\citeauthoryear{{The SPRINT Research
  Group}}{2015}]{SPRINT2015}
\begin{barticle}[author]
\bauthor{\bsnm{{The SPRINT Research Group}}}
(\byear{2015}).
\btitle{{A Randomized Trial of Intensive versus Standard Blood-Pressure
  Control}}.
\bjournal{New England Journal of Medicine}
\bvolume{373}
\bpages{2103--2116}.
\bdoi{10.1056/nejmoa1511939}
\end{barticle}
\endbibitem

\bibitem[\protect\citeauthoryear{Hajjem, Bellavance and
  Larocque}{2014}]{Hajjem2014}
\begin{barticle}[author]
\bauthor{\bsnm{Hajjem},~\bfnm{Ahlem}\binits{A.}},
  \bauthor{\bsnm{Bellavance},~\bfnm{François}\binits{F.}} \AND
  \bauthor{\bsnm{Larocque},~\bfnm{Denis}\binits{D.}}
(\byear{2014}).
\btitle{Mixed-effects random forest for clustered data}.
\bjournal{Journal of Statistical Computation and Simulation}
\bvolume{84}
\bpages{1313--1328}.
\bdoi{10.1080/00949655.2012.741599}
\end{barticle}
\endbibitem

\bibitem[\protect\citeauthoryear{Hajjem, Larocque and
  Bellavance}{2017}]{Hajjem2017}
\begin{barticle}[author]
\bauthor{\bsnm{Hajjem},~\bfnm{A.}\binits{A.}},
  \bauthor{\bsnm{Larocque},~\bfnm{D.}\binits{D.}} \AND
  \bauthor{\bsnm{Bellavance},~\bfnm{F.}\binits{F.}}
(\byear{2017}).
\btitle{Generalized mixed effects regression trees}.
\bjournal{Statistics and Probability Letters}
\bvolume{126}
\bpages{114--118}.
\bdoi{10.1016/j.spl.2017.02.033}
\end{barticle}
\endbibitem

\bibitem[\protect\citeauthoryear{Hand}{1992}]{Hand1992}
\begin{barticle}[author]
\bauthor{\bsnm{Hand},~\bfnm{David~J.}\binits{D.~J.}}
(\byear{1992}).
\btitle{{On comparing two treatments}}.
\bjournal{The American Statistician}
\bvolume{46}
\bpages{190--192}.
\bdoi{10.1080/00031305.1992.10475881}
\end{barticle}
\endbibitem

\bibitem[\protect\citeauthoryear{Hern{\'{a}}n}{2004}]{Hernan2004b}
\begin{barticle}[author]
\bauthor{\bsnm{Hern{\'{a}}n},~\bfnm{Miguel~A}\binits{M.~A.}}
(\byear{2004}).
\btitle{{A definition of causal effect for epidemiological research}}.
\bjournal{Journal of the Royal Statistical Society: Series B (Statistical
  Methodology)}
\bvolume{58}
\bpages{265--271}.
\bdoi{10.1136/jech.2002.006361}
\end{barticle}
\endbibitem

\bibitem[\protect\citeauthoryear{Hern{\'{a}}n and Robins}{2020}]{Hernan2019}
\begin{bbook}[author]
\bauthor{\bsnm{Hern{\'{a}}n},~\bfnm{Miguel~A}\binits{M.~A.}} \AND
  \bauthor{\bsnm{Robins},~\bfnm{James~M.}\binits{J.~M.}}
(\byear{2020}).
\btitle{Causal Inference: What If.},
\bedition{1st} ed.
\bpublisher{Boca Raton: Chapman {\&} Hall/CRC}, \baddress{Boca Raton, Florida}.
\end{bbook}
\endbibitem

\bibitem[\protect\citeauthoryear{Hill}{2011}]{Hill2011}
\begin{barticle}[author]
\bauthor{\bsnm{Hill},~\bfnm{Jennifer~L.}\binits{J.~L.}}
(\byear{2011}).
\btitle{{Bayesian nonparametric modeling for causal inference}}.
\bjournal{Journal of Computational and Graphical Statistics}
\bvolume{20}
\bpages{217--240}.
\bdoi{10.1198/jcgs.2010.08162}
\end{barticle}
\endbibitem

\bibitem[\protect\citeauthoryear{Holland}{1986}]{Holland1986}
\begin{barticle}[author]
\bauthor{\bsnm{Holland},~\bfnm{Paul~W.}\binits{P.~W.}}
(\byear{1986}).
\btitle{{Statistics and causal inference}}.
\bjournal{Journal of the American Statistical Association}
\bvolume{81}
\bpages{945--960}.
\bdoi{10.1080/01621459.1986.10478354}
\end{barticle}
\endbibitem

\bibitem[\protect\citeauthoryear{Jiang}{2017}]{Jiang2017}
\begin{bbook}[author]
\bauthor{\bsnm{Jiang},~\bfnm{J.}\binits{J.}}
(\byear{2017}).
\btitle{Asymptotic analysis of mixed effects models: Theory, applications, and
  open problems},
\bedition{1st} ed.
\bpublisher{CRC Press}, \baddress{Boca Raton, Florida}.
\bdoi{10.1201/9781315119281}
\end{bbook}
\endbibitem

\bibitem[\protect\citeauthoryear{Kane, Bittlinger and
  Kimmelman}{2021}]{Kane2021}
\begin{barticle}[author]
\bauthor{\bsnm{Kane},~\bfnm{P.~B.}\binits{P.~B.}},
  \bauthor{\bsnm{Bittlinger},~\bfnm{M.}\binits{M.}} \AND
  \bauthor{\bsnm{Kimmelman},~\bfnm{J.}\binits{J.}}
(\byear{2021}).
\btitle{Individualized therapy trials: navigating patient care, research goals
  and ethics}.
\bjournal{Nature Medicine}
\bvolume{27}
\bpages{1679--1686}.
\bdoi{10.1038/s41591-021-01519-y}
\end{barticle}
\endbibitem

\bibitem[\protect\citeauthoryear{Kravitz, Duan and Braslow}{2004}]{Kravitz2004}
\begin{barticle}[author]
\bauthor{\bsnm{Kravitz},~\bfnm{R.~L.}\binits{R.~L.}},
  \bauthor{\bsnm{Duan},~\bfnm{N.}\binits{N.}} \AND
  \bauthor{\bsnm{Braslow},~\bfnm{J.}\binits{J.}}
(\byear{2004}).
\btitle{Evidence-based medicine, heterogeneity of treatment effects, and the
  trouble with averages}.
\bjournal{Milbank Quarterly}
\bvolume{82}
\bpages{661--687}.
\bdoi{10.1111/j.0887-378X.2004.00327.x}
\end{barticle}
\endbibitem

\bibitem[\protect\citeauthoryear{Li et~al.}{2020}]{Li2020}
\begin{barticle}[author]
\bauthor{\bsnm{Li},~\bfnm{Shuang}\binits{S.}},
  \bauthor{\bsnm{Psihogios},~\bfnm{Alexandra~M.}\binits{A.~M.}},
  \bauthor{\bsnm{McKelvey},~\bfnm{Elise~R.}\binits{E.~R.}},
  \bauthor{\bsnm{Ahmed},~\bfnm{Annisa}\binits{A.}},
  \bauthor{\bsnm{Rabbi},~\bfnm{Mashfiqui}\binits{M.}} \AND
  \bauthor{\bsnm{Murphy},~\bfnm{Susan~A.}\binits{S.~A.}}
(\byear{2020}).
\btitle{Microrandomized trials for promoting engagement in mobile health data
  collection: Adolescent/young adult oral chemotherapy adherence as an
  example}.
\bjournal{Current Opinion in Systems Biology}
\bvolume{21}
\bpages{1--8}.
\bdoi{10.1016/j.coisb.2020.07.002}
\end{barticle}
\endbibitem

\bibitem[\protect\citeauthoryear{Lillie et~al.}{2011}]{Lillie2011}
\begin{barticle}[author]
\bauthor{\bsnm{Lillie},~\bfnm{E.~O.}\binits{E.~O.}},
  \bauthor{\bsnm{Patay},~\bfnm{B.}\binits{B.}},
  \bauthor{\bsnm{Diamant},~\bfnm{J.}\binits{J.}},
  \bauthor{\bsnm{Issell},~\bfnm{B.}\binits{B.}},
  \bauthor{\bsnm{Topol},~\bfnm{E.~J.}\binits{E.~J.}} \AND
  \bauthor{\bsnm{Schork},~\bfnm{N.~J.}\binits{N.~J.}}
(\byear{2011}).
\btitle{The n-of-1 clinical trial: The ultimate strategy for individualizing
  medicine?}
\bjournal{Personalized Medicine}
\bvolume{8}
\bpages{161--173}.
\bdoi{10.2217/pme.11.7}
\end{barticle}
\endbibitem

\bibitem[\protect\citeauthoryear{Lok et~al.}{2004}]{Lok2004}
\begin{barticle}[author]
\bauthor{\bsnm{Lok},~\bfnm{Judith}\binits{J.}},
  \bauthor{\bsnm{Gill},~\bfnm{Richard}\binits{R.}}, \bauthor{\bsnm{{van der
  Vaart}},~\bfnm{Aad}\binits{A.}} \AND
  \bauthor{\bsnm{Robins},~\bfnm{James~M.}\binits{J.~M.}}
(\byear{2004}).
\btitle{Estimating the causal effect of a time-varying treatment on
  time-to-event using structural nested failure time models}.
\bjournal{Statistica Neerlandica}
\bvolume{58}
\bpages{271--295}.
\bdoi{10.1111/j.1467-9574.2004.00123.x}
\end{barticle}
\endbibitem

\bibitem[\protect\citeauthoryear{Lu et~al.}{2018}]{Lu2018}
\begin{barticle}[author]
\bauthor{\bsnm{Lu},~\bfnm{Min}\binits{M.}},
  \bauthor{\bsnm{Sadiq},~\bfnm{Saad}\binits{S.}},
  \bauthor{\bsnm{Feaster},~\bfnm{Daniel~J.}\binits{D.~J.}} \AND
  \bauthor{\bsnm{Ishwaran},~\bfnm{Hemant}\binits{H.}}
(\byear{2018}).
\btitle{{Estimating Individual Treatment Effect in Observational Data Using
  Random Forest Methods}}.
\bjournal{Journal of Computational and Graphical Statistics}
\bvolume{27}
\bpages{209--219}.
\bnote{PMID: 29706752}.
\bdoi{10.1080/10618600.2017.1356325}
\end{barticle}
\endbibitem

\bibitem[\protect\citeauthoryear{Miller}{1977}]{Miller1977}
\begin{barticle}[author]
\bauthor{\bsnm{Miller},~\bfnm{John~J.}\binits{J.~J.}}
(\byear{1977}).
\btitle{Asymptotic Properties of Maximum Likelihood Estimates in the Mixed
  Model of the Analysis of Variance}.
\bjournal{The Annals of Statistics}
\bvolume{5}
\bpages{746--762}.
\bdoi{10.1214/aos/1176343897}
\end{barticle}
\endbibitem

\bibitem[\protect\citeauthoryear{Murphy}{2003}]{Murphy2003}
\begin{barticle}[author]
\bauthor{\bsnm{Murphy},~\bfnm{Susan~A.}\binits{S.~A.}}
(\byear{2003}).
\btitle{{Optimal dynamic treatment regimes}}.
\bjournal{Journal of the Royal Statistical Society: Series B (Statistical
  Methodology)}
\bvolume{65}
\bpages{331--366}.
\bdoi{10.1111/1467-9868.00389}
\end{barticle}
\endbibitem

\bibitem[\protect\citeauthoryear{Naimi, Cole and Kennedy}{2017}]{Naimi2017}
\begin{barticle}[author]
\bauthor{\bsnm{Naimi},~\bfnm{Ashley~I.}\binits{A.~I.}},
  \bauthor{\bsnm{Cole},~\bfnm{Stephen~R.}\binits{S.~R.}} \AND
  \bauthor{\bsnm{Kennedy},~\bfnm{Edward~H.}\binits{E.~H.}}
(\byear{2017}).
\btitle{{An introduction to g methods}}.
\bjournal{International Journal of Epidemiology}
\bvolume{46}
\bpages{756--762}.
\bdoi{10.1093/ije/dyw323}
\end{barticle}
\endbibitem

\bibitem[\protect\citeauthoryear{Neyman}{1923}]{Neyman1990}
\begin{barticle}[author]
\bauthor{\bsnm{Neyman},~\bfnm{Jerzy}\binits{J.}}
(\byear{1923}).
\btitle{{On the Application of Probability Theory to Agricultural Experiments.
  Essay on Principles}}.
\bjournal{Statistical Science}
\bvolume{5}
\bpages{465--472}.
\bdoi{10.1214/ss/1177012031}
\end{barticle}
\endbibitem

\bibitem[\protect\citeauthoryear{Pearl}{1995}]{Pearl1995}
\begin{barticle}[author]
\bauthor{\bsnm{Pearl},~\bfnm{Judea}\binits{J.}}
(\byear{1995}).
\btitle{{Causal diagrams for empirical research}}.
\bjournal{Biometrika}
\bvolume{82}
\bpages{669--688}.
\bdoi{10.1093/biomet/82.4.669}
\end{barticle}
\endbibitem

\bibitem[\protect\citeauthoryear{Pearl}{2009}]{Pearl2009book}
\begin{bbook}[author]
\bauthor{\bsnm{Pearl},~\bfnm{Judea}\binits{J.}}
(\byear{2009}).
\btitle{{Causality: Models, reasoning, and inference}},
\bedition{2nd} ed.
\bpublisher{Cambridge University Press}.
\bdoi{10.1017/CBO9780511803161}
\end{bbook}
\endbibitem

\bibitem[\protect\citeauthoryear{Peters, Janzing and
  Sch{\"o}lkopf}{2018}]{Peters2017}
\begin{bbook}[author]
\bauthor{\bsnm{Peters},~\bfnm{Jonas}\binits{J.}},
  \bauthor{\bsnm{Janzing},~\bfnm{Dominik}\binits{D.}} \AND
  \bauthor{\bsnm{Sch{\"o}lkopf},~\bfnm{Bernhard}\binits{B.}}
(\byear{2018}).
\btitle{{Elements of causal inference: foundations and learning algorithms}},
\bedition{1st} ed.
\bpublisher{The MIT Press}, \baddress{Cambridge}.
\bdoi{10.1080/00949655.2018.1505197}
\end{bbook}
\endbibitem

\bibitem[\protect\citeauthoryear{Qian, Klasnja and Murphy}{2020}]{Qian2020}
\begin{barticle}[author]
\bauthor{\bsnm{Qian},~\bfnm{Tianchen}\binits{T.}},
  \bauthor{\bsnm{Klasnja},~\bfnm{Predrag}\binits{P.}} \AND
  \bauthor{\bsnm{Murphy},~\bfnm{Susan~A.}\binits{S.~A.}}
(\byear{2020}).
\btitle{{Linear mixed models with endogenous covariates : modeling sequential
  treatment effects with application to a mobile health study}}.
\bjournal{Statistical Science}
\bvolume{35}
\bpages{375--390}.
\bdoi{10.1214/19-sts720}
\end{barticle}
\endbibitem

\bibitem[\protect\citeauthoryear{Raman et~al.}{2018}]{Raman2018}
\begin{barticle}[author]
\bauthor{\bsnm{Raman},~\bfnm{G.}\binits{G.}},
  \bauthor{\bsnm{Balk},~\bfnm{E.~M.}\binits{E.~M.}},
  \bauthor{\bsnm{Lai},~\bfnm{L.}\binits{L.}},
  \bauthor{\bsnm{Shi},~\bfnm{J.}\binits{J.}},
  \bauthor{\bsnm{Chan},~\bfnm{J.}\binits{J.}},
  \bauthor{\bsnm{Lutz},~\bfnm{J.~S.}\binits{J.~S.}},
  \bauthor{\bsnm{Dubois},~\bfnm{R.~W.}\binits{R.~W.}},
  \bauthor{\bsnm{Kravitz},~\bfnm{R.~L.}\binits{R.~L.}} \AND
  \bauthor{\bsnm{Kent},~\bfnm{David~M}\binits{D.~M.}}
(\byear{2018}).
\btitle{Evaluation of person-level heterogeneity of treatment effects in
  published multiperson N-of-1 studies: Systematic review and reanalysis}.
\bjournal{BMJ Open}
\bvolume{8}.
\bdoi{10.1136/bmjopen-2017-017641}
\end{barticle}
\endbibitem

\bibitem[\protect\citeauthoryear{Robins, Hern{\'{a}}n and
  Brumback}{2000}]{Robins2000}
\begin{barticle}[author]
\bauthor{\bsnm{Robins},~\bfnm{James~M.}\binits{J.~M.}},
  \bauthor{\bsnm{Hern{\'{a}}n},~\bfnm{Miguel~A}\binits{M.~A.}} \AND
  \bauthor{\bsnm{Brumback},~\bfnm{Babette}\binits{B.}}
(\byear{2000}).
\btitle{{Marginal structural models and causal inference in epidemiology}}.
\bjournal{Epidemiology}
\bvolume{11}
\bpages{550--560}.
\bdoi{10.1097/00001648-200009000-00011}
\end{barticle}
\endbibitem

\bibitem[\protect\citeauthoryear{Robins, Orellana and
  Rotnitzky}{2008}]{Robins2008}
\begin{barticle}[author]
\bauthor{\bsnm{Robins},~\bfnm{James~M.}\binits{J.~M.}},
  \bauthor{\bsnm{Orellana},~\bfnm{Liliana}\binits{L.}} \AND
  \bauthor{\bsnm{Rotnitzky},~\bfnm{Andrea}\binits{A.}}
(\byear{2008}).
\btitle{{Estimation and extrapolation of optimal treatment and testing
  strategies}}.
\bjournal{Statistics in Medicine}
\bvolume{27}
\bpages{4678--4721}.
\bdoi{10.1002/sim.3301}
\end{barticle}
\endbibitem

\bibitem[\protect\citeauthoryear{Rothman, Greenland and
  Lash}{2008}]{Rothman2008}
\begin{bbook}[author]
\bauthor{\bsnm{Rothman},~\bfnm{Kenneth~J.}\binits{K.~J.}},
  \bauthor{\bsnm{Greenland},~\bfnm{Sander}\binits{S.}} \AND
  \bauthor{\bsnm{Lash},~\bfnm{Timothy~L.}\binits{T.~L.}}
(\byear{2008}).
\btitle{{Modern epidemiology}},
\bedition{3th} ed.
\bpublisher{Lippincott Williams \& Wilkins}, \baddress{Philadelphia,
  Pennsylvania}.
\end{bbook}
\endbibitem

\bibitem[\protect\citeauthoryear{Rubin}{1974}]{Rubin1974}
\begin{barticle}[author]
\bauthor{\bsnm{Rubin},~\bfnm{Donald~B.}\binits{D.~B.}}
(\byear{1974}).
\btitle{{Estimating causal effects of treatments in randomized and
  nonrandomized studies}}.
\bjournal{Journal of Educational Psychology}
\bvolume{66}
\bpages{688--701}.
\bdoi{10.1037/h0037350}
\end{barticle}
\endbibitem

\bibitem[\protect\citeauthoryear{Senarathne, Overstall and
  McGree}{2020}]{Senarathne2020}
\begin{barticle}[author]
\bauthor{\bsnm{Senarathne},~\bfnm{S.~G.~J.}\binits{S.~G.~J.}},
  \bauthor{\bsnm{Overstall},~\bfnm{A.~M.}\binits{A.~M.}} \AND
  \bauthor{\bsnm{McGree},~\bfnm{J.~M.}\binits{J.~M.}}
(\byear{2020}).
\btitle{Bayesian adaptive N-of-1 trials for estimating population and
  individual treatment effects}.
\bjournal{Statistics in Medicine}
\bvolume{39}
\bpages{4499--4518}.
\bdoi{10.1002/sim.8737}
\end{barticle}
\endbibitem

\bibitem[\protect\citeauthoryear{Shardell and Ferrucci}{2018}]{Shardell2018}
\begin{barticle}[author]
\bauthor{\bsnm{Shardell},~\bfnm{Michelle}\binits{M.}} \AND
  \bauthor{\bsnm{Ferrucci},~\bfnm{Luigi}\binits{L.}}
(\byear{2018}).
\btitle{{Joint mixed-effects models for causal inference with longitudinal
  data}}.
\bjournal{Statistics in Medicine}
\bvolume{37}
\bpages{829--846}.
\bdoi{10.1002/sim.7567}
\end{barticle}
\endbibitem

\bibitem[\protect\citeauthoryear{Shpitser and Pearl}{2007}]{Shpitser2007}
\begin{binproceedings}[author]
\bauthor{\bsnm{Shpitser},~\bfnm{Ilya}\binits{I.}} \AND
  \bauthor{\bsnm{Pearl},~\bfnm{Judea}\binits{J.}}
(\byear{2007}).
\btitle{What Counterfactuals Can Be Tested}.
In \bbooktitle{Proceedings of the Twenty-Third Conference on Uncertainty in
  Artificial Intelligence}.
\bseries{UAI'07}
\bpages{352–359}.
\bpublisher{AUAI Press}, \baddress{Arlington, Virginia, USA}.
\bdoi{10.5555/3020488.3020531}
\end{binproceedings}
\endbibitem

\bibitem[\protect\citeauthoryear{Steyer}{2005}]{Steyer2005}
\begin{barticle}[author]
\bauthor{\bsnm{Steyer},~\bfnm{Rolf}\binits{R.}}
(\byear{2005}).
\btitle{{Analyzing Individual and Average Causal Effects via Structural
  Equation Models}}.
\bjournal{Methodology}
\bvolume{1}
\bpages{39--54}.
\bdoi{10.1027/1614-1881.1.1.39}
\end{barticle}
\endbibitem

\bibitem[\protect\citeauthoryear{{van~der~Vaart}}{1998}]{vanderVaart1998}
\begin{bbook}[author]
\bauthor{\bsnm{{van~der~Vaart}},~\bfnm{A.~W.}\binits{A.~W.}}
(\byear{1998}).
\btitle{Asymptotic Statistics}.
\bseries{Cambridge Series in Statistical and Probabilistic Mathematics}.
\bpublisher{Cambridge University Press}.
\bdoi{10.1017/CBO9780511802256}
\end{bbook}
\endbibitem

\bibitem[\protect\citeauthoryear{VanderWeele}{2016}]{VanderWeele2016}
\begin{barticle}[author]
\bauthor{\bsnm{VanderWeele},~\bfnm{Tyler~J.}\binits{T.~J.}}
(\byear{2016}).
\btitle{{Explanation in causal inference: Developments in mediation and
  interaction}}.
\bjournal{International Journal of Epidemiology}
\bvolume{45}
\bpages{1904--1908}.
\bdoi{10.1093/ije/dyw277}
\end{barticle}
\endbibitem

\bibitem[\protect\citeauthoryear{{VanderWeele} and
  Knol}{2014}]{VanDerWeele2014}
\begin{barticle}[author]
\bauthor{\bsnm{{VanderWeele}},~\bfnm{Tyler~J.}\binits{T.~J.}} \AND
  \bauthor{\bsnm{Knol},~\bfnm{Mirjam~J.}\binits{M.~J.}}
(\byear{2014}).
\btitle{{A tutorial on interaction}}.
\bjournal{Epidemiologic Methods}
\bvolume{3}
\bpages{33--72}.
\bdoi{10.1515/em-2013-0005}
\end{barticle}
\endbibitem

\bibitem[\protect\citeauthoryear{VanderWeele and
  Robins}{2007}]{VanderWeele2007}
\begin{barticle}[author]
\bauthor{\bsnm{VanderWeele},~\bfnm{Tyler~J.}\binits{T.~J.}} \AND
  \bauthor{\bsnm{Robins},~\bfnm{James~M.}\binits{J.~M.}}
(\byear{2007}).
\btitle{{Four types of effect modification: A classification based on directed
  acyclic graphs}}.
\bjournal{Epidemiology}
\bvolume{18}
\bpages{561--568}.
\bdoi{10.1097/EDE.0b013e318127181b}
\end{barticle}
\endbibitem

\bibitem[\protect\citeauthoryear{Vansteelandt and
  Joffe}{2014}]{Vansteelandt2015}
\begin{barticle}[author]
\bauthor{\bsnm{Vansteelandt},~\bfnm{Stijn}\binits{S.}} \AND
  \bauthor{\bsnm{Joffe},~\bfnm{Marshall}\binits{M.}}
(\byear{2014}).
\btitle{{Structural nested models and G-estimation: The partially realized
  promise}}.
\bjournal{Statistical Science}
\bvolume{29}
\bpages{707--731}.
\bdoi{10.1214/14-STS493}
\end{barticle}
\endbibitem

\bibitem[\protect\citeauthoryear{Verbeke and Molenberghs}{2000}]{Verbeke2000}
\begin{bbook}[author]
\bauthor{\bsnm{Verbeke},~\bfnm{Geert.}\binits{G.}} \AND
  \bauthor{\bsnm{Molenberghs},~\bfnm{Geert}\binits{G.}}
(\byear{2000}).
\btitle{Linear mixed models for longitudinal data},
\bedition{1st} ed.
\bpublisher{Springer}, \baddress{New York}.
\bdoi{10.1007/b98969}
\end{bbook}
\endbibitem

\bibitem[\protect\citeauthoryear{Vonesh, Chinchilli and Pu}{1996}]{Vonesh1996}
\begin{barticle}[author]
\bauthor{\bsnm{Vonesh},~\bfnm{Edward~F.}\binits{E.~F.}},
  \bauthor{\bsnm{Chinchilli},~\bfnm{Vernon~M.}\binits{V.~M.}} \AND
  \bauthor{\bsnm{Pu},~\bfnm{Kewei}\binits{K.}}
(\byear{1996}).
\btitle{Goodness-of-Fit in Generalized Nonlinear Mixed-Effects Models}.
\bjournal{Biometrics}
\bvolume{52}
\bpages{572--587}.
\bdoi{10.2307/2532896}
\end{barticle}
\endbibitem

\bibitem[\protect\citeauthoryear{Wager and Athey}{2018}]{Wager2018}
\begin{barticle}[author]
\bauthor{\bsnm{Wager},~\bfnm{Stefan}\binits{S.}} \AND
  \bauthor{\bsnm{Athey},~\bfnm{Susan}\binits{S.}}
(\byear{2018}).
\btitle{{Estimation and Inference of Heterogeneous Treatment Effects using
  Random Forests}}.
\bjournal{Journal of the American Statistical Association}
\bvolume{113}
\bpages{1228--1242}.
\bdoi{10.1080/01621459.2017.1319839}
\end{barticle}
\endbibitem

\bibitem[\protect\citeauthoryear{Weinberg}{2007}]{Weinberg2007}
\begin{barticle}[author]
\bauthor{\bsnm{Weinberg},~\bfnm{Clarice~R.}\binits{C.~R.}}
(\byear{2007}).
\btitle{{Can DAGs clarify effect modification?}}
\bjournal{Epidemiology}
\bvolume{18}
\bpages{569--572}.
\bdoi{10.1097/EDE.0b013e318126c11d}
\end{barticle}
\endbibitem

\bibitem[\protect\citeauthoryear{Wendling et~al.}{2018}]{Wendling2018}
\begin{barticle}[author]
\bauthor{\bsnm{Wendling},~\bfnm{Thierry}\binits{T.}},
  \bauthor{\bsnm{Jung},~\bfnm{K.}\binits{K.}},
  \bauthor{\bsnm{Callahan},~\bfnm{A.}\binits{A.}},
  \bauthor{\bsnm{Schuler},~\bfnm{A.}\binits{A.}},
  \bauthor{\bsnm{Shah},~\bfnm{N.~H.}\binits{N.~H.}} \AND
  \bauthor{\bsnm{Gallego},~\bfnm{B.}\binits{B.}}
(\byear{2018}).
\btitle{{Comparing methods for estimation of heterogeneous treatment effects
  using observational data from health care databases}}.
\bjournal{Statistics in Medicine}
\bvolume{37}
\bpages{3309--3324}.
\bdoi{10.1002/sim.7820}
\end{barticle}
\endbibitem

\end{thebibliography}

\newpage
\begin{appendix}

\section{Proofs}\label{CH4proofsCH4}
	\subsection{Proof of Lemma	\ref{CH4l1}}
	\begin{proof} 
		Trivially, as 	$Y_{ji}^{\overline{a}}~{:}{=}~f_{Y_{j}}(\boldsymbol{U}_{0i},\boldsymbol{U}_{\text{AY}i},\overline{\boldsymbol{M}}_{j-1,i},   \overline{a}_{j-1},\overline{\boldsymbol{L}}_{j-1,i},\overline{N}_{Yji})$,
		$$ Y_{j}^{\overline{a}}\mid \boldsymbol{U}_{0}, \boldsymbol{U}_{\text{AY}},\overline{\boldsymbol{M}}_{h},\overline{\boldsymbol{L}}_{h}, \overline{Y}_{h},\overline{N}_{Yj}~\overset{d}=~
		Y_{j}^{\overline{a}}\mid \boldsymbol{U}_{0}, \boldsymbol{U}_{\text{AY}},\overline{\boldsymbol{M}}_{j-1},\overline{\boldsymbol{L}}_{j-1}, \overline{N}_{Yj},$$
		and $Y_{j}^{\overline{a}}\mid \boldsymbol{U}_{0}, \boldsymbol{U}_{\text{AY}},\overline{\boldsymbol{M}}_{j-1},\overline{\boldsymbol{L}}_{j-1}, \overline{N}_{Yj}$
		is a degenerate random variable independent of $\overline{A}_{j-1}$ and $Y_{j}^{\overline{b}}$ for an arbitrary exposure strategy $\overline{b}$. 
	\end{proof} 
	
	\subsection{Proof of Theorem \ref{CH4th4.2}}
	\begin{proof} By the law of total probability,
		$$f\left((Y_{j}^{\overline{a}},Y_{j}^{\overline{0}})\mid \mathcal{H}_{h}\right) ~{=}~ \int f\left((Y_{j}^{\overline{a}},Y_{j}^{\overline{0}})\mid  \boldsymbol{U}, \mathcal{H}_{h}, \overline{N}_{Yj}\right) dF_{(\boldsymbol{U},\overline{N}_{Yj}) \mid \mathcal{H}_{h}}.$$
		By the law of conditional probability,
		$$ ~{=}~ \int f\left(Y_{j}^{\overline{a}}\mid \boldsymbol{U}, \mathcal{H}_{h}, \overline{N}_{Yj}, Y_{j}^{\overline{0}}\right)f\left(Y_{j}^{\overline{0}}\mid  \boldsymbol{U}, \mathcal{H}_{h}, \overline{N}_{Yj}\right) dF_{(\boldsymbol{U},\overline{N}_{Yj}) \mid \mathcal{H}_{h}}.
		$$
		By Lemma \ref{CH4l1}, for $\overline{b}~{=}~\overline{0}$, 
		$$ ~{=}~ \int f\left(Y_{j}^{\overline{a}}\mid \boldsymbol{U}, \overline{\boldsymbol{M}}_{h}, \overline{\boldsymbol{L}}_{h}, \overline{Y}_{h}, \overline{A}_{h}, \overline{N}_{Yj}\right)f\left(Y_{j}^{\overline{0}}\mid \boldsymbol{U}, \overline{\boldsymbol{M}}_{h}, \overline{\boldsymbol{L}}_{h}, \overline{Y}_{h}, \overline{A}_{h}, \overline{N}_{Yj}\right) dF_{(\boldsymbol{U},\overline{N}_{Yj}) \mid \mathcal{H}_{h}},$$ and by SCM \eqref{CH4indSCM},
    $$ ~{=}~ \int f\left(Y_{j}^{\overline{a}}\mid \boldsymbol{U}, \overline{\boldsymbol{M}}_{j-1}, \overline{\boldsymbol{L}}_{j-1}, \overline{N}_{Yj}\right)f\left(Y_{j}^{\overline{0}}\mid \boldsymbol{U}, \overline{\boldsymbol{M}}_{j-1}, \overline{\boldsymbol{L}}_{j-1}, \overline{N}_{Yj}\right) dF_{(\boldsymbol{U},\overline{N}_{Yj}) \mid \mathcal{H}_{h}}.$$
		Again by Lemma \ref{CH4l1}, the density equals
		$$ \resizebox{\linewidth}{!}{$\int f\left(Y_{j}^{\overline{a}}\mid \boldsymbol{U}, \overline{\boldsymbol{M}}_{j-1}, \overline{\boldsymbol{L}}_{j-1}, \overline{N}_{Yj}, \overline{A}_{j-1}{=}\overline{a}_{j-1}\right)f\left(Y_{j}^{\overline{0}}\mid \boldsymbol{U}, \overline{\boldsymbol{M}}_{j-1}, \overline{\boldsymbol{L}}_{j-1}, \overline{N}_{Yj},\overline{A}_{j-1}~{=}~\overline{0}_{j-1}\right) dF_{(\boldsymbol{U},\overline{N}_{Yj}) \mid \mathcal{H}_{h}}.$}$$
		By Assumption \ref{CH4A2}, the joint density is thus equal to
		$$ \resizebox{\linewidth}{!}{$\int f\left(Y_{j} \mid \boldsymbol{U}, \overline{\boldsymbol{M}}_{j-1}, \overline{\boldsymbol{L}}_{j-1}, \overline{N}_{Yj}, \overline{A}_{j-1}{=}\overline{a}_{j-1}\right)f\left(Y_{j} \mid \boldsymbol{U}, \overline{\boldsymbol{M}}_{j-1}, \overline{\boldsymbol{L}}_{j-1}, \overline{N}_{Yj}.\overline{A}_{j-1}~{=}~\overline{0}_{j-1}\right) dF_{(\boldsymbol{U},\overline{N}_{Yj}) \mid \mathcal{H}_{h}}.$}
		$$
	\end{proof} 
		
	\subsection{Proof of Corollary \ref{CH4cor:331}}
	\begin{proof}
		By Theorem \ref{CH4th4.2}, $f\left(Y_{j}^{\overline{a}},Y_{j}^{\overline{0}}\mid \mathcal{H}_{h}\right) $ is equal to
		$$ \resizebox{\linewidth}{!}{$\int f\left(Y_{j}^{\overline{a}}\mid \boldsymbol{U}, \boldsymbol{M}, \overline{\boldsymbol{L}}_{j-1}, \overline{N}_{Yj}, \overline{A}_{j-1}{=}\overline{a}_{j-1}\right)f\left(Y_{j}^{\overline{0}}\mid \boldsymbol{U}, \overline{\boldsymbol{M}}_{j-1}, \overline{\boldsymbol{L}}_{j-1}, \overline{N}_{Yj}.\overline{A}_{j-1}~{=}~\overline{0}_{j-1}\right) dF_{(\boldsymbol{U},\overline{N}_{Yj}) \mid \mathcal{H}_{h}}.$}$$
		By the law of conditional probability this equals,
		$$ \resizebox{\linewidth}{!}{$\int \int f\left(Y_{j}^{\overline{a}}\mid \boldsymbol{U}, \overline{\boldsymbol{M}}_{j-1}, \overline{\boldsymbol{L}}_{j-1}, \overline{N}_{Yj}, \overline{A}_{j-1}{=}\overline{a}_{j-1}\right)f\left(Y_{j}^{\overline{0}}\mid \boldsymbol{U}, \overline{\boldsymbol{M}}_{j-1}, \overline{\boldsymbol{L}}_{j-1}, \overline{N}_{Yj}.\overline{A}_{j-1}~{=}~\overline{0}_{j-1}\right) dF_{\overline{N}_{Yj} \mid \boldsymbol{U}, \mathcal{H}_{h}}dF_{\boldsymbol{U} \mid \mathcal{H}_{h}}.$}$$
		By definition of the SCM \eqref{CH4indSCM}, the support of $\overline{N}_{Yj} \mid \mathcal{H}_{h}, \boldsymbol{U}$ equals	$$ \left\{ \overline{N}_{Yj}{:}~\forall 1{\leq}k{\leq}j~f_{Y_{k}}(\boldsymbol{U}_{0}, \boldsymbol{U}_{\text{AY}}, \overline{\boldsymbol{M}}_{j-1},\overline{\boldsymbol{L}}_{k-1},\overline{A}_{k-1},\overline{N}_{Yk}
		)~{=}~Y_{k}\right\}. 
		$$ 
		If $f_{Y_{k}}$ is an injective function of $N_{Yk}$ $\forall 1{\leq}k{\leq}j$, then $\overline{N}_{Yj} \mid \mathcal{H}_{h}, \boldsymbol{U}$ is a degenerate random vector with
		$$N_{Yk}~{=}~f_{Y_{k}}^{-1}(\boldsymbol{U}, \overline{\boldsymbol{M}}_{k-1},\overline{\boldsymbol{L}}_{k-1},\overline{A}_{k-1}, \overline{N}_{Y,k-1}, \circ)(Y_{k}),$$
		where $f_{Y_{k}}^{-1}(\boldsymbol{U}, \overline{\boldsymbol{M}}_{k-1},\overline{\boldsymbol{L}}_{k-1},\overline{A}_{k-1}, \overline{N}_{k-1}, \circ)({Y}_{k})$ is the inverse function of $f_{Y_{k}}$ w.r.t. to $N_{Yk}$. Then, the joint pdf equals
		\begin{align*}	
		& \scalebox{0.825}{$\int f\left(Y_{j}\mid \overline{\boldsymbol{M}}_{j-1},\overline{\boldsymbol{L}}_{j-1} ,\boldsymbol{U}, 
  \forall k{\leq} j{:}~~ N_{Yk}~{=}~f_{Y_{k}}^{-1}(\boldsymbol{U}, \overline{\boldsymbol{M}}_{k-1},\overline{\boldsymbol{L}}_{k-1},\overline{A}_{k-1}, \overline{N}_{Y,k-1}, \circ)(Y_{k}), \overline{A}_{j-1}{=}\overline{a}_{j-1} \right)$}\\
		&\scalebox{0.825}{$f\left(Y_{j}\mid \overline{\boldsymbol{M}}_{j-1},\overline{\boldsymbol{L}}_{j-1} ,\boldsymbol{U}, 
   \forall k{\leq} j{:}~~ N_{Yk}~{=}~f_{Y_{k}}^{-1}(\boldsymbol{U}, \overline{\boldsymbol{M}}_{k-1},\overline{\boldsymbol{L}}_{k-1},\overline{A}_{k-1}, \overline{N}_{Y,k-1}, \circ)(Y_{k}), \overline{A}_{j-1}{=}\overline{0} \right)dF_{\boldsymbol{U}\mid \mathcal{H}_{h}}.$} 
		\end{align*}
	\end{proof}
		
	\subsection{Proof Theorem \ref{CH4th:latent}}
	\begin{proof}
	By the law of total probability, and by the independence between $\boldsymbol{U}$ and $\overline{N}_{Y}$,
		\begin{equation*}\resizebox{1\linewidth}{!}{$f\left(\overline{Y}_{j}^{\overline{a}}\mid \overline{\boldsymbol{M}}_{j-1}, \overline{\boldsymbol{L}}_{j-1} \right)~{=}~ 	\int f\left(\overline{Y}_{j}^{\overline{a}}\mid \boldsymbol{U}_{0},\boldsymbol{U}_{\text{AY}}, \overline{\boldsymbol{M}}_{j-1},\overline{\boldsymbol{L}}_{j-1},  \overline{N}_{Yj} \right)
		dF_{(\boldsymbol{U}_{0}, \boldsymbol{U}_{\text{AY}},\overline{N}_{Yj}) \mid \overline{\boldsymbol{M}}_{j-1},\overline{\boldsymbol{L}}_{j-1}},$}\end{equation*}
	\noindent which, by Lemma \ref{CH4l1}, equals
	$$\int f\left(\overline{Y}_{j}^{\overline{a}}\mid \boldsymbol{U}_{0},\boldsymbol{U}_{\text{AY}}, \overline{\boldsymbol{M}}_{j-1},\overline{\boldsymbol{L}}_{j-1},  \overline{N}_{Yj}, \overline{A}_{j-1}{=}\overline{a}_{j-1} \right)
		dF_{(\boldsymbol{U}_{0}, \boldsymbol{U}_{\text{AY}},\overline{N}_{Yj}) \mid \overline{\boldsymbol{M}}_{j-1},\overline{\boldsymbol{L}}_{j-1}},$$
		\noindent and,  by Assumption \ref{CH4A2}, is equal to
		$$\int f\left(\overline{Y}_{j}\mid \boldsymbol{U}_{0},\boldsymbol{U}_{\text{AY}}, \overline{\boldsymbol{M}}_{j-1},\overline{\boldsymbol{L}}_{j-1},  \overline{N}_{Yj}, \overline{A}_{j-1}{=}\overline{a}_{j-1} \right)
		dF_{(\boldsymbol{U}_{0}, \boldsymbol{U}_{\text{AY}},\overline{N}_{Yj}) \mid \overline{\boldsymbol{M}}_{j-1},\overline{\boldsymbol{L}}_{j-1}}. \qedhere$$
	\end{proof} 
	
	\subsection{Proof of Lemma \ref{CH4lemma:fUN}}
	\begin{proof} Let $(\boldsymbol{U})_{\overline{a}_{j-1}}$ equal those elements of $\boldsymbol{U}$ that affect $\overline{Y}_{j}^{\overline{a}}$ and $(\boldsymbol{U})_{\overline{a}_{j-1}} \backslash (\boldsymbol{U})_{\overline{a}_{j-2}}$ is referred to as $(\boldsymbol{U})_{a_{j-1}}$. By the law of conditional probability, $$\resizebox{\linewidth}{!}{$f\left((\boldsymbol{U})_{\overline{a}_{j-1}}, \overline{N}_{Yj} \mid \overline{\boldsymbol{M}}_{j-1},\overline{\boldsymbol{L}}_{j-1}\right) ~{=}~
  \prod_{k~{=}~1}^{j} f\left((\boldsymbol{U})_{a_{k-1}},N_{Yk} \mid (\boldsymbol{U})_{\overline{a}_{k-2}},\overline{N}_{Y,k-1}, \overline{\boldsymbol{M}}_{j-1},\overline{\boldsymbol{L}}_{j-1}\right),$}$$ where $(\boldsymbol{U})_{\overline{a}_{0}}~{=}~(\boldsymbol{U})_{a_{0}}~{=}~\boldsymbol{U}_{0}$ and $(\boldsymbol{U})_{\overline{a}_{-1}}~{=}~\emptyset$. 
  By Assumption \ref{CH4A1}, 
				$N_{Y2},(\boldsymbol{U})_{a_{1}} \independent A_{1} \mid \boldsymbol{U}_{0}, N_{Y1}, \overline{\boldsymbol{M}}_{j-1},\overline{\boldsymbol{L}}_{j-1}$, 
    where $(\boldsymbol{U})_{a_{1}}=(\boldsymbol{U})_{\overline{a}_{1}}{\backslash}\boldsymbol{U}_{0}$. 
  Moreover, for $k>2$
		$$(\boldsymbol{U})_{a_{k-1}},N_{Yk} \independent A_{k-1}\mid (\boldsymbol{U})_{\overline{a}_{k-2}},\overline{N}_{Y,k-1}, \overline{A}_{k-2},\overline{\boldsymbol{M}}_{j-1},\overline{\boldsymbol{L}}_{j-1},$$ since $A_{k-1,i}~{:}{=}~f_{A_{k-1}}(\overline{\boldsymbol{L}}_{k-1,i},\overline{Y}_{k-1,i},\overline{A}_{k-2,i},N_{A,k-1,i})$, where\\ $\overline{Y}_{k-1,i} \mid (\boldsymbol{U})_{\overline{a}_{k-2,i}}, \overline{N}_{Y,k-1,i}, \overline{A}_{k-2,i}, \overline{\boldsymbol{L}}_{k-2,i}, \overline{\boldsymbol{M}}_{k-2,i}$ is degenerate, and Assumption \ref{CH4A1} implies $\overline{N}_{A} \independent \overline{N}_{Y}, \boldsymbol{U}$ (absence of direct unmeasured confounding). Thus,
		$$\resizebox{\linewidth}{!}{$f\left((\boldsymbol{U})_{\overline{a}_{j-1}}, \overline{N}_{Yj} \mid \overline{\boldsymbol{M}}_{j-1},\overline{\boldsymbol{L}}_{j-1}\right) ~{=}~
  \prod_{k~{=}~1}^{j} f\left((\boldsymbol{U})_{a_{k-1}},N_{Yk} \mid (\boldsymbol{U})_{\overline{a}_{k-2}},\overline{N}_{Y,k-1}, \overline{A}_{k-1}{=}\overline{a}_{k-1},\overline{\boldsymbol{M}}_{j-1},\overline{\boldsymbol{L}}_{j-1}\right), $} $$ where $\overline{A}_{0}=\emptyset$.	\end{proof}

 \subsection{Proof of Corollary \ref{CH4th:latent2}}
	\begin{proof}
 	By Theorem \ref{CH4th:latent} the pdf equals 
		$$ \int f\left(\overline{Y}_{j}\mid \boldsymbol{U}_{0},\boldsymbol{U}_{\text{AY}}, \overline{\boldsymbol{M}}_{j-1},\overline{\boldsymbol{L}}_{j-1},  \overline{N}_{Yj}, \overline{A}_{j-1}{=}\overline{a}_{j-1} \right)
		dF_{(\boldsymbol{U}_{0}, \boldsymbol{U}_{\text{AY}},\overline{N}_{Yj})\mid \overline{\boldsymbol{M}}_{j-1},\overline{\boldsymbol{L}}_{j-1}},
		$$ that is by defintion of $(\boldsymbol{U})_{\overline{a}_{j-1}}$ equal to
		$$\int f\left(\overline{Y}_{j}\mid (\boldsymbol{U})_{\overline{a}_{j-1}}, \overline{\boldsymbol{M}}_{j-1},\overline{\boldsymbol{L}}_{j-1},  \overline{N}_{Yj}, \overline{A}_{j-1}{=}\overline{a}_{j-1} \right)
		dF_{((\boldsymbol{U})_{\overline{a}_{j-1}},\overline{N}_{Yj})\mid \overline{\boldsymbol{M}}_{j-1},\overline{\boldsymbol{L}}_{j-1}}$$ Moreover, by the definition of SCM \eqref{CH4indSCM},
  	$$\int \prod_{k=1}^{j} \mathbbm{1}_{\{ \scalebox{0.75}{$f_{Y_{k}}((\boldsymbol{U})_{\overline{a}_{k-1}},\overline{\boldsymbol{M}}_{k-1},   \overline{a}_{k-1},\overline{\boldsymbol{L}}_{k-1},\overline{N}_{Yk})$}~{=}~y_{k}\}}		dF_{((\boldsymbol{U})_{\overline{a}_{j-1}},\overline{N}_{Yj})\mid \overline{\boldsymbol{M}}_{j-1},\overline{\boldsymbol{L}}_{j-1}}.$$
   That can be rewritten as, 
   $$\resizebox{\linewidth}{!}{$f \left(\scalebox{0.75}{$f_{Y_{1}}(\boldsymbol{U}_{0},N_{Y1})$}{=}y_{1},~ \scalebox{0.75}{$ f_{Y_{2}}((\boldsymbol{U})_{\overline{a}_{1}},\boldsymbol{M}_{1},   a_{1},\boldsymbol{L}_{1},\overline{N}_{Y2})$}{=}y_{2},~ \hdots,~ \scalebox{0.75}{$f_{Y_{j}}((\boldsymbol{U})_{\overline{a}_{j-1}},\overline{\boldsymbol{M}}_{j-1},   \overline{a}_{j-1},\overline{\boldsymbol{L}}_{j-1},\overline{N}_{Yj})$}{=}y_{j}\mid \overline{\boldsymbol{M}}_{j-1},\overline{\boldsymbol{L}}_{j-1}\right).$}$$ By the law of conditional probability this equals 
   $$  
   \resizebox{\linewidth}{!}{$\prod_{k=1}^{j} f \left(\scalebox{0.75}{$f_{Y_{k}}((\boldsymbol{U})_{\overline{a}_{k-1}},\overline{\boldsymbol{M}}_{k-1},   \overline{a}_{k-1},\overline{\boldsymbol{L}}_{k-1},\overline{N}_{Yk})$}{=}y_{k} \mid ((\boldsymbol{U})_{\overline{a}_{k-2}}, \overline{N}_{Y,k-1}) \in \mathcal{B}_{k-1}(\overline{a}), \overline{\boldsymbol{M}}_{j-1},\overline{\boldsymbol{L}}_{j-1}\right),$}$$
    where $\mathcal{B}_{k-1}(\overline{a})=\left\{ (\boldsymbol{u}, \boldsymbol{n}_{Y}){:}~ \forall m{<}k{:}~ \scalebox{0.75}{$f_{Y_{m}}(\boldsymbol{u}, \overline{\boldsymbol{M}}_{m-1}, \overline{a}_{m-1},\overline{\boldsymbol{L}}_{m-1},\boldsymbol{n}_{Y})$}{=}y_{m} \right\}$ and $\mathcal{B}_{0}(\overline{a})=\emptyset$. By SCM \eqref{CH4indSCM}, $A_{ki}^{\overline{a}} ~{:}{=}~f_{A_{k}}(\overline{\boldsymbol{L}}_{ki},\overline{Y}_{ki}^{\overline{a}},\overline{a}_{k-1},N_{Aki})$, and by Assumption \ref{CH4A2} $\overline{N}_{A} \independent \boldsymbol{U}, \overline{N}_{Y}$ (no direct unmeasured confounding), so that
   $$A_{k-1}^{\overline{a}} \independent \boldsymbol{U}, \overline{N}_{Y} \mid  \overline{Y}_{k-1}^{\overline{a}},   \overline{\boldsymbol{L}}_{k-1}, \overline{\boldsymbol{M}}_{k-1}$$ where
   $Y_{ki}^{\overline{a}}~{:}{=}~f_{Y_{k}}(\boldsymbol{U}_{0i},\boldsymbol{U}_{\text{AY}i},\overline{\boldsymbol{M}}_{k-1,i},   \overline{a}_{k-1},\overline{\boldsymbol{L}}_{k-1,i},\overline{N}_{Yki}).$ So, 
$$
\resizebox{\linewidth}{!}{$\prod_{k=1}^{j} f \left(\scalebox{0.75}{$f_{Y_{k}}((\boldsymbol{U})_{\overline{a}_{k-1}},\overline{\boldsymbol{M}}_{k-1},   \overline{a}_{k-1},\overline{\boldsymbol{L}}_{k-1},\overline{N}_{Yk})$}{=}y_{k} \mid ((\boldsymbol{U})_{\overline{a}_{k-2}}, \overline{N}_{Y,k-1}) \in \mathcal{B}_{k-1}(\overline{a}), \overline{\boldsymbol{M}}_{j-1},\overline{\boldsymbol{L}}_{j-1}, \overline{A}^{\overline{a}}_{k-1}=\overline{a}_{k-1} \right),$}$$ thus equal to
   $$
   \resizebox{\linewidth}{!}{$\prod_{k=1}^{j} f \left(\scalebox{0.75}{$f_{Y_{k}}((\boldsymbol{U})_{\overline{a}_{k-1}},\overline{\boldsymbol{M}}_{k-1},  \overline{A}^{\overline{a}}_{k-1},\overline{\boldsymbol{L}}_{k-1},\overline{N}_{Yk})$}{=}y_{k} \mid ((\boldsymbol{U})_{\overline{a}_{k-2}}, \overline{N}_{Y,k-1}) \in \mathcal{B}_{k-1}(\overline{A}^{\overline{a}}), \overline{\boldsymbol{M}}_{j-1},\overline{\boldsymbol{L}}_{j-1},\overline{A}^{\overline{a}}_{k-1}=\overline{a}_{k-1} ) \right).$}$$ By causal consistency (Assumption \ref{CH4A1}), 
    $$
    \resizebox{\linewidth}{!}{$\prod_{k=1}^{j} f \left(\scalebox{0.75}{$f_{Y_{k}}((\boldsymbol{U})_{\overline{a}_{k-1}},\overline{\boldsymbol{M}}_{k-1},  \overline{A}_{k-1},\overline{\boldsymbol{L}}_{k-1},\overline{N}_{Yk})$}{=}y_{k} \mid ((\boldsymbol{U})_{\overline{a}_{k-2}}, \overline{N}_{Y,k-1}) \in \mathcal{B}_{k-1}(\overline{A}), \overline{\boldsymbol{M}}_{j-1},\overline{\boldsymbol{L}}_{j-1},\overline{A}_{k-1}=\overline{a}_{k-1} ) \right).$}$$ Finally, since $Y_{ki}~{:}{=}~f_{Y_{k}}((\boldsymbol{U})_{\overline{a}_{k-1}i},\overline{\boldsymbol{M}}_{k-1,i},   \overline{A}_{k-1,i},\overline{\boldsymbol{L}}_{k-1,i},\overline{N}_{Yki})$, 
   the g-formula is recovered
   $$\prod_{k~{=}~1}^{j} f(Y_{k} \mid \overline{Y}_{k-1}, \overline{A}_{k-1} ~{=}~ \overline{a}_{k-1}, \overline{\boldsymbol{M}}_{j-1}, \overline{\boldsymbol{L}}_{j-1}).$$ If $((\boldsymbol{U})_{\overline{a}_{k-1}}, \overline{N}_{Yk}) \independent \overline{\boldsymbol{M}}_{j-1}, \overline{\boldsymbol{L}}_{j-1} \mid \overline{\boldsymbol{M}}_{k-1}, \overline{\boldsymbol{L}}_{k-1}$, the g-formula simplifies to $$\prod_{k~{=}~1}^{j} f(Y_{k} \mid \overline{Y}_{k-1}, \overline{A}_{k-1} ~{=}~ \overline{a}_{k-1}, \overline{\boldsymbol{M}}_{k-1}, \overline{\boldsymbol{L}}_{k-1}). \qedhere$$
 \end{proof}

	\subsection{Proof of Theorem \ref{CH4th41}}
	\begin{proof}
		First of all, Assumption \ref{CH4A3} should apply for identifiablity of the observed conditional distributions. 
  Let $ \mathcal{X}(\tilde{{}\boldsymbol{x}})$ equal the set, so that for each $(\boldsymbol{u}, \boldsymbol{n}_{Y}) \in \mathcal{X}(\tilde{{}\boldsymbol{x}})$, $F_{\boldsymbol{U},\overline{N}_{Y} }(\boldsymbol{u}, \boldsymbol{n}_{Y})$ is continuous and $(\boldsymbol{u}, \boldsymbol{n}_{Y})\leq \tilde{{}\boldsymbol{x}}$. Therefore, $\mathcal{X}(\boldsymbol{\infty})$ contains all $(\boldsymbol{u}, \boldsymbol{n}_{Y})$ where $F_{\boldsymbol{U},\overline{N}_{Y} }(\boldsymbol{u}, \boldsymbol{n}_{Y})$ is continuous. Furthermore, 	let
		\begin{equation*}
		\mathcal{A}(\mathcal{H}_{h})~{=}~\left \{(\boldsymbol{u},\boldsymbol{n}_{Y}):\forall 1{\leq} j {\leq} h{:}~f_{Y_{j}}(\boldsymbol{u}, \overline{\boldsymbol{M}}_{j-1}, \overline{A}_{j-1}, \overline{\boldsymbol{L}}_{j-1}, \boldsymbol{n}_{Y}) ~{=}~ Y_{j} \right \},
		\end{equation*} and similarly 
		\begin{equation*}
		\mathcal{A}_{n}(\mathcal{H}_{h})~{=}~\left \{(\boldsymbol{u},\boldsymbol{n}_{Y}):\forall 1{\leq} j {\leq} h{:}~\hat{f}_{Y_{j}}(\boldsymbol{u}, \overline{\boldsymbol{M}}_{j-1}, \overline{A}_{j-1}, \overline{\boldsymbol{L}}_{j-1}, \boldsymbol{n}_{Y}) ~{=}~ Y_{j} \right \}.
		\end{equation*} The distribution function of $(\boldsymbol{U},\overline{N}_{Y})_{\overline{a}_{h-1}}$ given $\mathcal{H}_{h}$, $ F_{(\boldsymbol{U}, \overline{N}_{Y})_{\overline{a}_{h-1}}\mid \mathcal{H}_{h}}(\tilde{{}\boldsymbol{x}})$ equals
		
		$$\frac{\int_{\mathcal{X}(\tilde{{}\boldsymbol{x}})\cap \mathcal{A}(\mathcal{H}_{h})} 1 d	F_{(\boldsymbol{U}, \overline{N}_{Y})_{\overline{a}_{h-1}}}
   }{\int_{\mathcal{X}(\boldsymbol{\infty})\cap \mathcal{A}(\mathcal{H}_{h})} 1 d	F_{(\boldsymbol{U}, \overline{N}_{Y})_{\overline{a}_{h-1}}}
   }, $$ if $\int_{\mathcal{X}(\tilde{{}\boldsymbol{x}})\cap \mathcal{A}(\mathcal{H}_{h})} 1 d	F_{(\boldsymbol{U}, \overline{N}_{Y})_{\overline{a}_{h-1}}}{>}0$ and $0$ otherwise.  Similarly,  the estimated distribution function, $ \hat{F}_{(\boldsymbol{U}, \overline{N}_{Y})_{\overline{a}_{h-1}}\mid \mathcal{H}_{h}}(\tilde{{}\boldsymbol{x}})$, equals
		$$\frac{\int_{\mathcal{X}(\tilde{{}\boldsymbol{x}})\cap \mathcal{A}_{n}(\mathcal{H}_{h})} 1 d\hat{F}_{(\boldsymbol{U},\overline{N}_{Y})_{\overline{a}_{h-1}}}}{\int_{\mathcal{X}(\boldsymbol{\infty})\cap \mathcal{A}_{n}(\mathcal{H}_{h})} 1 d\hat{F}_{(\boldsymbol{U},\overline{N}_{Y})_{\overline{a}_{h-1}}}}, $$if $\int_{\mathcal{X}(\tilde{{}\boldsymbol{x}})\cap \mathcal{A}_{n}(\mathcal{H}_{h})} 1 d\hat{F}_{(\boldsymbol{U},\overline{N}_{Y})_{\overline{a}_{h-1}}}{>}0$ and $0$ otherwise. For all $\tilde{{}\boldsymbol{x}}$ where $F_{(\boldsymbol{U},\overline{N}_{Y})_{\overline{a}_{h-1}}}(\tilde{{}\boldsymbol{x}})$ is continuous, $\mathcal{X}(\tilde{{}x}){\cap}\mathcal{A}(\mathcal{H}_{h})$ is a continuity set (no probability on the boundary of this set). Furthermore, 
		\begin{align*}
		&\int_{\mathcal{X}(\tilde{{}\boldsymbol{x}})\cap \mathcal{A}(\mathcal{H}_{h})} 1 d	F_{(\boldsymbol{U}, \overline{N}_{Y})_{\overline{a}_{h-1}}}
		 - \int_{\mathcal{X}(\tilde{{}\boldsymbol{x}})\cap \mathcal{A}_{n}(\mathcal{H}_{h})} 1 d\hat{F}_{(\boldsymbol{U}, \overline{N}_{Y})_{\overline{a}_{h-1}}}\\
		&~{=}~ \int_{\mathcal{X}(\tilde{{}\boldsymbol{x}})\cap \mathcal{A}(\mathcal{H}_{h})} 1 dF_{(\boldsymbol{U}, \overline{N}_{Y})_{\overline{a}_{h-1}}}-
		\int_{\mathcal{X}(\tilde{{}\boldsymbol{x}})\cap \mathcal{A}(\mathcal{H}_{h})} 1\hat{F}_{(\boldsymbol{U}, \overline{N}_{Y})_{\overline{a}_{h-1}}}\\
		&-  \int_{\mathcal{X}(\tilde{{}\boldsymbol{x}}) \cap (\mathcal{A}_{n}(\mathcal{H}_{h}) \backslash \mathcal{A}(\mathcal{H}_{h}))} 1d\hat{F}_{(\boldsymbol{U}, \overline{N}_{Y})_{\overline{a}_{h-1}}} +  \int_{\mathcal{X}(\tilde{{}\boldsymbol{x}}) \cap (\mathcal{A}(\mathcal{H}_{h}) \backslash \mathcal{A}_{n}(\mathcal{H}_{h}))}  1d\hat{F}_{(\boldsymbol{U}, \overline{N}_{Y})_{\overline{a}_{h-1}}}. 
		\end{align*} As $\mathcal{X}(\tilde{{}\boldsymbol{x}}) \cap \mathcal{A}(\mathcal{H}_{h})$ is a continuity set and requirement $i$ applies, the difference of the first two integrals converges to $0$ by the Portmanteau theorem \citep{vanderVaart1998}. Moreover, by requirement $ii$, $\mathcal{A}_{n}(\mathcal{H}_{h}) \xrightarrow[n \to \infty]{} \mathcal{A}(\mathcal{H}_{h})$ so that the last two integrals converge pointwise to $0$. Thus, 
		$$ \int_{\mathcal{X}(\tilde{{}\boldsymbol{x}})\cap \mathcal{A}_{n}(\mathcal{H}_{h})} 1 d\hat{F}_{(\boldsymbol{U}, \overline{N}_{Y})_{\overline{a}_{h-1}}} \xrightarrow[n \to \infty]{} \int_{\mathcal{X}(\tilde{{}\boldsymbol{x}})\cap \mathcal{A}(\mathcal{H}_{h})} 1 dF_{(\boldsymbol{U}, \overline{N}_{Y})_{\overline{a}_{h-1}}}. $$ Based on the same arguments
		$$ \int_{\mathcal{X}(\boldsymbol{\infty})\cap \mathcal{A}_{n}(\mathcal{H}_{h})} 1 d\hat{F}_{(\boldsymbol{U}, \overline{N}_{Y})_{\overline{a}_{h-1}}} \xrightarrow[n \to \infty]{} \int_{\mathcal{X}(\boldsymbol{\infty})\cap \mathcal{A}(\mathcal{H}_{h})} 1 dF_{(\boldsymbol{U}, \overline{N}_{Y})_{\overline{a}_{h-1}}}. $$ As a result, for all $\tilde{{}\boldsymbol{x}}$ where $F_{(\boldsymbol{U}, \overline{N}_{Y})_{\overline{a}_{h-1}}\mid \mathcal{H}_{h}}(\tilde{{}\boldsymbol{x}})$ is continuous, pointwise $$ \hat{F}_{(\boldsymbol{U}, \overline{N}_{Y})_{\overline{a}_{h-1}}\mid \mathcal{H}_{h}}(\tilde{{}\boldsymbol{x}})  \xrightarrow[n \to \infty]{} F_{(\boldsymbol{U}, \overline{N}_{Y})_{\overline{a}_{h-1}}\mid \mathcal{H}_{h}}(\tilde{{}\boldsymbol{x}}).$$ Furthermore, let $$ \mathcal{D}(d,j,\overline{a},\mathcal{H}_{h}) ~{=}~ \{ (\boldsymbol{u},\boldsymbol{n}_{Y}){:}~f_{Y_{j}}(\boldsymbol{u}, \overline{\boldsymbol{M}}_{j-1},  \overline{a}, \overline{\boldsymbol{L}}_{j-1}, \boldsymbol{n}_{Y}) - f_{Y_{j}}(\boldsymbol{u}, \overline{\boldsymbol{M}}_{j-1}, \overline{0}, \overline{\boldsymbol{L}}_{j-1}, \boldsymbol{n}_{Y}) {\leq} d\},$$ so that under assumptions \ref{CH4A2} and \ref{CH4A1}, the distribution function of the CWCE is equal to $$ F_{Y_{j}^{\overline{a}}-Y_{j}^{\overline{0}} \mid \mathcal{H}_{h}}(d) ~{=}~ \int_{\mathcal{D}(d,j,\overline{a},\mathcal{H}_{h})} 1 d F_{\boldsymbol{U}, \overline{N}_{Y} \mid \mathcal{H}_{h}},$$ by Theorem \ref{CH4th4.2}. Similarly, let
		$$ \mathcal{D}_{n}(d,j,\overline{a},\mathcal{H}_{h}) ~{=}~ \{ (\boldsymbol{u},\boldsymbol{n}_{Y}){:}~\hat{f}_{Y_{j}}(\boldsymbol{u}, \overline{\boldsymbol{M}}_{j-1}, \overline{a}, \overline{\boldsymbol{L}}_{j-1}, \boldsymbol{n}_{Y}) - \hat{f}_{Y_{j}}(\boldsymbol{u}, \boldsymbol{M}, \overline{0}, \overline{\boldsymbol{L}}_{j-1}, \boldsymbol{n}_{Y}) {\leq} d\},$$ so that our estimated CWCE distribution function is equal to
		$$ \hat{F}_{Y_{j}^{\overline{a}}-Y_{j}^{\overline{0}} \mid \mathcal{H}_{h}}(d) ~{=}~ \int_{\mathcal{D}_{n}(d,j,\overline{a},\mathcal{H}_{h})} 1 d \hat{F} _{(\boldsymbol{U}, \overline{N}_{Y})_{\overline{a}_{h-1}}\mid \mathcal{H}_{h}}.$$ We can repeat the steps used before in this proof, 
		\begin{align*}
		& \int_{\mathcal{D}(d,j,\overline{a},\mathcal{H}_{h})} 1 d F_{(\boldsymbol{U}, \overline{N}_{Y})_{\overline{a}_{h-1}}\mid \mathcal{H}_{h}}
		- \int_{\mathcal{D}_{n}(d,j,\overline{a},\mathcal{H}_{h})} 1 d \hat{F}_{(\boldsymbol{U}, \overline{N}_{Y})_{\overline{a}_{h-1}}\mid \mathcal{H}_{h}} ~{=}~\\
		& \int_{\mathcal{D}(d,j,\overline{a},\mathcal{H}_{h})} 1 d F_{(\boldsymbol{U}, \overline{N}_{Y})_{\overline{a}_{h-1}}\mid \mathcal{H}_{h}} - \int_{\mathcal{D}(d,j,\overline{a},\mathcal{H}_{h})} 1 d \hat{F}_{(\boldsymbol{U}, \overline{N}_{Y})_{\overline{a}_{h-1}}\mid \mathcal{H}_{h}}\\
		&- \int_{\mathcal{D}_{n}(d,j,\overline{a},\mathcal{H}_{h})\backslash \mathcal{D}(d,j,\overline{a},\mathcal{H}_{h}) } 1 d \hat{F}_{(\boldsymbol{U}, \overline{N}_{Y}) \mid \mathcal{H}_{h}} + \int_{\mathcal{D}(d,j,\overline{a},\mathcal{H}_{h})\backslash \mathcal{D}_{n}(d,j,\overline{a},\mathcal{H}_{h}) } 1 d \hat{F}_{(\boldsymbol{U}, \overline{N}_{Y}) \mid \mathcal{H}_{h}} 
		\end{align*}
		By requirement $ii$, $\mathcal{D}_{n}(d,j,\overline{a},\mathcal{H}_{h}) \xrightarrow[n \to \infty]{} \mathcal{D}(d,j,\overline{a},\mathcal{H}_{h})$, while\\ $ \hat{F}_{(\boldsymbol{U}, \overline{N}_{Y})_{\overline{a}_{h-1}}\mid \mathcal{H}_{h}}(\tilde{{}\boldsymbol{x}})$  $\xrightarrow[n \to \infty]{}$  $F_{(\boldsymbol{U}, \overline{N}_{Y})_{\overline{a}_{h-1}}\mid \mathcal{H}_{h}}(\tilde{{}\boldsymbol{x}})$ and $\mathcal{D}(d,j,\overline{a},\mathcal{H}_{h})$ is a continuity set for $d$ where $F_{Y_{j}^{\overline{a}}-Y_{j}^{\overline{0}} \mid \mathcal{H}_{h}}(d)$ is continuous, so
		
		$$  \int_{\mathcal{D}_{n}(d,j,\overline{a},\mathcal{H}_{h})} 1 d \hat{F}_{((\boldsymbol{U}, \overline{N}_{Y})_{\overline{a}_{h-1}} \mid \mathcal{H}_{h}} \xrightarrow[n \to \infty]{}  \int_{\mathcal{D}_{n}(d,j,\overline{a},\mathcal{H}_{h})} 1 d F_{((\boldsymbol{U}, \overline{N}_{Y})_{\overline{a}_{h-1}} \mid \mathcal{H}_{h}}. $$ Finally, by Assumption \ref{CH4A4}, there exists an exposure strategy  $ \overline{b}$ for which, $$ 
  \int_{\mathcal{D}(d,j,\overline{a},\mathcal{H}_{h})} 1 d F_{\boldsymbol{U}, \overline{N}_{Y} \mid \mathcal{H}_{h}}~{=}~ \int_{\mathcal{D}(d,j,\overline{a},\mathcal{H}_{h})} 1 d F_{((\boldsymbol{U}, \overline{N}_{Y})_{\overline{b}_{h-1}} \mid \mathcal{H}_{h}},$$
		so that
		
		$$ \hat{F}_{Y_{j}^{\overline{a}}-Y_{j}^{\overline{0}} \mid \mathcal{H}_{h}}(d) 
		\xrightarrow[n \to \infty]{} F_{Y_{j}^{\overline{a}}-Y_{j}^{\overline{0}} \mid \mathcal{H}_{h}}(d) 
		$$ pointwise for all $d$ where $F_{Y_{j}^{\overline{a}}-Y_{j}^{\overline{0}} \mid \mathcal{H}_{h}}(d)$ is continuous. 	\end{proof}

\subsection{Proof of Theorem \ref{CH4prop1}}
	
	\begin{proof}
By the definition of SCM \eqref{CH4indSCM} and Assumption \ref{CH4A2}, 
  	$$Y_{ji}~{=}~Y_{ji}^{\overline{A}_{i}}~{=}~f_{Y_{j}}(\boldsymbol{U}_{0i},\boldsymbol{U}_{\text{AY}i},\overline{\boldsymbol{M}}_{j-1,i},   \overline{A}_{j-1,i},\overline{\boldsymbol{L}}_{j-1,i},\overline{N}_{Yji}).$$ So,
\begin{align*}&f_{Y_{j}}(\boldsymbol{U}_{0},\boldsymbol{U}_{\text{AY}},\overline{\boldsymbol{M}}_{j-1},   \overline{A}_{j-1},\overline{\boldsymbol{L}}_{j-1},\overline{N}_{Yj})~\overset{d}=~\\
&\beta_{0}+Z_{0}+\beta_{L}{L}_{j-1} + \epsilon_{j} + (\beta_{1}+Z_{1}) A_{j-1} + (\beta_{2}+Z_{2}) A_{j-2},\end{align*}
Which is equivalent to,    
\begin{align*}&f_{Y_{j}}(\boldsymbol{U}_{0i},\boldsymbol{U}_{\text{AY}i},\overline{\boldsymbol{M}}_{j-1,i},   \overline{A}_{j-1,i},\overline{\boldsymbol{L}}_{j-1,i},\overline{N}_{Yji})=\\
&\beta_{0}+Z_{0i}^{\overline{A}}+\beta_{L}{L}_{j-1,i} + \epsilon_{ji}^{\overline{A}}+ (\beta_{1}+Z_{1i}^{\overline{A}}) A_{j-1,i} + (\beta_{2}+Z_{2i}^{\overline{A}}) A_{j-2,i},\end{align*}
where $\forall \overline{a}, \overline{b}{:}~Z_{1i}^{\overline{a}} ~\overset{d}=~ Z_{1i}^{\overline{b}},~ Z_{2i}^{\overline{a}} ~\overset{d}=~ Z_{2i}^{\overline{b}}$ and $\forall \overline{a}, j{:}~Z_{0}^{\overline{a}}+\epsilon_{ji}^{\overline{a}} ~\overset{d}=~ Z_{0i}+\epsilon_{ji}$. Since $Y_{ji}^{\overline{A}_{i}}$ equals the sum of $Y_{ji}^{\overline{0}}$ and the ICE of exposure strategy $\overline{A}_{j-1,i}$ $(Y^{\overline{A}_{i}}_{ji}~{=}~Y^{\overline{0}}_{ji}+\text{ICE}(\overline{A}_{j-1,i}))$, 
\begin{align*}&f_{Y_{j}}(\boldsymbol{U}_{0i},\boldsymbol{U}_{\text{AY}i},\overline{\boldsymbol{M}}_{j-1,i},   \overline{A}_{j-1,i},\overline{\boldsymbol{L}}_{j-1,i},\overline{N}_{Yji})\\
&=\beta_{0}+Z_{0i}+\beta_{L}{L}_{j-1,i} + \epsilon_{ji}+ (\beta_{1}+Z_{1i}^{\overline{A}}) A_{j-1,i} + (\beta_{2}+Z_{2i}^{\overline{A}}) A_{j-2,i}, \end{align*}
where still $\forall \overline{a}, \overline{b}{:}~Z_{1}^{\overline{a}} ~\overset{d}=~ Z_{1}^{\overline{b}}, \text{ and } Z_{2}^{\overline{a}} ~\overset{d}=~ Z_{2}^{\overline{b}}$. 

Under Assumption \ref{CH4A1}, by the absence of direct unmeasured confounding,\\ $\boldsymbol{U}_{0}=Z_{0}$, $(\boldsymbol{U}_{\text{AY}})_{\overline{a}}=\left(Z_{1}^{\overline{a}}, Z_{2}^{\overline{a}}\right)$, and $\overline{N}_{Y}=\overline{\epsilon}$. 

\noindent Since $Y_{j}^{\overline{a}_{j-1}}\independent \overline{A}_{j-1} \mid \boldsymbol{U}_{0},(\boldsymbol{U}_{\text{AY}})_{\overline{{a}}}, \overline{\boldsymbol{L}}_{j-1}, \overline{\boldsymbol{M}}_{j-1}, N_{Yj}$, and by Assumption \ref{CH4A2}, 
\begin{align*}	& f_{Y_{j}}(\boldsymbol{U}_{0i},\boldsymbol{U}_{\text{AY}i},\overline{\boldsymbol{M}}_{j-1,i},   \overline{a}_{j-1,i},\overline{\boldsymbol{L}}_{j-1,i},\overline{N}_{Yji}) \mid \boldsymbol{U}_{0},(\boldsymbol{U}_{\text{AY}})_{\overline{{a}}}, \overline{\boldsymbol{L}}_{j-1}, N_{Yj}
~\overset{d}=~\\ &f_{Y_{j}}(\boldsymbol{U}_{0i},\boldsymbol{U}_{\text{AY}i},\overline{\boldsymbol{M}}_{j-1,i},   \overline{A}_{j-1,i},\overline{\boldsymbol{L}}_{j-1,i},\overline{N}_{Yji}) \mid \boldsymbol{U}_{0},(\boldsymbol{U}_{\text{AY}})_{\overline{{a}}}, \overline{\boldsymbol{L}}_{j-1}, N_{Yj}, \overline{A}_{j-1}{=}\overline{a}_{j-1}.\end{align*} Thus, 
\begin{align*}&f_{Y_{j}}(\boldsymbol{U}_{0i},\boldsymbol{U}_{\text{AY}i},\overline{\boldsymbol{M}}_{j-1,i},   \overline{a}_{j-1,i},\overline{\boldsymbol{L}}_{j-1,i},\overline{N}_{Yji})\\
&=\beta_{0}+Z_{0i}+\beta_{L}{L}_{j-1,i} + \epsilon_{ji}+ (\beta_{1}+Z_{1i}^{\overline{a}}) a_{j-1} + (\beta_{2}+Z_{2i}^{\overline{a}}) a_{j-2}, \end{align*}
where $\forall \overline{a}, \overline{b}{:}~Z_{1}^{\overline{a}} ~\overset{d}=~ Z_{1}^{\overline{b}}, \text{ and } Z_{2}^{\overline{a}} ~\overset{d}=~ Z_{2}^{\overline{b}}$. 
As a result, for timepoints $j_{1}$, $j_{2}$, $\hdots$, $j_{q}$, 
\begin{align*}&(Y_{j_{1}}^{\overline{a}}-Y_{j_{1}}^{\overline{0}}, \hdots, Y_{j_{q}}^{\overline{a}}-Y_{j_{q}}^{\overline{0}})  ~\overset{d}=~\\ 
&\left((\beta_{1}+Z_{1})a_{j_{1}-1} + 
(\beta_{2}+Z_{2})a_{j_{1}-2}, \hdots, (\beta_{1}+Z_{1})a_{j_{q}-1} + 
(\beta_{2}+Z_{2})a_{j_{q}-2}\right). \qedhere\end{align*}
	\end{proof}

\end{appendix}

\newpage
\beginsupplement
\textbf{Supplement to “Individual causal effects from observational longitudinal studies with time-varying exposures” by Post et al.}

\newpage
\section{Supplementary Figures}\label{supp:fig}
Supplementary figures that are referred to in the main text. \vspace{-0.75cm}
\begin{figure}[H]
	\centering
	\begin{subfigure}{.325\textwidth}
		\resizebox{1\textwidth}{!}{\includegraphics[page=1]{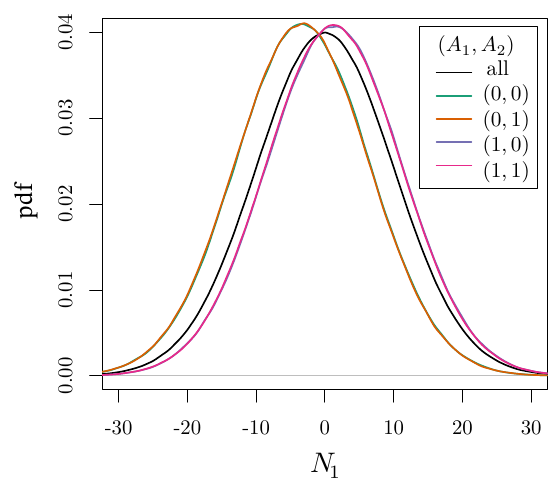}}
		\caption{}\label{CH4TVCNa}	
	\end{subfigure} \hfill
	\begin{subfigure}{.325\textwidth}
		\resizebox{1\textwidth}{!}{\includegraphics[page=1]{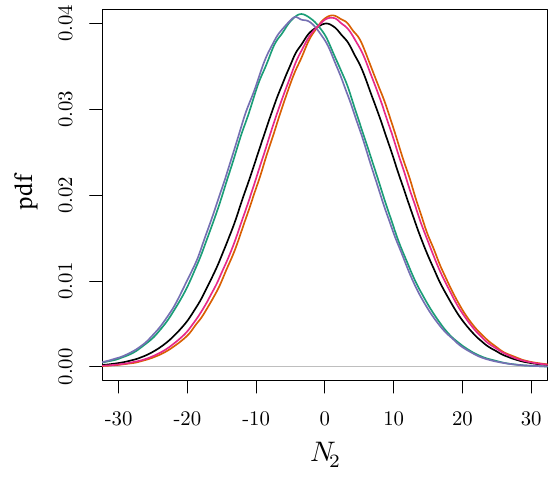}}
		\caption{}\label{CH4TVCNb}	
	\end{subfigure} \hfill
	\begin{subfigure}{.325\textwidth}
		\resizebox{1\textwidth}{!}{\includegraphics[page=1]{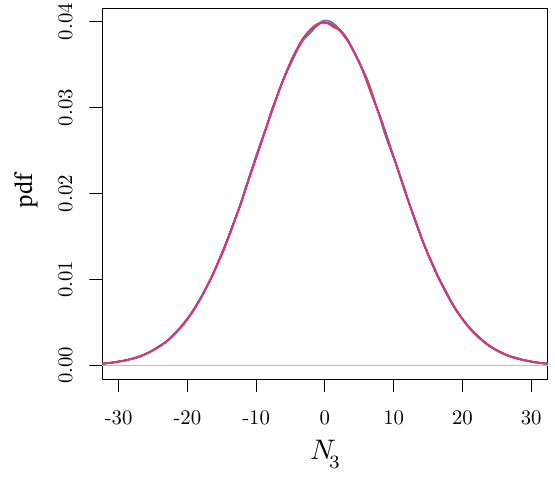}}
		\caption{}\label{CH4TVCNc}	
	\end{subfigure}
	\captionsetup{width=\textwidth}
	\caption{\footnotesize{Conditional distributions of $N_{Y1} {\mid} A_{1}, A_{2}$ (a), $N_{Y2} {\mid} A_{1}, A_{0}$ (b) and $N_{Y3} {\mid} A_{1}, A_{0}$ (c) and the population distribution of these latent variables (black lines) for the Gaussian linear mixed assignment example. The exposure levels can be found in the legend. The lines for $(0,0)$ and $(0,1)$ as well as for $(1,0)$ and $(1,1)$ in (a), for $(0,0)$ and $(1,0)$ as well as for $(0,1)$ and $(1,1)$ in (b) and all lines in (c) do overlap.}}\label{CH4TVCN}
\end{figure} \vspace{-1.1cm}
\begin{figure}[H]
		\centering
		\resizebox{0.9\textwidth}{!}{\includegraphics{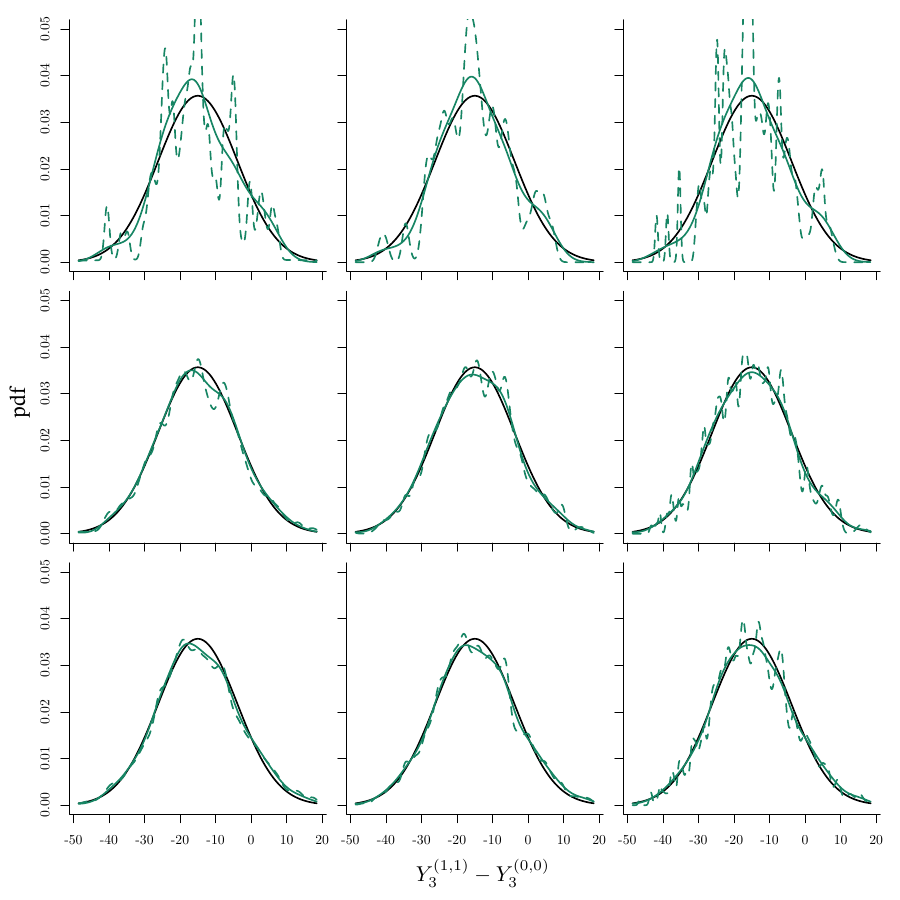}}
		\captionsetup{width=\textwidth}
	\caption{\footnotesize{Average CWCE of $\overline{a}{=}\overline{1}$ density (dashed green) and the kernel density of $\widehat{{}\mathbb{E}}[Y_{3i}^{(1,1)}-Y_{3i}^{(0,0)} {\mid}\mathcal{H}_{hi}]$ (solid green, using the \texttt{density()} function in \texttt{R} with default settings), based on different subsets of the data, and the true ICE density (black). The 
 rows correspond to the sample sizes $(100, 500, 1000)$ and the columns to the number of repeated measurements $(3,10,100)$.}}
	\label{CH4fig:ex1marginalicedistribution}
\end{figure}

	\section{Log-normal linear mixed assignment}\label{CH4ex2}
\begin{figure}[H]
		\centering
		\includegraphics[width=0.6\textwidth]{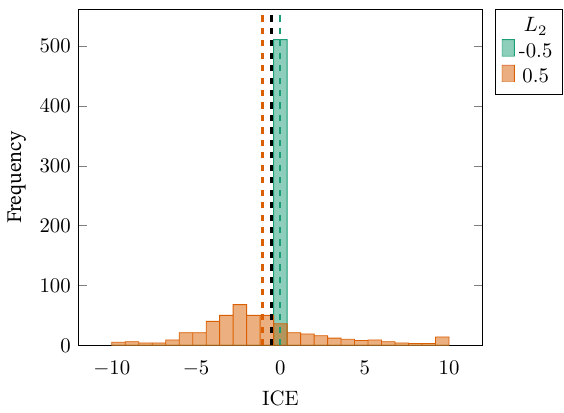}
		\captionsetup{width=\linewidth}
		\caption{\footnotesize{ICE distribution from the causal log-normal linear mixed assignment for $j~{=}~3$ and the different levels of $L_{2}$. Moreover, the dotted vertical lines represent the ACE and the two CACEs. }}\label{CH4fig:ICE2a}
	\end{figure}

	Let us consider a setting in which $f_{Y_{j}}$ in SCM \eqref{CH4indSCM} is a non-linear injective function of $N_{Yj}$. The cause-effect relations are parameterized as in SCM \eqref{CH4SCM:ex1} of the main text. However, now the outcome of interest is  \begin{equation}\widetilde{{}Y}_{j}^{\overline{a}}~{:}{=}~\exp\left(Y_{j}^{\overline{a}}\right),\end{equation} so that $N_{Yj}$ and $U_{0}$ in the outcome assignment are no longer additive. In this case $U_{0}$, the $N_{Yj}$ and $L_{j-1}$ become receptiveness factors and a modifier respectively of $\widetilde{{}Y}_{j}^{\overline{a}}-\widetilde{{}Y}_{j}^{\overline{0}}$ as the ICE is equal to
	$$\exp\left(\theta_{0}+\theta_{L}L_{j-1}+U_{0}+N_{Yj}\right) \left(\exp\left(\left(\theta_{1}+U_{1}\right)a_{j-1}+\left(\theta_{1}+U_{2}\right)a_{j-2}\right) - 1 \right).$$
	In this example, the confounder is also a modifier and the $L_{j-1}$-CACE is equal to
	$$\resizebox{\linewidth}{!}{$\exp\left(\theta_{0}+L_{j-1}\theta_{L}\right)\exp\left(\tfrac{\sigma_{0}^2}{2}\right)\exp\left(\tfrac{\sigma^2}{2}\right)\left(\exp\left(\left(\theta_{1}+\tfrac{\sigma_{1}^2}{2}\right)a_{j-1}\right)\exp\left(\left(\theta_{2}+\tfrac{\sigma_{2}^2}{2}\right)a_{j-2}\right)-1\right),$}$$
	which is different from the ACE equal to
	$$\resizebox{\linewidth}{!}{$\sum_{\ell}\exp\left(\theta_{0}\right)\exp\left(\theta_{\ell}\ell\right)\exp\left(\tfrac{\sigma_{0}^2}{2}\right)\exp\left(\tfrac{\sigma^2}{2}\right)\left(\exp\left(\left(\theta_{1}+\tfrac{\sigma_{1}^2}{2}\right)a_{j-1}\right)\exp\left(\left(\theta_{2}+\tfrac{\sigma_{2}^2}{2}\right)a_{j-2}\right)-1\right)\mathbb{P}(L_{j-1}~{=}~\ell).$}$$ We have adjusted the parameters introduced in Section \ref{CH4ex1} so that the outcome of interest has the same order of magnitude as in the first example. Now $\theta_{0}~{=}~0$, $\theta_{1}~{=}~-0.2$, $\theta_{2}~{=}~-0.1$, $\sigma_{0}~{=}~0.25$, $\sigma_{1}~{=}~0.5$, $\sigma_{2}~{=}~0.25$, $\sigma~{=}~0.25$, $\theta_{L}~{=}~4$, $\alpha_{0}~{=}~-0.5$, $\alpha_{1}~{=}~0.01$, $\alpha_{2}~{=}~1$, $\alpha_{3}~{=}~0.7$ and $L_{j}~{=}~0.5$ with probability $0.5$ and $L_{j}~{=}~ -0.5$ otherwise. The histogram of the ICE of $\overline{a}{=}\overline{1}$, at the third repeated measure, of $1000$ simulated individuals is presented together with the sample ACE (equal to $-0.54$, dashed black) in Figure \ref{CH4fig:ICE2a}. 
 This causal-effect distribution is less variable than the distribution in the previous example, as presented in Figure \ref{CH4fig:ICE1}. The CACE equals $-0.02$ for $L_{2}~{=}~-0.5$ and $-1.05$ for $L_{2}~{=}~0.5$.   \noindent Since the noise variable $N_{Yj}$ is now a time-varying receptiveness factor it affects the ICE at time $j$, so that $Y_{1}^{1}-Y_{1}^{0}$ differs from $Y_{2}^{(0,1)}-Y_{2}^{(0,0)}$. Which variables are modifiers or receptiveness factors also depends on the causal measure of interest. This dependence is in line with the interpretation of modification for different measures of effect discussed in the epidemiological literature (see, e.g.~\citet{VanDerWeele2014}). For example, a factor could be a modifier for a risk ratio but not for a risk difference. In this work, the ICE is defined as the difference of potential outcomes. However, if the ICE would be defined as the ratio of potential outcomes, then the confounder $L_{j-1}$ would not be a modifier and the $U_{0}$ nor $N_{Yj}$ would be receptiveness factors. The framework is also applicable for other measures since the main theorems presented in Section \ref{CH4CWIT} address the joint distribution of potential outcomes. 

	\subsection{Cross-world causal effect}\label{CH4sec:32}
\begin{figure}[H]
		\centering
		\resizebox{0.7\textwidth}{!}{\includegraphics{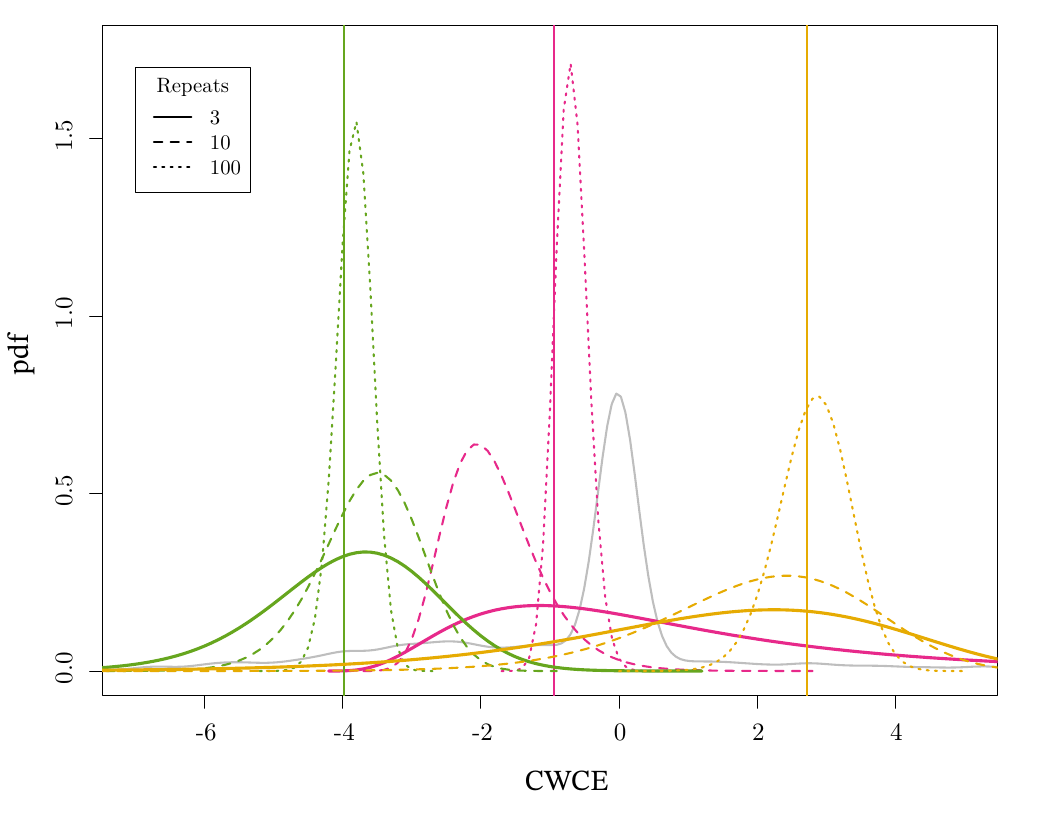}}
		\captionsetup{width=\textwidth}
		\caption{\footnotesize{Distribution of the CWCE for $\overline{a}{=}\overline{1}$, at the third repeat, for three individuals, based on information from $3$ (solid), $10$ (dashed) and $100$ (dotted) repeats respectively for the log-normal example. The exposure assignment at the first two time points equal $(1,0)$, $(0,0)$ and $(1,1)$ for the green, pink and yellow curves, respectively. Furthermore, the actual ICE for each individual (vertical lines) and the population ICE distribution (grey) are presented.}}\label{CH4fig8b}
	\end{figure}

	For the causal log-normal linear mixed assignment, the $f_{Y_{j}}$ are also injective functions of $N_{Yj}$. The pdf of the CWCE, $f\left(\widetilde{{}Y}_{j}^{\overline{a}}-\widetilde{{}Y}_{j}^{\overline{0}}~{=}~d \mid \mathcal{H}_{h} \right)$, can thus be derived by applying Corollary \ref{CH4cor:331} as
	\begin{align*}	
	& \int_{y~{=}~0}^{\infty}\int \mathbbm{1}_{\left\{\exp\left((\theta_{1}+U_{1})a_{j-1}+(\theta_{2}+U_{2})a_{j-2}+\left(Y_{j}-(\theta_{1}+U_{1})A_{j-1}-(\theta_{2}+U_{2})A_{j-2}\right)\right)~{=}~y\right\}}\\
	&\mathbbm{1}_{\left\{\exp\left(Y_{j}-(\theta_{1}+U_{1})A_{j-1}-(\theta_{2}+U_{2})A_{j-2}\right)~{=}~y-d\right\}}dF_{(U_{1},U_{2})\mid \mathcal{H}_{h}} dy.
	\end{align*} Note that 
	\begin{equation*}
	(U_{1},U_{2})\mid \mathcal{H}_{h} ~\overset{d}=~ (U_{1}, U_{2})\mid \overline{Y}_{h}, \overline{\boldsymbol{L}}_{h}, 
 \overline{A}_{h}, 
	\end{equation*} whose distribution was derived at the start of Section \ref{CH4sec:31} of the main text. The integral cannot be further simplified for arbitrary parameter values and should be evaluated numerically. For the parameter choices introduced in the previous subsection, the CWCE distributions for different individuals based on a varying number of repeats are presented in Figure \ref{CH4fig8b}. Contrary to the Gaussian linear mixed example, the CWCE distribution at $j~{=}~3$ for an unexposed (at the first two time points) individual (pink curves) is now not equal to the marginal ICE distribution since $U_{0}$ and $N_{Y3}$ are receptiveness factors that can be informed on by $\overline{Y}_{3}$. The uncertainty on the level of this time-varying receptiveness factor is also why the CWCE of the individual with exposure at both time points (yellow curves) is not the lowest in variability. 
\noindent The ICE distribution in the sample is equal to the ICE distribution of the population under the assumption that the modifier distribution in the sample is the same as the distribution in the population. In this log-normal linear mixed assignment, the confounder $L_{2}$ is also a modifier. Therefore, the confounder distribution in the population should be known to derive the ICE distribution in the population.
	
\subsection{Inference}\label{CH4s5LNL}
\begin{figure}[H]
		\centering
		\begin{subfigure}{.325\textwidth}
			\resizebox{1\textwidth}{!}{\includegraphics[page=1]{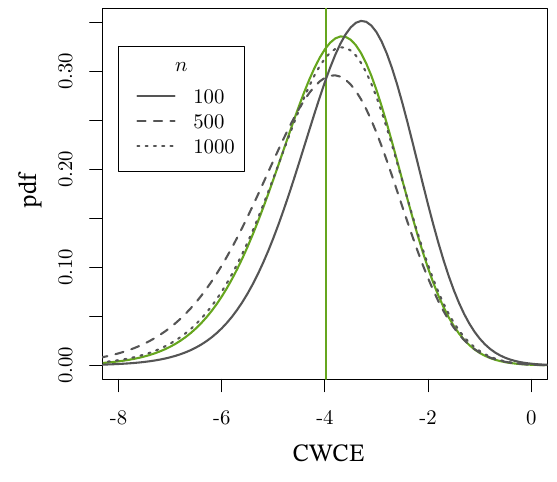}}
			\caption{}\label{CH4CWCEaS}	
		\end{subfigure}
		\begin{subfigure}{.325\textwidth}
			\resizebox{1\textwidth}{!}{\includegraphics[page=1]{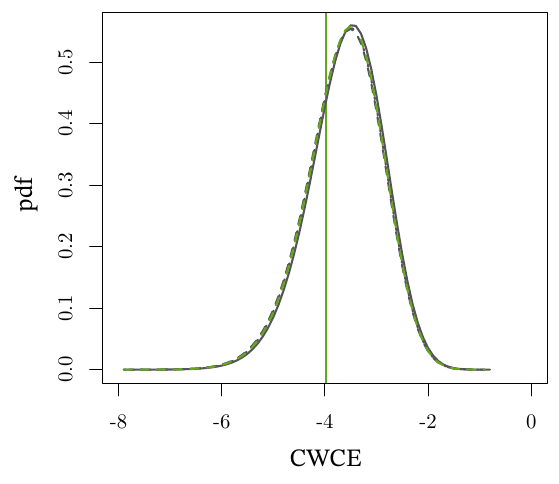}}
			\caption{}\label{CH4CWCEbS}	
		\end{subfigure}
		\begin{subfigure}{.325\textwidth}
			\resizebox{1\textwidth}{!}{\includegraphics[page=1]{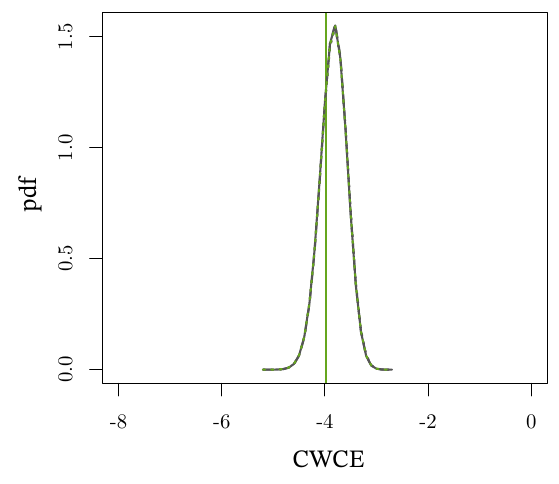}}
			\caption{}\label{CH4CWCEcS}	
		\end{subfigure}
		\captionsetup{width=\textwidth}
		\caption{\footnotesize{Estimated CWCE of $\overline{a}{=}\overline{1}$ distribution, at the third repeat, for a specific individual, based on a sample containing $100$ (solid grey), $500$ (dashed grey) and $1000$ (dotted grey) individuals	for $3$ (a), $10$ (b) and $100$ (c) repeats respectively. The actual CWCE distribution (green lines) and ICE (vertical lines) were already presented in Figure \ref{CH4fig8b}.}}\label{CH4figS8est}
	\end{figure}

\begin{figure}[H]
			\centering
			\resizebox{\textwidth}{!}{\includegraphics{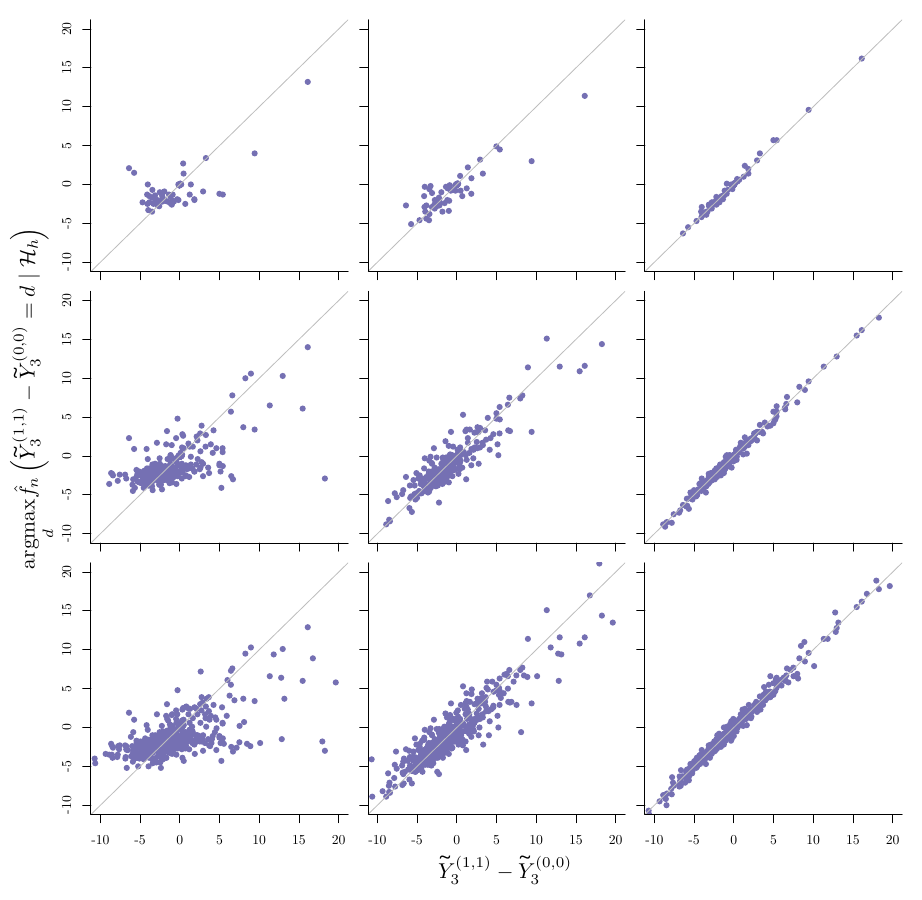}}
		\captionsetup{width=\textwidth}
		\caption{\footnotesize{The actual ICE of $\overline{a}{=}\overline{1}$ at the time of the third repeat versus the estimated ICE based on different subsets of the data. The rows correspond to the sample sizes $(100, 500, 1000)$ and the columns to the number of repeats $(3,10,100)$.}}\label{CH4LNICE}
	\end{figure}

\begin{figure}[H]
			\centering
			\resizebox{0.9\textwidth}{!}{\includegraphics{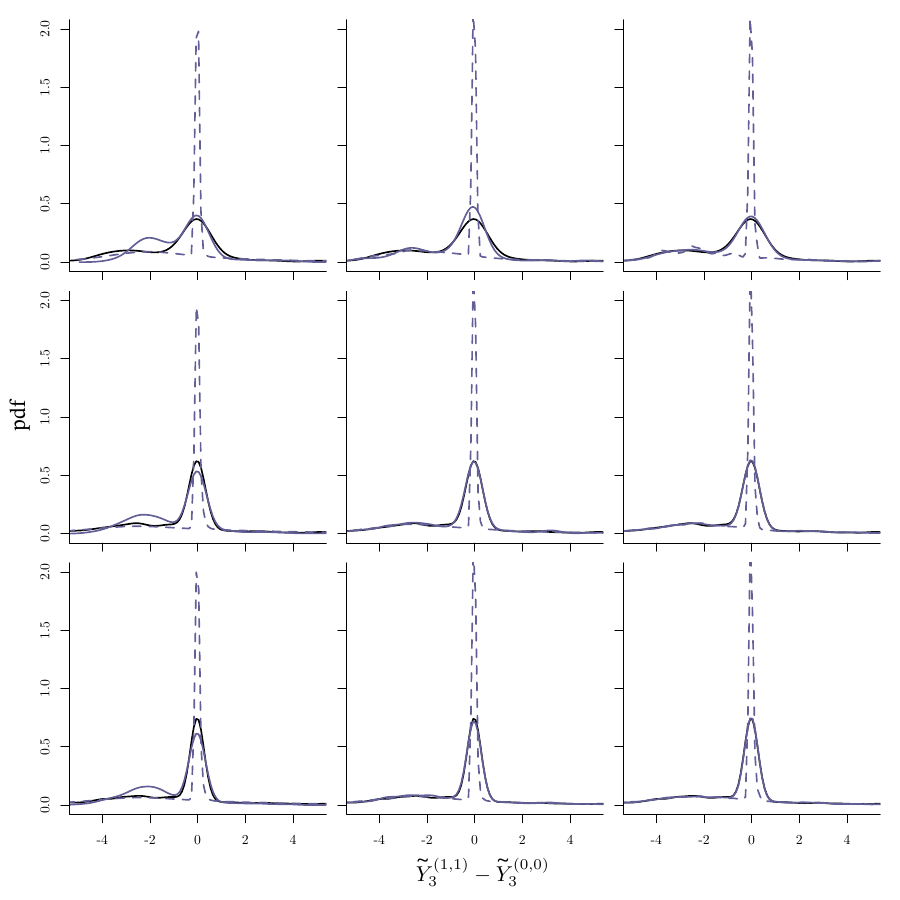}}
		\captionsetup{width=\textwidth}
		\caption{\footnotesize{Average estimated CWCE of $\overline{a}{=}\overline{1}$ density (dashed blue) and the kernel density of the mode of the estimated CWCE per individual (solid blue, using the \texttt{density()} function in \texttt{R} with default settings), based on different subsets of the data, and the kernel density of the actual ICE in the subsets (black).The rows correspond to the sample sizes $(100, 500, 1000)$ and the columns to the number of repeats $(3,10,100)$.}}
		\label{CH4fig:ex2marginalicedistribution}
	\end{figure}

	Inference based on data generated by the log-normal linear mixed assignment as presented in Section \ref{CH4ex2} is based on the REML model fit of $\log (\widetilde{{}Y}_{j})$, for $1{\leq}j{\leq}h$, with which also $\sigma_{0}^{2}$ was estimated. We have estimated the distribution of the CWCE of $\overline{a}{=}\overline{1}$ at the time of the third repeat $(j~{=}~3)$. For a specific individual (with $A_{1}~{=}~1$ and $A_{2}~{=}~1$), which CWCE of $\overline{a}{=}\overline{1}$ was also shown in Figure \ref{CH4fig8b}, the estimated pdf of the CWCE is presented for a varying number of individuals and number of repeats in Figure \ref{CH4figS8est}. For $j~{=}~3$, the convergence presented in Theorem \ref{CH4th41} is again applicable for $n~{=}~1000$ and $h~{=}~3$ and $n~{=}~100$ and $h~{=}~10$.   
 \noindent Moreover, we estimated the ICE for all individuals with the mode of the estimated CWCE distributions. In Figure \ref{CH4LNICE}, we present the actual versus estimated ICE based on different subsets of the data. With only one hundred individuals, but one hundred repeats, the ICE can be estimated quite accurately for this setting. However, in contrast to the first example, we see that three repeats are insufficient to make precise individual inferences, regardless of the number of individuals. From Figure \ref{CH4fig:ex2marginalicedistribution}, we observe that also the density of the causal effects cannot be estimated accurately with only three repeats. In comparison, the estimate is rather accurate for ten repeats. 
 
 \newpage
   \null
    \vfill
    \vfill
     \null

\newpage
 
\section{Logistic linear mixed assignment}\label{CH4ex3}
\begin{figure}[H]
		\centering
		\begin{minipage}[b]{.45\linewidth}
				\centering
					\resizebox{1\textwidth}{!}{\includegraphics{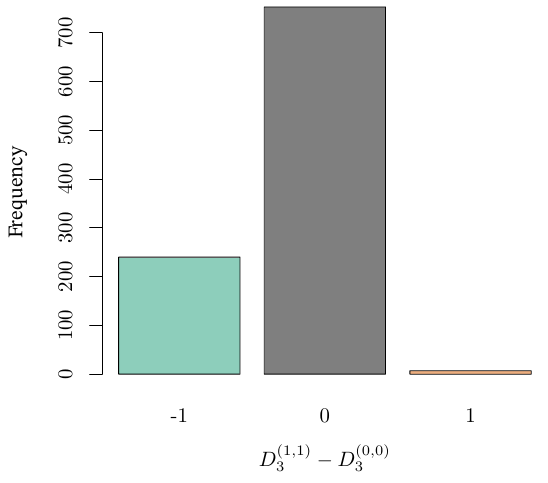}}
				\captionof{figure}{\footnotesize{Distribution of $D_{3}^{(1,1)}-D_{3}^{(0,0)}$ for the $1000$ individuals as described in this section.}}\label{CH4figSM31}
		\end{minipage}
		\hfill
		\begin{minipage}[b]{.45\linewidth}
				\centering
				\resizebox{1\textwidth}{!}{\includegraphics{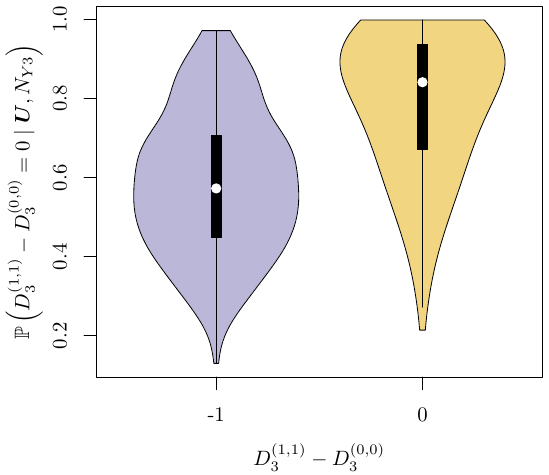}}
					\captionof{figure}{\footnotesize{Density estimate and boxplot of \scalebox{0.8}{$\mathbb{P}\left(D_{3}^{(1,1)}-D_{3}^{(0,0)}~{=}~0 \mid \boldsymbol{U}, N_{Y3}\right)$} stratified by ICE.}}\label{CH4figSM32}
		\end{minipage}
	\end{figure}

	Finally, we present a setting in which the $f_{Y_{j}}$ in SCM \eqref{CH4indSCM} is not an injective function of $N_{Yj}$. Again, the cause-effect relations extend the system introduced in Section \ref{CH4ex1}, now with the binary outcome of interest equal to $D_{j}^{\overline{a}}$. For all $j$,
	\begin{equation}
	D_{ji}^{\overline{a}}~{:}{=}~\mathbbm{1}_{\{\text{logit}^{-1}\left(\gamma_{0}+U_{3i}+\gamma_{1}Y_{ji}^{\overline{a}}\right)>N_{Dji}\}},
	\end{equation} where $U_{3i}\sim \mathcal{N}(0,\sigma_{3}^{2})$, $N_{Dji} \sim \text{Uni}[0,1]$, $\forall i \forall j,k {:}~ N_{Dki} \independent N_{Aji}$, and for $j{\neq}k$, $N_{Dki} \independent N_{Dji}$. As in the log-normal example, the $U_{0}$ and $N_{Yj}$ are (time-varying) receptiveness factors. Also,  $N_{Dj}$ is a time-varying receptiveness factor since for all $j$,
	\begin{equation}
	D_{ji}^{\overline{a}}~{:}{=}~\mathbbm{1}_{\{\text{logit}^{-1}\left(\gamma_{0}+U_{3i}+\gamma_{1}\left(\theta_{0}+U_{i0}+L_{j-1,i}\beta_{L}+(\beta_{1}+U_{1i})a_{j-1}+(\beta_{1}+U_{2i})a_{j-2}+N_{Yji}\right)\right)>N_{Dji}\}}.
	\end{equation} The `noise' variables ($(N_{Yj}, N_{Dj})$) in the SCM are now two dimensional,  and $f_{D_{j}}$ is not injective in $(N_{Yj}, N_{Dj})$. For this example, we assume that the mediator $Y_{j}$ is observed and could be used to inform on $U_{0}$, $U_{1}$, $U_{2}$ and $N_{Yj}$. Notice that even when $N_{Yj}$ is known, $f_{D_{j}}$ is not injective in $N_{Dj}$. The ICE, $D_{j}^{\overline{a}}-D_{j}^{\overline{0}}$ is either $-1$, $0$ or $1$ and the corresponding probabilities are individual-specific and time-varying as a result of the receptiveness factors. As for the log-normal example, the confounder $L_{j-1}$ is a modifier, and the $L_{j-1}$-CACE equals 
	\begin{align*}
	\scalebox{0.9}{$\int \bigg{(}$} & \scalebox{0.9}{$
\text{logit}^{-1}\left(\gamma_{0}+U_{3}+\gamma_{1}\left(\theta_{0}+U_{0}+L_{j-1}\beta_{L}+(\beta_{1}+U_{1})a_{j-1}+(\beta_{1}+U_{2})a_{j-2}+N_{Yj}\right)\right)$}
\\
	& \scalebox{0.9}{$-\text{logit}^{-1}\left(\gamma_{0}+U_{3}+\gamma_{1}(\theta_{0}+U_{0}+L_{j-1}\beta_{L}+N_{Yj})\right)\bigg{)} dF_{(U_{0},U_{1},U_{2},U_{3},N_{Yj})}.$}\end{align*} For $1000$ individuals simulated using the values introduced in Section \ref{CH4ex1}, additionaly  $\gamma_{0}~{=}~-10$, $\gamma_{1}~{=}~0.1$ and $\sigma_{3}~{=}~1$, the CACE for $L_{j-1}~{=}~0.7$ (equal to $-0.230$) and $-0.3$ ($-0.234$) are  similar. The ICE distribution is presented in Figure \ref{CH4figSM31}. 
\noindent To further illustrate the heterogeneity, we present a Violin plot of the values of $\mathbb{P}(D_{3}^{(1,1)}-D_{3}^{(0,0)}~{=}~0 \mid \boldsymbol{U}, N_{Y3})$, which is equal to the probability that $N_{D3}$ is realized between the values of $\mathbb{P}(D_{3}^{(1,1)}~{=}~0\mid \boldsymbol{U}, N_{Y3})$ and $\mathbb{P}(D_{3}^{(1,1)}~{=}~1\mid \boldsymbol{U}, N_{Y3})$, i.e.~
	$$ 1- |\mathbb{P}(D_{3}^{(1,1)}~{=}~1\mid \boldsymbol{U}, N_{Y3})-\mathbb{P}(D_{3}^{(0,0)}~{=}~1\mid \boldsymbol{U}, N_{Y3})|,$$ grouped by the ICE in Figure \ref{CH4figSM32}. The impact of the receptiveness factor $N_{D3}$ becomes clear as there are individuals with an ICE equal to $-1$ which $\mathbb{P}(D_{3}^{(1,1)}-D_{3}^{(0,0)}~{=}~0 \mid \boldsymbol{U}, N_{Y3})$ is larger than for some individuals with an ICE equal to $0$. We did not present the probabilities for the $7$ individuals with an ICE equal to $1$.

	\subsection{Cross-world causal effect}
\begin{figure}[H]
		\centering
		\begin{subfigure}{.32\textwidth}
			\resizebox{1\textwidth}{!}{\includegraphics[page=1]{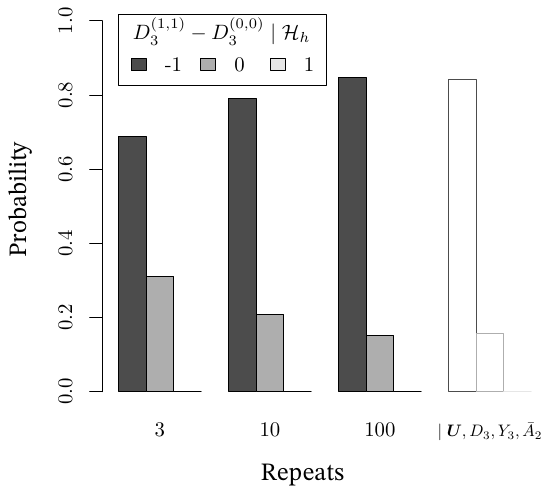}}
			\caption{}\label{CH4SM3fig3a}	
		\end{subfigure}
		\begin{subfigure}{.32\textwidth}
			\resizebox{1\textwidth}{!}{\includegraphics[page=2]{TikZFig/SM3Fig3b.pdf}}
			\caption{}\label{CH4SM3fig3b}	
		\end{subfigure}
		\begin{subfigure}{.32\textwidth}
			\resizebox{1\textwidth}{!}{\includegraphics[page=3]{TikZFig/SM3Fig3b.pdf}}
			\caption{}\label{CH4SM3fig3c}	
		\end{subfigure}
		\captionsetup{width=\textwidth}
		\caption{\footnotesize{CWCE of $\overline{a}{=}\overline{1}$ distribution of the effect on the third observation, given $3$, $10$ and $100$ repeated measurements, for three different individuals having an ICE equal to $-1$ (a), $0$ (b) and $1$ (c). Furthermore, $\mathbb{P}(D_{3}^{(1,1)}-D_{3}^{(0,0)}~{=}~d \mid \boldsymbol{U}, D_{3}, Y_{3}, \overline{A}_{2})$ is presented for each individual.}}\label{CH4SM3fig3}
	\end{figure}

	When $U_{3}$ would be known in addition to the observations of $Y_{j}$ and $D_{j}$, 
	\begin{equation*}
	N_{Dj} \mid U_{3}, D_{j}, Y_{j}, \overline{A}_{j-1} \sim \begin{cases}
	\text{Uni}\left(\text{logit}^{-1}\left(\gamma_{0}+U_{3}+\gamma_{1}Y_{j}\right),1 \right] &\text{if $D_{j}~{=}~0$} \\
	\text{Uni}\left[0, \text{logit}^{-1}\left(\gamma_{0}+U_{3i}+\gamma_{1}Y_{j}\right) \right]&\text{if $D_{j}~{=}~1$}
	\end{cases}. 
	\end{equation*} While more repeats can inform on the levels of $\boldsymbol{U}$ (and as such $N_{Yj}$), they do not provide more information on the level of $N_{Dj}$. For this reason, the CWCE will, in many cases, not converge to the ICE, as we will discuss next. Let $$p(u,y)~{=}~\text{logit}^{-1}\left(\gamma_{0}+u+\gamma_{1}y\right), \text{ and }$$ 
 $$\resizebox{\linewidth}{!}{$p_{j}^{\overline{a}}(\boldsymbol{u},Y_{j},\overline{A}_{j-1})~{=}~ \text{logit}^{-1}\left(\gamma_{0}+u_{3}+\gamma_{1} \left(Y_{j} + (a_{j-1}-A_{j-1})(\beta_{1}+u_{1}) + (a_{j-2}-A_{j-2})(\beta_{2}+u_{2})\right)\right).$}$$ 
	Note that $p_{j}^{\overline{a}}(\boldsymbol{U},Y_{j},\overline{A}_{j-1})~{=}~ \mathbb{P}(D_{j}^{\overline{a}}~{=}~1 \mid \mathcal{H}_{h},\boldsymbol{U})$. Then, $p_{d}~{=}~\mathbb{P}(D_{j}^{\overline{a}} - D_{j}^{\overline{a}}~{=}~d \mid \mathcal{H}_{h},\boldsymbol{U})$, depends on the values of the event indicators $A$, $B$ and $C$ defined as
	\begin{align*}
	A &~{=}~ \mathbbm{1}_{\{p_{j}^{\overline{0}}(\boldsymbol{U},Y_{j}, \overline{A}_{j-1}) {\geq} p(U_{3},Y_{j})\}},\\
	B &~{=}~ \mathbbm{1}_{\{p_{j}^{\overline{a}}(\boldsymbol{U},Y_{j}, \overline{A}_{j-1}) {\geq} p(U_{3},Y_{j})\}},\\
	C &~{=}~ \mathbbm{1}_{\{p_{j}^{\overline{a}}(\boldsymbol{U},Y_{j}, \overline{A}_{j-1}) {\geq} p_{j}^{\overline{0}}(\boldsymbol{U},Y_{j}, \overline{A}_{j-1})\}},
	\end{align*} and are presented in Table \ref{CH4tab:sc}. 	\begin{table}[H]
		\caption{Black, green and magenta points represents $ p(U_{3},Y_{j})$, $ p_{j}^{\overline{a}}(\boldsymbol{U},Y_{j}, \overline{A}_{j-1})$ and $ p_{j}^{\overline{0}}(\boldsymbol{U},Y_{j}, \overline{A}_{j-1})$ respectively, while $N_{Dj} \mid U_{3}, D_{j}, Y_{j}, \overline{A}_{j-1}$ follows a uniform distribution over the shaded area. }\label{CH4tab:sc}
		\centering
		\resizebox{\textwidth}{!}{\begin{tabular}{l||lll|lll|c}
				$D_{j}$ & $A$ & $B$ & $C$ & $p_{-1}$ & $p_{0}$ & $p_{1}$ & Situation 
				\\ \hline
				1 & 1 & 1 & 1 & 0 & 1 & 0 & \begin{tikzpicture}
				\draw [line width=10] [lightgray] (0,0) -- (1,0);
				\draw (0,0) -- (3,0);
				\draw [fill][black] (1,0) circle [radius=0.05];
				\draw [fill][clr4] (1.5,0) circle [radius=0.05];
				\draw [fill][clr5] (2,0) circle [radius=0.05];
				
				\end{tikzpicture}\\
				1 & 1 & 1 & 0 & 0 & 1 & 0 & \begin{tikzpicture}
				\draw [line width=10] [lightgray] (0,0) -- (1,0);
				\draw (0,0) -- (3,0);
				\draw [fill][black] (1,0) circle [radius=0.05];
				\draw [fill][clr4] (2,0) circle [radius=0.05];
				\draw [fill][clr5] (1.5,0) circle [radius=0.05];
				
				\end{tikzpicture}\\
				1 & 1 & 0 & 1 & $\frac{p(U_{3},Y_{j})-p_{j}^{\overline{a}}(\boldsymbol{U},Y_{j}, \overline{A}_{j-1})}{p(U_{3},Y_{j})}$& $1-p_{-1}$ & 0 & \begin{tikzpicture}
				\draw [line width=10] [lightgray] (0,0) -- (1,0);
				\draw (0,0) -- (3,0);
				\draw [fill][black] (1,0) circle [radius=0.05];
				\draw [fill][clr4] (0.5,0) circle [radius=0.05];
				\draw [fill][clr5] (2,0) circle [radius=0.05];
				
				\end{tikzpicture}\\
				1 & 0 & 1 & 0 & 0 & $1-p_{1}$ & $\frac{p(U_{3},Y_{j})-p_{j}^{\overline{0}}(\boldsymbol{U},Y_{j}, \overline{A}_{j-1})}{p(U_{3},Y_{j})}$ & \begin{tikzpicture}
				\draw [line width=10] [lightgray] (0,0) -- (1,0);
				\draw (0,0) -- (3,0);
				\draw [fill][black] (1,0) circle [radius=0.05];
				\draw [fill][clr4] (2,0) circle [radius=0.05];
				\draw [fill][clr5] (0.5,0) circle [radius=0.05];
				
				\end{tikzpicture}\\
				1 & 0 & 0 & 1 & $\frac{p_{j}^{\overline{a}}(\boldsymbol{U},Y_{j}, \overline{A}_{j-1})-p_{j}^{\overline{0}}(\boldsymbol{U},Y_{j}, \overline{A}_{j-1})}{p(U_{3},Y_{j})}$ & $1-p_{-1}$ & 0 & \begin{tikzpicture}
				\draw [line width=10] [lightgray] (0,0) -- (2,0);
				\draw (0,0) -- (3,0);
				\draw [fill][black] (2,0) circle [radius=0.05];
				\draw [fill][clr4] (0.5,0) circle [radius=0.05];
				\draw [fill][clr5] (1,0) circle [radius=0.05];
				
				\end{tikzpicture}\\
				1 & 0 & 0 & 0 & 0 & $1-p_{1}$ & $\frac{p_{j}^{\overline{0}}(\boldsymbol{U},Y_{j}, \overline{A}_{j-1})-p_{j}^{\overline{a}}(\boldsymbol{U},Y_{j}, \overline{A}_{j-1})}{p(U_{3},Y_{j})}$ & \begin{tikzpicture}
				\draw [line width=10] [lightgray] (0,0) -- (2,0);
				\draw (0,0) -- (3,0);
				\draw [fill][black] (2,0) circle [radius=0.05];
				\draw [fill][clr4] (1,0) circle [radius=0.05];
				\draw [fill][clr5] (0.5,0) circle [radius=0.05];
				
				\end{tikzpicture}\\
				0 & 1 & 1 & 1 & $\frac{p_{j}^{\overline{0}}(\boldsymbol{U},Y_{j}, \overline{A}_{j-1})-p_{j}^{\overline{a}}(\boldsymbol{U},Y_{j}, \overline{A}_{j-1})}{1-p(U_{3},Y_{j})}$  & $1-p_{-1}$ & 0 & \begin{tikzpicture}
				\draw [line width=10] [lightgray] (1,0) -- (3,0);
				\draw (0,0) -- (3,0);
				\draw [fill][black] (1,0) circle [radius=0.05];
				\draw [fill][clr4] (1.5,0) circle [radius=0.05];
				\draw [fill][clr5] (2,0) circle [radius=0.05];
				
				\end{tikzpicture}\\
				0 & 1 & 1 & 0 & 0 & $1-p_{1}$ & $\frac{p_{j}^{\overline{a}}(\boldsymbol{U},Y_{j}, \overline{A}_{j-1})-p_{j}^{\overline{0}}(\boldsymbol{U},Y_{j}, \overline{A}_{j-1})}{1-p(U_{3},Y_{j})}$  & \begin{tikzpicture}
				\draw [line width=10] [lightgray] (1,0) -- (3,0);
				\draw (0,0) -- (3,0);
				\draw [fill][black] (1,0) circle [radius=0.05];
				\draw [fill][clr4] (2,0) circle [radius=0.05];
				\draw [fill][clr5] (1.5,0) circle [radius=0.05];
				
				\end{tikzpicture}\\
				0 & 1 & 0 & 1 & $\frac{p_{j}^{\overline{0}}(\boldsymbol{U},Y_{j}, \overline{A}_{j-1})-p(U_{3},Y_{j})}{1-p(U_{3},Y_{j})}$  & $1-p_{-1}$ & 0 & \begin{tikzpicture}
				\draw [line width=10] [lightgray] (1,0) -- (3,0);
				\draw (0,0) -- (3,0);
				\draw [fill][black] (1,0) circle [radius=0.05];
				\draw [fill][clr4] (0.5,0) circle [radius=0.05];
				\draw [fill][clr5] (2,0) circle [radius=0.05];
				
				\end{tikzpicture}\\
				0 & 0 & 1 & 0 & 0 & $1-p_{1}$ & $\frac{p_{j}^{\overline{a}}(\boldsymbol{U},Y_{j}, \overline{A}_{j-1})-p(U_{3},Y_{j})}{1-p(U_{3},Y_{j})}$ & \begin{tikzpicture}
				\draw [line width=10] [lightgray] (1,0) -- (3,0);
				\draw (0,0) -- (3,0);
				\draw [fill][black] (1,0) circle [radius=0.05];
				\draw [fill][clr4] (2,0) circle [radius=0.05];
				\draw [fill][clr5] (0.5,0) circle [radius=0.05];
				
				\end{tikzpicture}\\
				0 & 0 & 0 & 1 & 0 & 1 & 0 & \begin{tikzpicture}
				\draw [line width=10] [lightgray] (2,0) -- (3,0);
				\draw (0,0) -- (3,0);
				\draw [fill][black] (2,0) circle [radius=0.05];
				\draw [fill][clr4] (0.5,0) circle [radius=0.05];
				\draw [fill][clr5] (1,0) circle [radius=0.05];
				
				\end{tikzpicture}\\
				0 & 0 & 0 & 0 & 0 & 1 & 0 & \begin{tikzpicture}
				\draw [line width=10] [lightgray] (2,0) -- (3,0);
				\draw (0,0) -- (3,0);
				\draw [fill][black] (2,0) circle [radius=0.05];
				\draw [fill][clr4] (1,0) circle [radius=0.05];
				\draw [fill][clr5] (0.5,0) circle [radius=0.05];			
				\end{tikzpicture}
		\end{tabular}}
	\end{table} \noindent By applying Theorem \ref{CH4th41}, we can derive the CWCE by integrating $p_{-1}$, $p_{0}$ or $p_{1}$ with respect to the prior distribution $F_{(U_{1}, U_{2}, U_{3}\mid \mathcal{H}_{h})}$ since $U_{0}+N_{Yj}$ is known given $(Y_{j}, A_{j-1}, A_{j-2}, L_{j-1}, U_{1}, U_{2})$ as was the case for the log-normal example. Moreover, the expression presented in table \ref{CH4tab:sc} were already integrated with respect to the distribution of $F_{(N_{Dj} \mid \boldsymbol{U}, \overline{N}_{Yj}, \mathcal{H}_{h})}~{=}~F_{(N_{Dj} \mid U_{3}, D_{j}, Y_{j}, \overline{A}_{j-1})}$. In this example $U_{3} \independent U_{0}, U_{1}, U_{2}, \overline{N}_{Y}$ which remains true conditionally on $\overline{D}_{j}, \overline{Y}_{j}, \overline{L}_{j}, \overline{A}_{j}$, so that 
	$$ F_{(U_{1}, U_{2}, U_{3} \mid \mathcal{H}_{h})} ~{=}~ F_{(U_{1}, U_{2} \mid \mathcal{H}_{h})}F_{(U_{3} \mid \overline{D}_{j}, \overline{Y}_{j})}.$$ At the start of Section \ref{CH4sec:31} of the main text we have derived $F_{(U_{1}, U_{2} \mid \mathcal{H}_{h})}$. Since $\forall j\neq k{:}~D_{j} \independent D_{k} \mid U_{3}, \overline{Y}_{h}$ we apply Bayes rule to derive the conditional pdf equal to
$$f_{U_{3}\mid \overline{D}_{h}, \overline{Y}_{h}}(u)~{=}~ \frac{f_{U_{3}}(u)\prod_{j~{=}~1}^{h}\mathbb{P}(D_{j}|U_{3}~{=}~u, Y_{j})}{\int f_{U_{3}}(k) \prod_{j~{=}~1}^{h}\mathbb{P}(D_{j}|U_{3}~{=}~k,  Y_{j}) dk }. $$ \newpage So for example, for $j \leq h$,  $\mathbb{P}(D_{j}^{\overline{a}}-D_{j}^{\overline{0}}~{=}~-1 \mid \mathcal{H}_{h})$ equals  

 	\begin{align*}
	 \scalebox{1.2}{$\int \int \int$} & 	\scalebox{1.2}{$\Bigg{(}
	\mathbbm{1}_{\{D_{ji}~{=}~1\}} \frac{\left((p(u_{3},Y_{j})-p_{j}^{\overline{a}}\left(\boldsymbol{u},Y_{j},\overline{A}_{j-1})\right)^{+}- (p(u_{3},Y_{j})-p_{j}^{\overline{0}}\left(\boldsymbol{u},Y_{j},\overline{A}_{j-1})\right)^{+}\right)^{+}}{p(u_{3},Y_{j})}$}\\
	&\scalebox{1.2}{$+ \mathbbm{1}_{\{D_{ji}~{=}~0\}} \frac{\left(\left(p_{j}^{\overline{0}}(\boldsymbol{u},Y_{j},\overline{A}_{j-1})-p(u_{3},Y_{j})\right)^{+}- \left(p_{j}^{\overline{a}}(\boldsymbol{u},Y_{j},\overline{A}_{j-1})-p(u_{3},Y_{j})\right)^{+}\right)^{+}}{1-p(u_{3},Y_{j})} \Bigg{)}$}\\
	&\scalebox{1.2}{$f_{U_{3}\mid \overline{D}_{h}, \overline{Y}_{h}    }(u_{3})f_{U_{1}, U_{2} \mid \overline{Y}_{h},\overline{L}_{h}, \overline{A}_{h}}\left(u_{1},u_{2}\right) du_{3}du_{2}du_{1}$},
	\end{align*} where $(x)^{+}~{=}~\text{max}(x,0)$. The CWCE can be computed using numerical integration for the parameter values introduced before. 
 For three individuals with different ICEs, the CWCE for varying $h$ is presented in Figure \ref{CH4SM3fig3c}. For this setting, the CWCE already mimics  $\mathbb{P}(D_{3}^{(1,1)}-D_{3}^{(0,0)} \mid \boldsymbol{U}, D_{3}, Y_{3},\overline{A}_{2})$ quite well when $h~{=}~3$. 
\noindent	In Figure \ref{CH4SM3fig3c}, we have presented the CWCE for one of the six individuals (out of the simulated $1000$) with an ICE equal to $1$ even though $\mathbb{P}(D_{3}^{(1,1)}-D_{3}^{(0,0)}~{=}~1 \mid \boldsymbol{U}, D_{3}, Y_{3}, \overline{A}_{2})$ is much smaller than $\mathbb{P}(D_{3}^{(1,1)}-D_{3}^{(0,0)}~{=}~0 \mid \boldsymbol{U}, D_{3}, Y_{3}, \overline{A}_{2})$. Even when the CWCE is known, it will be impossible for some individuals to get to their ICE as it is impossible to learn more about the $N_{D3}$.

	\subsection{Inference}
\begin{figure}[H]
		\centering
		\begin{subfigure}{.32\textwidth}
			\resizebox{1\textwidth}{!}{\includegraphics[page=1]{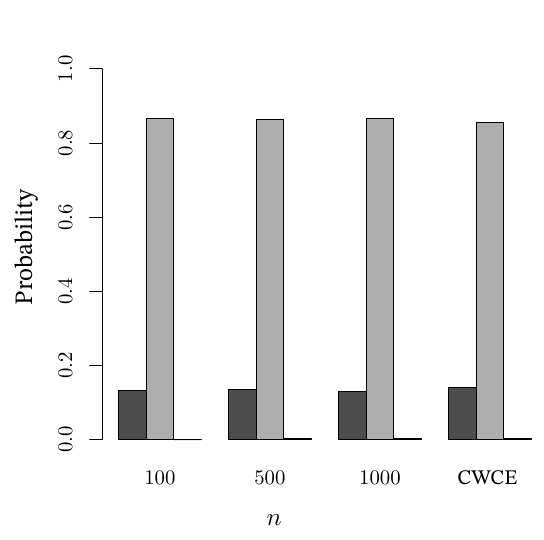}}
			\caption{}\label{CH4SM3fig4a}	
		\end{subfigure}
		\begin{subfigure}{.32\textwidth}
			\resizebox{1\textwidth}{!}{\includegraphics[page=2]{TikZFig/SM3Fig4.pdf}}
			\caption{}\label{CH4SM3fig4b}	
		\end{subfigure}
		\begin{subfigure}{.32\textwidth}
			\resizebox{1\textwidth}{!}{\includegraphics[page=3]{TikZFig/SM3Fig4.pdf}}
			\caption{}\label{CH4SM3fig4c}	
		\end{subfigure}
		\captionsetup{width=\textwidth}
		\caption{\footnotesize{Estimated CWCE of $\overline{a}{=}\overline{1}$ distribution of the effect on the third observation for an individual with an ICE equal to 0, given $3$ (a), $10$ (b) and $100$ (c) repeated measurements, for a varying sample size ($n \in \{100, 500, 1000\})$. Moreover, the CWCE distributions are presented for reference, which are the same as presented in Figure \ref{CH4SM3fig3b}.}}\label{CH4SM3fig4}
	\end{figure}

\begin{figure}[H]
			\centering
			\resizebox{\textwidth}{!}{\includegraphics{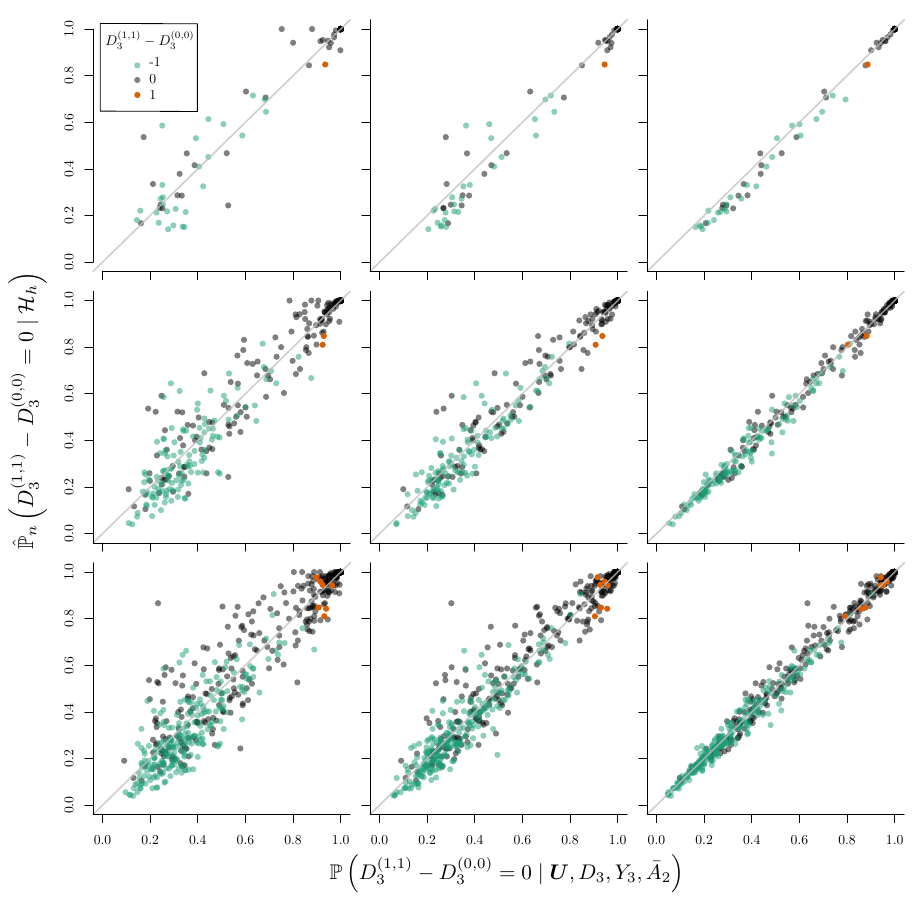}}
		\captionsetup{width=\textwidth}
		\caption{\footnotesize{The $\mathbb{P}\left(D_{3}^{(1,1)}-D_{3}^{(0,0)}~{=}~0 \mid \boldsymbol{U}, D_{3}, Y_{3}, \overline{A}_{2}\right)$, versus the estimated $\hat{\mathbb{P}}_{n}\left(D_{3}^{(1,1)},D_{3}^{(0,0)}~{=}~0 \mid \mathcal{H}_{h}\right)$ based on different subsets of the data. The rows correspond to the sample sizes $(100, 500, 1000)$ and the columns to the number of repeated measurements $(3,10,100)$.}}\label{CH4SM3fig5}
	\end{figure}

	Inference based on data generated by the logistic linear mixed assignment is again based on the REML model fit of $Y_{j}$ as discussed in Section \ref{CH4Ex1I} of the main text. Furthermore, to estimate the parameters $\gamma_{0}, \gamma_{1}$ and $\sigma_{2}$ of the data-generating mechanism, we fit a generalized linear mixed model (GLMM) for the outcomes $D_{1}, D_{2}$ up to $D_{h}$,
	\begin{equation}\label{CH4model:latentglm}
	\text{logit}(\mathbb{P}(D_{ji}~{=}~1)) ~{=}~ Z_{3i}+\beta_{D}+\beta_{Y}Y_{ji},
	\end{equation} where $Z_{3i}\sim \mathcal{N}(0,\tau_{3}^2)$. maximum-likelihood estimation based on a Laplace approximation of the parameters of \eqref{CH4model:latentglm} was performed in \texttt{SAS} using \texttt{PROC GLIMMIX}. 
 For this GLMM, maximum-likelihood estimation based on a Laplace approximation gives rise to consistent estimates of $\gamma_{0}, \gamma_{1}$ and $\sigma_{3}$ for $h\rightarrow \infty$ \citep{Vonesh1996}, for which case Theorem \ref{CH4th41} applies. 
	We have estimated the distribution of the CWCE at the time of the third repeat $(j~{=}~3)$ for all individuals in subsets of the simulated data. For the individual with an ICE equal to $0$ that was already highlighted in Figure \ref{CH4SM3fig3b}, the estimated CWCE for different $h$ and a varying number of individuals in the sample is presented in Figure \ref{CH4SM3fig4}. For the parameter choices in this example, we can conclude that the asymptotic approximation of Theorem \ref{CH4th41} works very well already for $n~{=}~100$ (and $h~{=}~3$).
	 \noindent In Figure \ref{CH4SM3fig5} we have plotted the $\mathbb{P}(D_{3}^{(1,1)}-D_{3}^{(0,0)}~{=}~0 \mid \boldsymbol{U}, D_{3}, Y_{3}, \overline{A}_{2})$ versus the estimated $\hat{\mathbb{P}}_{n}(D_{3}^{\overline{a}}-D_{3}^{\overline{0}}~{=}~0 \mid \mathcal{H}_{h})$ for each individual for varying $n$ and $h$. The estimated probability is more precise for an increased number of repeats as the receptiveness factors $(U_{1}, U_{2}, U_{3})$ can be estimated better.  As discussed before (see, e.g.~Figure \ref{CH4SM3fig3c}), some of the individuals have an ICE that differs from the mode of $\mathbb{P}(D_{3}^{(1,1)}-D_{3}^{(0,0)}~{=}~0 \mid \boldsymbol{U}, D_{3}, Y_{3}, \overline{A}_{2})$. Considering $100$, $500$ and $1000$ of the simulated individuals, this applies to $21 \%$, $12.8 \%$ and $12.7 \%$ of the individuals, respectively. 
 In Table \ref{CH4SM3tab1}, the total error of the classification using the mode of the estimated CWCE as ICE estimator, is presented for this simulated dataset for all settings considered. 

 \begin{table}[H]
		\caption{Fraction of individuals where the mode of the estimated CWCE distribution deviates from their ICE of $\overline{a}{=}\overline{1}$.}\label{CH4SM3tab1}
		\centering
		\resizebox{0.4\linewidth}{!}{\begin{tabular}{c||ccc}
			\backslashbox{$n$}{$h$} & 3      & 10     & 100    \\ \hline
			100   & 17.0\%   & 19.0\%   & 20.0\%   \\
			500   & 12.0\%   & 13.8\% & 14.0\%   \\
			1000  & 13.4\% & 13.6\% & 13.3\%
		\end{tabular}}%
	\end{table}

\end{document}